\documentclass[aps,prd,twocolumn,floatfix,superscriptaddress,nofootinbib]{revtex4-1}

\newcommand\mdp{m_{A^\prime}}
\newcommand\Adp{A^\prime}
\newcommand\mdm{m_\chi}

\usepackage{slashed}
\usepackage{amsmath}
\usepackage{graphicx,bm}
\usepackage{enumitem}
\usepackage{hhline}
\usepackage{multirow}
\usepackage{subcaption}
\usepackage[utf8]{inputenc} 
\usepackage[dvipsnames]{xcolor}
\usepackage[colorlinks,citecolor=blue]{hyperref}
\begin{document}

\title{Supernova constraint on self-interacting dark sector particles}

\author{Allan Sung}
\email{as111@princeton.edu}
\affiliation{Institute of Physics, Academia Sinica, 
  Taipei, 11529, Taiwan}
\affiliation{Department of Physics, Princeton University, Princeton, New Jersey 08544, USA}

\author{Gang Guo}
\email{gangg23@gmail.com}
\affiliation{Institute of Physics, Academia Sinica, 
  Taipei, 11529, Taiwan}
\affiliation{Key Laboratory of Dark Matter and Space Astronomy,
Purple Mountain Observatory, Chinese Academy of Sciences, Nanjing 210033, China}

\author{Meng-Ru Wu}
\email{mwu@gate.sinica.edu.tw}
\affiliation{Institute of Physics, Academia Sinica, 
  Taipei, 11529, Taiwan}
\affiliation{Institute of Astronomy and Astrophysics, Academia Sinica, 
  Taipei, 10617, Taiwan}

\begin{abstract}
We examine the constraints on sub-GeV dark sector particles set by the proto-neutron star cooling associated with the core-collapse supernova event SN1987a.
Considering explicitly a dark photon portal dark sector model, we compute the relevant interaction rates of dark photon ($A'$) and dark fermion ($\chi$) with the Standard Model particles as well as their self-interaction inside the dark sector. 
We find that even with a small dark sector fine structure constant $\alpha_D\ll 1$, dark sector self-interactions can easily lead to their own self-trapping. 
This effect strongly limits the energy luminosity carried away by dark sector particles from the supernova core and thus drastically affects the parameter space that can be constrained by SN1987a.
We consider specifically two mass ratios $m_{A'}=3m_\chi$ and $3m_{A'}=m_\chi$ which represent scenarios where the decay of $A'$ to $\chi\bar\chi$ is allowed or not.
For $m_{A'}=3m_\chi$, we show that this effect can completely evade the supernova bounds on widely-examined dark photon parameter space for a dark sector with $\alpha_D\gtrsim 10^{-7}$. 
In particular, for the mass range $m_\chi\lesssim 20$~MeV, supernova bounds can only be applied to weakly self-interacting dark sector with $\alpha_D\lesssim 10^{-15}$. 
For $3m_{A'}=m_\chi$, bounds in regions where $\alpha_D\gtrsim 10^{-7}$ for $m_\chi\lesssim 20$~MeV can be evaded similarly.
Our findings thus imply that the existing supernova bounds on light dark particles can be generally eluded by a similar self-trapping mechanism. 
This also implies that nonstandard strongly self-interacting neutrino is not consistent with the SN1987a observation. 
Same effects can also take place for other known stellar bounds on dark sector particles. 
\end{abstract}

\date{\today}
\maketitle


\section{Introduction}\label{sec:intro}

The detection of 20 electron antineutrinos emitted from the core-collapse supernova (SN) explosion event, SN1987a~\cite{Hirata:1987hu,Bionta:1987qt,Alekseev:1988gp}, not only broadly confirmed the prevalent SN theory, but also led to several important consequences to fundamental physics, including, e.g., bounds on the neutrino decay lifetime, the absolute masses of neutrinos, and nonstandard neutrino interactions~\cite{Sato:1987rd,Spergel:1987ch,Bahcall:1987,Burrows:1987,Schramm:1990pf,Loredo:2001rx,Frieman:1987as,Kolb:1988pe,Berezhiani:1989za,Farzan:2002wx}.
In particular, important constraints on a variety of particles beyond the Standard Model (SM) including the axion, sterile neutrino, dark photon, etc., whose masses are sub-GeV, were derived~\cite{Raffelt:1987yt,Turner:1987by,Mayle:1987as,Brinkmann:1988vi,Janka:1995ir,Keil:1996ju,Fischer:2016cyd,Carenza:2019pxu,Bollig:2020xdr,Lucente:2020whw,Raffelt:2011nc,Arguelles:2016uwb,Suliga:2019bsq,Syvolap:2019dat,Mastrototaro:2019vug,Suliga:2020vpz,Dent:2012mx,Rrapaj:2015wgs,Mahoney:2017jqk,Chang:2016ntp,Hardy:2016kme,Chang:2018rso,Guha:2015kka,Guha:2018mli,Ishizuka:1989ts,Arndt:2002yg,Keung:2013mfa,Tu:2017dhl,Dev:2020eam,Croon:2020lrf,Camalich:2020wac,Farzan:2002wx,Hanhart:2000er,Hanhart:2001fx,Hannestad:2007ys,Freitas:2007ip}, which complement ongoing experimental searches for those particles.
These constraints were based on the requirement that the exotic particles should not carry away an amount of energy from the cooling proto-neutron star (PNS) more than the inferred total energy carried by neutrinos, $E_\nu \simeq 3\times 10^{53}$~erg (see, however, a caution from Ref.~\cite{Bar:2019ifz}).
In addition to the SN cooling (more precisely, the PNS cooling) constraint, recent works also proposed new constraints on light dark photon or dark photon portal light dark matter (DM) based on other SN-related observables, such as the measured SN explosion energy~\cite{Sung:2019xie}, the $\gamma$-rays~\cite{Kazanas:2014mca,DeRocco:2019njg}, or the produced (semi-)relativistic dark matter flux arriving at the terrestrial detectors~\cite{DeRocco:2019jti}.

One important aspect in deriving the SN constraint on light dark sector (DS) particles is that their interaction with SM particles cannot be too strong for them being trapped inside the PNS.
We note that previous studies always ignored the \emph{self-interactions} between dark sector particles when deriving the SN bounds. However, if the abundance of dark sector particles inside the SN core can be as large as SM particles, and if the self-interaction cross section can be as large as the neutrino-nucleon scattering cross section $\sim O(10^{-41})$~cm$^2$ for neutrinos of $\sim \mathcal{O}(10)$~MeV, dark sector particles can trap themselves inside the PNS.
Consequently, SN bounds on self-interacting dark sector particles can be largely evaded (see also a very recent study discussing SN bound on axion-like particle portal light DM \cite{Darme:2020sjf}).
Note that self-interacting dark matter has been considered to be a viable option to 
resolve a number of tensions in the scale of galaxies or galaxy clusters, e.g., the core-cusp, too-big-to-fail, and the missing satellites problems; see, e.g., Refs.~\cite{Spergel:1999mh,Vogelsberger2012,Rocha:2012jg,Tulin:2017ara}.
Extensive efforts investigating consequences of self-interacting or annihilating dark matter on various cosmological and astrophysical signatures have been pursued in recent years, e.g., Refs.~\cite{Tulin:2013teo,Bernal:2015ova,Kawasaki:2015yya,Bringmann:2016din,Elor:2015bho,Ackermann:2015tah,Abdallah:2016ygi,Feng:2016ijc,Leane:2017vag,Kouvaris:2016ltf,Essig:2018pzq,Chang:2018bgx,Depta:2019lbe,Bernal:2019uqr,Foot:2014uba}.

In this work, we aim to address the issue of SN constraints on self-interacting dark sector particles by taking into account their self-trapping effect in a systematic way for the first time.
We use a widely examined dark photon portal dark sector model explicitly to compute all the relevant interaction cross sections and the decay rates.
In principle, to determine precisely the dark sector particle fluxes emerging from the PNS requires solving full Boltzmann transport equations in a way similar to the neutrino transport problem in SNe (see e.g., a recent review~\cite{Mezzacappa:2020oyq} and references therein).\footnote{Reference~\cite{DeRocco:2019jti} adopted a Monte-Carlo based particle transport scheme to compute the light dark matter flux emitted from the PNS, without considering their potential self-interactions.}
Such approach demands intensive computational power to fully incorporate the scattering kernels and particle annihilation.
Instead of directly pursuing full numerical simulations, we adopt an approximated approach to estimate the energy fluxes carried 
by dark photons and dark fermions evaluated in the nondiffuse regime and diffuse regime separately, and formulate a physically motivated criterion to switch from one regime to another.
This approach allows us to estimate the effect of dark sector self-trapping on SN bounds for a wide range of parameter space, which turns out to be very important even for small couplings in the dark sector.

The rest of the paper is organized as follows.
In Sec.~\ref{sec:model}, we describe the underlying dark photon portal dark sector model, the considered SN model, and list the relevant interactions and decay processes that we included in this work. 
In Sec.~\ref{sec:lum}, we compute the energy luminosity of dark sector particles leaving the PNS in the nondiffuse and the diffuse regime, respectively, and formulate the criterion to switch from one regime to another.
We apply this method to derive SN bounds on self-interacting dark sector particles in Sec.~\ref{sec:res}.
Our conclusion and discussions of potential caveats as well as other implications are given in Sec.~\ref{sec:con}.
All detailed derivations of the cross sections, the decay rates, and the diffusion luminosity of dark sector particles are given in the Appendices.
We adopt natural units with $\hbar=c=1$ throughout the paper unless explicitly specified.

\section{Models}\label{sec:model}

\subsection{Dark sector model}

\begin{table}[]
    \centering
\begin{tabular}{|c|c|c|c|c|}
    \hline
    Mass & Interaction & Type & Particle & Coupling \\
    \hhline{|=|=|=|=|=|}
    \multirow{5}{6em}{$\mdp < 2\mdm$} & $\Adp np \rightarrow np$ & Abs. & SM & $\epsilon^2$ \\
    & $\Adp e^-\rightarrow e^- \gamma$ & Abs. & SM & $\epsilon^2$\\
    & $\Adp\rightarrow e^- e^+$ & Abs. & SM & $\epsilon^2$\\
    & $\Adp\Adp\rightarrow\chi\bar{\chi}$ & Abs. & DS & $\alpha_D^2$\\
    & $\Adp\chi\rightarrow\chi\Adp$ & Sca. & DS & $\alpha_D^2$\\
    \hline
    \multirow{4}{6em}{$\mdp > 2\mdm$} & $\Adp np \rightarrow np$ & Abs. & SM & $\epsilon^2$ \\
    & $\Adp e^-\rightarrow e^- \gamma$ & Abs. & SM & $\epsilon^2$\\
    & $\Adp\rightarrow e^- e^+$ & Abs. & SM & $\epsilon^2$\\
    & $\Adp\rightarrow\chi\bar{\chi}$ & Abs. & DS & $\alpha_D$\\
    \hline
\end{tabular}
    \caption{Relevant processes of dark photon interactions considered in this work.
    ``Abs.'' refers to a process which absorbs dark photon(s) or decay of dark photon.
    ``Sca.'' refers to a scattering process of a dark photon with a Standard Model (SM) particle or a dark sector (DS) particle.
    Only leading-order processes are included here.
    Note that for $\mdp>2\mdm$, $\Adp\rightarrow\chi\bar{\chi}$ also accounts for contribution from $\Adp\chi\rightarrow\Adp\chi$ [see Eq.~\eqref{eq:A17} and text below for details].
    }
    \label{tab:dp_imfp}
\end{table}

\begin{table}[]
    \centering
\begin{center}\begin{tabular}{|c|c|c|c|c|}
    \hline
    Mass & Interaction & Type & Particle & Coupling \\
    \hhline{|=|=|=|=|=|}
    \hline
    \multirow{9}{6em}{$\mdp < 2\mdm$} & $\chi\bar{\chi}np \rightarrow np$ & Abs. & SM & $\epsilon^2 \alpha_D$ \\
    & $\chi\bar{\chi} \rightarrow e^- e^+$ & Abs. & SM & $\epsilon^2 \alpha_D$\\
    & $\chi\bar{\chi} \rightarrow \gamma^{\ast}$ & Abs. & SM & $\epsilon^2 \alpha_D$\\
    & $\chi p \rightarrow \chi p$ & Sca. & SM & $\epsilon^2 \alpha_D$\\
    & $\chi e^- \rightarrow \chi e^-$ & Sca. & SM & $\epsilon^2 \alpha_D$\\
    & $\chi\bar{\chi} \rightarrow \Adp\Adp$ & Abs. & DS & $\alpha_D^2$\\
    & $\chi\chi \rightarrow \chi\chi$ & Sca. & DS & $\alpha_D^2$\\
    & $\chi\bar{\chi} \rightarrow \chi\bar{\chi}$ & Sca. & DS & $\alpha_D^2$\\
    & $\chi\Adp \rightarrow \Adp\chi$ & Sca. & DS & $\alpha_D^2$\\
    \hline
    \multirow{3}{6em}{$\mdp > 2\mdm$} & $\chi p \rightarrow \chi p$ & Sca. & SM & $\epsilon^2 \alpha_D$\\
    & $\chi e^- \rightarrow \chi e^-$ & Sca. & SM & $\epsilon^2 \alpha_D$\\
    & $\chi\bar{\chi} \rightarrow \Adp$ & Abs. & DS & $\alpha_D$\\
    \hline
\end{tabular}\end{center}
    \caption{Relevant processes of dark fermion interactions. Notations are the same as in Table~\ref{tab:dp_imfp}.
    Only leading-order processes are included here.
    Note that for $\mdp>2\mdm$, $\chi\bar{\chi}\rightarrow\Adp$ also accounts for DS processes of $\Adp\chi\rightarrow\Adp\chi$ and $\chi\bar\chi\rightarrow\chi\bar\chi$, as well as SM processes involving a pair of dark fermions [see Eq.~\eqref{eq:A17} and text below for details].
    }
    \label{tab:df_imfp}
\end{table}

We consider a dark photon portal DS model wherein the Dirac dark fermion $\chi$ couples to dark photon $\Adp$ and the dark photon kinetically mixes with the SM photon~\cite{Holdom:1985ag,Okun:1982xi,Pospelov:2007mp}.  
The corresponding Lagrangian of the dark sector is given by
\begin{equation}\begin{split}
    \mathcal{L} &\supset -\frac{1}{4}F^\prime_{\mu\nu}F^{\prime\mu\nu}-\frac{\epsilon}{2}F^\prime_{\mu\nu}F^{\mu\nu} + \frac{1}{2} \mdp^2 A^\prime_{\mu} A^{\prime\mu}\\
    & + \bar{\chi}\left( i\slashed{\partial} - m_{\chi} \right) \chi + g_D \bar{\chi}\slashed{\Adp}\chi,
\end{split}\end{equation}
where $\epsilon$ is the mixing parameter, $\mdp$ is the dark photon mass, $\mdm$ is the mass of the dark fermion, and $g_D$ is the DS coupling constant. We define the DS fine structure constant $\alpha_D \equiv g_D^2/4\pi$ analogous to the electromagnetic fine structure constant $\alpha_e \equiv e^2/4\pi$.

Through the mixing of dark photon with the SM photon, dark photons and dark fermions can be produced via processes analogous to the SM electromagnetic ones.
We follow Refs.~\cite{Chang:2016ntp,Chang:2018rso} to consider the following production
channels for light dark photons and dark fermions inside the hot and dense interior of a PNS with a temperature $\simeq 30$~MeV and a core mass density $\gtrsim 10^{14}$~g~cm$^{-3}$. 
For the dark photon, we include the nucleon-nucleon bremsstrahlung $np \rightarrow np\Adp$, Compton-like interaction $\gamma e^- \rightarrow e^- \Adp$, and electron-positron annihilation $e^- e^+ \rightarrow \Adp$. 
For the dark fermion, we consider three pair-production channels including nucleon-nucleon bremsstrahlung $np \rightarrow np\chi\bar{\chi}$, electron-positron annihilation $e^- e^+ \rightarrow \chi\bar{\chi}$, and plasmon decay $\gamma^\ast \rightarrow \chi\bar{\chi}$.

When the PNS interior is optically thin to dark photons or dark fermions, the rates of the above production channels directly determine the energy luminosity carried away by the dark particles. 
However, when the interactions between dark particles and the SM medium, as well as the self-interactions in the DS, are strong enough, dark photons and fermions can be trapped in the PNS. 
These interactions include the inverse processes of the above production channels, 
the dark photon (fermion) pair-annihilation $\Adp\Adp\leftrightarrow\chi\bar{\chi}$, the DS Compton scattering $\Adp\chi\leftrightarrow\chi\Adp$,
the scattering of dark fermions $\chi\chi \rightarrow \chi\chi$ and $\chi\bar\chi \rightarrow \chi\bar\chi$,
as well as the dark fermion scattering with SM charged fermions $\chi p \rightarrow \chi p$ and $\chi e^- \rightarrow \chi e^-$.
These interactions determine the mean free path of the DS particles and thus the energy loss rate through the neutrinosphere in the diffuse regime.
Moreover, when $\mdp > 2\mdm$, the (inverse) decay of dark photon $\Adp\leftrightarrow\chi\bar{\chi}$ needs to be considered. 
In this scenario, the inverse decay process dominates all the dark fermion pair-absorption processes and the DS self-interactions due to the resonances.\footnote{$\chi\bar{\chi}\rightarrow\chi\bar{\chi}$ has a dark photon resonance and $\Adp\chi\rightarrow\Adp\chi$ has a dark fermion resonance.; 
see Eq.~\eqref{eq:A17} and text below in the Appendix~\ref{app:nwa} for details.}
Other DS self-interactions without resonances are suppressed by an extra factor of $\alpha_D$ compared to the (inverse) decay rate and thus can be neglected for $\mdp > 2\mdm$.
We list all the included interactions in this work in Tables~\ref{tab:dp_imfp} and~\ref{tab:df_imfp} for 
dark photon and dark fermion, respectively.

Throughout this work, we will use perturbative calculation for the interaction rates. 
Since the temperature in the PNS is $\sim30$ MeV, we only consider DS particles of masses below $\simeq 1$~GeV.

\begin{figure*}[htb]
    \centering
    \begin{subfigure}[b]{0.45\textwidth}
        \centering
        \includegraphics[width=\textwidth]{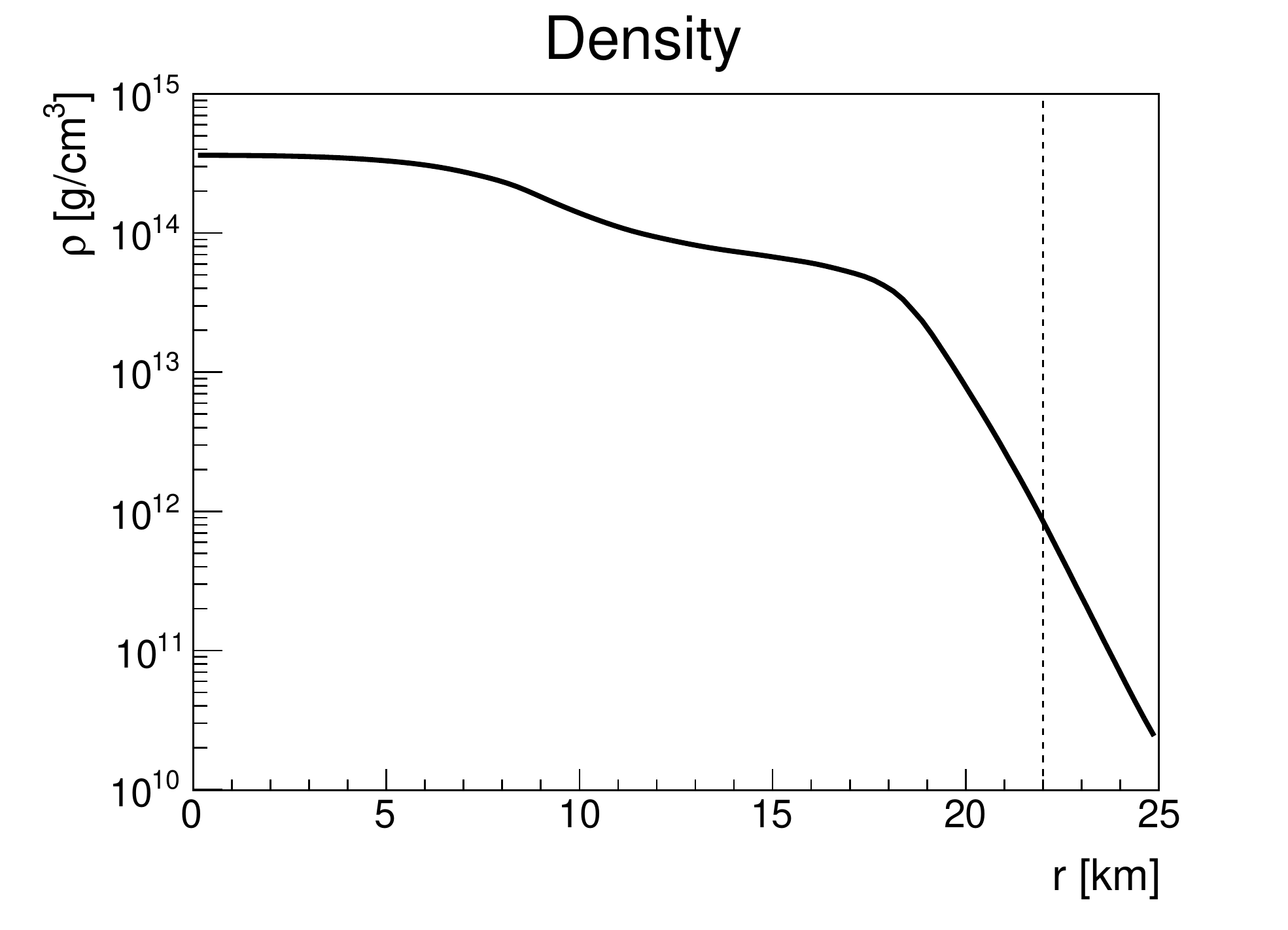}
    \end{subfigure}%
    \begin{subfigure}[b]{0.45\textwidth}
        \centering
        \includegraphics[width=\textwidth]{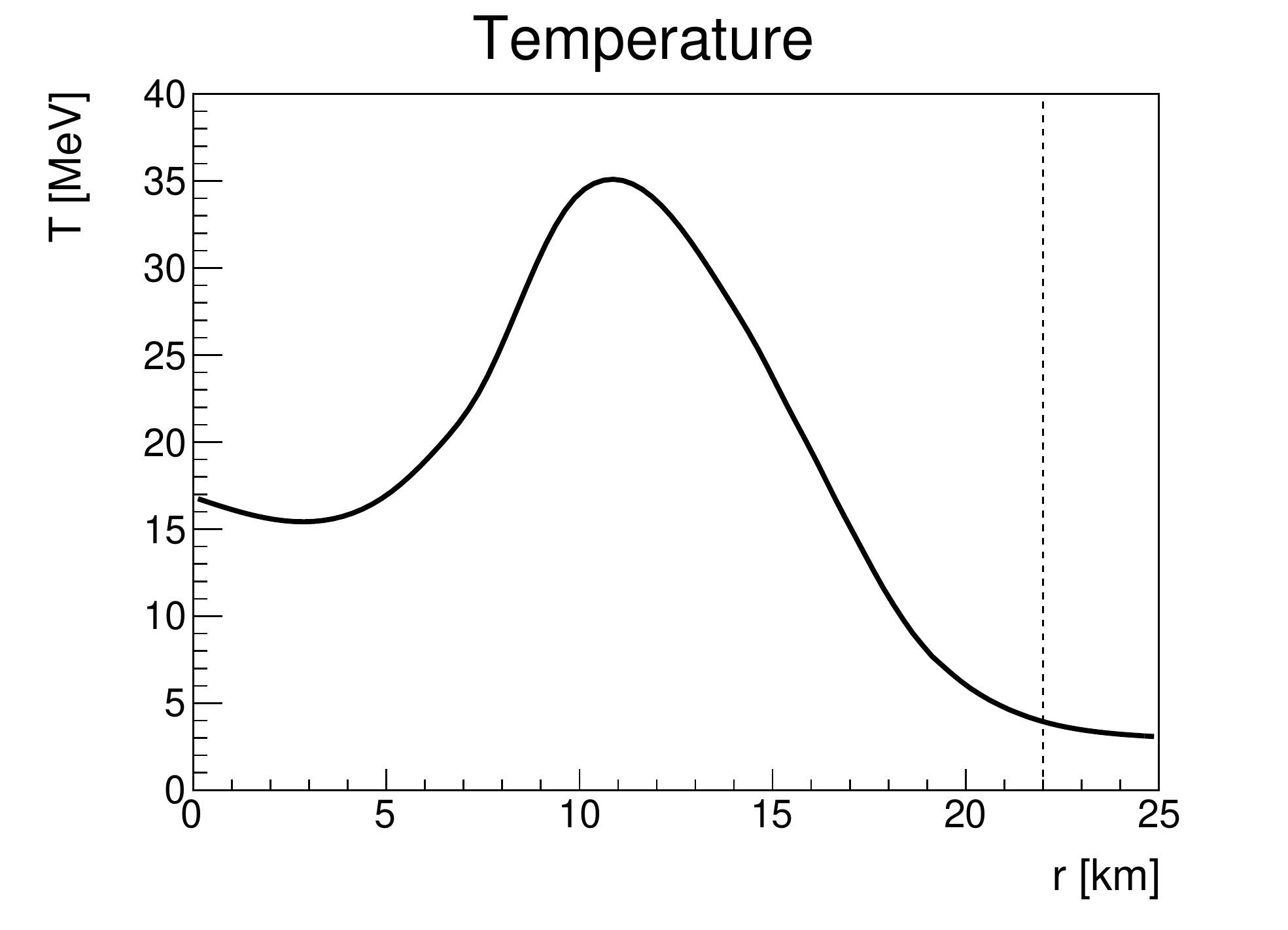}
    \end{subfigure}
    \begin{subfigure}[b]{0.45\textwidth}
        \centering
        \includegraphics[width=\textwidth]{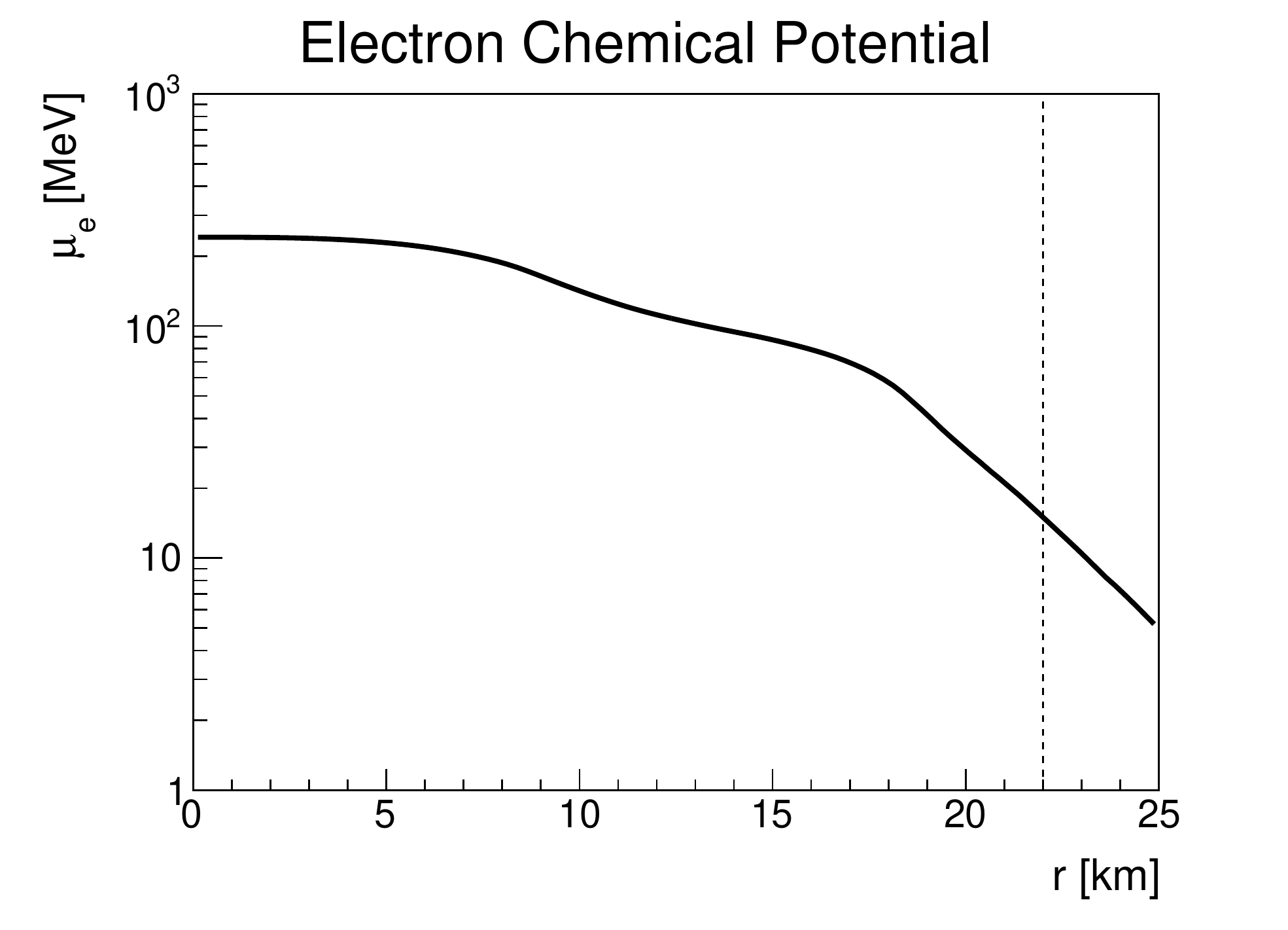}
    \end{subfigure}%
    \begin{subfigure}[b]{0.45\textwidth}
        \centering
        \includegraphics[width=\textwidth]{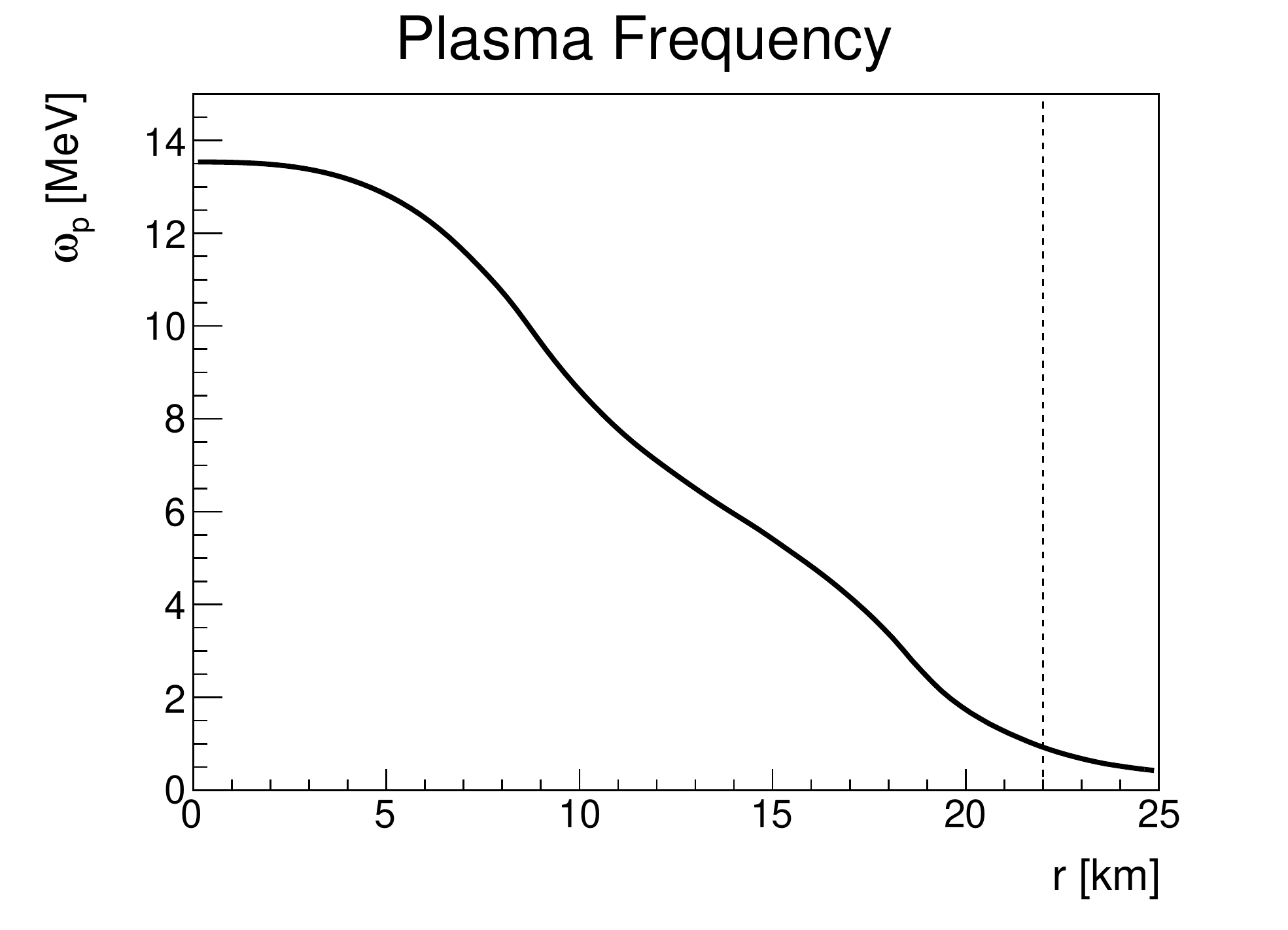}
    \end{subfigure}
    \caption{Supernova profiles (density, temperature, electron chemical potential, and plasma frequency) used in this work from the $18$~$M_\odot$ progenitor model in Ref.~\cite{Fischer:2009af}, extracted at the time 1~s post the core bounce.
    The vertical dash line indicates where the neutrinosphere is located.}
    \label{fig:SN_prof}
\end{figure*}

\subsection{PNS cooling constraint and SN model}
Without the presence of dark particles, long-term SN simulations predicted that 
the PNS cools by emitting neutrinos of all flavors in $\sim 10$~s.
The total energy carried away by neutrinos is $\simeq 3\times 10^{53}$~erg, which is fixed by the available gravitational energy released to form a compact neutron star.
The potential emission of exotic particles that may be produced inside the PNS will thus reduce the duration of the neutrino emission and can be constrained by the
observed neutrino events from SN1987a.
Based on comparisons with simulations including the emission of axions from the PNS, 
a well-known criterion (Raffelt's criterion) was formulated to constrain the
maximal energy luminosity carried by any exotic particles~\cite{Raffelt:1990yz},
\begin{equation}\label{eq:L_crit}
    L_D\leq L_\nu\equiv 3\times 10^{52}~{\rm erg~s}^{-1},
\end{equation}
where $L_D$ denotes the energy luminosity of dark emission, evaluated during the early PNS cooling phase, and $L_\nu$ is approximately the time-averaged neutrino energy luminosity in SM.

In this work, we use the PNS density and temperature profile at 1~s post the SN core bounce obtained in Ref.~\cite{Fischer:2009af} to compute the emission of dark photons and dark fermions.
This PNS profile was widely used recently for similar purposes (see e.g., Refs.~\cite{Chang:2016ntp,Sung:2019xie}).
Note that the choice of a particular SN model may introduce uncertainties of a factor of a few for the derived bounds~\cite{Chang:2016ntp,Mahoney:2017jqk}.
Figure~\ref{fig:SN_prof} shows the radial evolution of the density $\rho$, temperature $T$, the electron chemical potential $\mu_e$, and the plasma frequency $\omega_p$ [see Eq.~\eqref{eq:plasmafreq}].
The position of the spectral-averaged neutrino decoupling sphere, i.e., neutrinosphere, is indicated by the vertical dash line at $R_\nu\simeq 22$~km.
The density profile shows a monotonically decreasing behavior as a function of radius, while the temperature profile exhibits a peak at $r\simeq 11$~km, due to the inefficient compression heating at the densest core region.
The electrons are highly degenerate inside the PNS.
The plasma frequency $\omega_p\simeq 14$~MeV at the PNS center and decreases at larger radii.
The plasma effect effectively alters the mixing between the dark photon and the SM photon differently for the transverse and the longitudinal polarizations~\cite{An:2013yfc,Hardy:2016kme,Chang:2016ntp}.
We have included this effect throughout this work and give the details in Appendix~\ref{app:plas}.

\section{Luminosity of dark sector particles}\label{sec:lum}

In this section, we first describe how we compute the energy luminosity of DS particles leaving the PNS for the scenario where the DS self-trapping can be ignored (Sec.~\ref{subsec:free}) and for cases where they can be considered as diffusive due to self-trapping (Sec.~\ref{subsec:trap}).
We formulate the criteria that determine if a DS particle species is in diffuse regime or not in Sec.~\ref{subsec:crit}. 
In Sec.~\ref{subsec:num}, we then apply our formalism to the adopted PNS profile to compute the total luminosity carried away by DS particles from the PNS interior.

In the rest of the paper, we denote the 4-momentum of $\Adp$ by $k = (\omega, \vec{k})$, that of $\chi$ by $p = (E, \vec{p})$ and that of $\bar{\chi}$ by $p^\prime = (E^\prime, \vec{p}^\prime)$, unless noted otherwise.

\subsection{Nondiffuse regime}\label{subsec:free}
In the nondiffuse regime, we consider the bulk emission rates of DS particles inside the neutrinosphere and the attenuation due to absorption and decay, following Refs.~\cite{Chang:2016ntp,Sung:2019xie}. 
The luminosity of the dark photon is given by
\begin{equation}\begin{split}\label{eq:LA'emis}
    L_{\Adp} &= \sum_{L,T} \int_0^{R_\nu} 4\pi r^2 dr \int \frac{d^3 k}{(2\pi)^3}\\
    &\quad \times g_{L,T} \omega \Gamma^{L,T}_{\Adp, \rm prod}(\omega, r) e^{-\tau_{L,T}(\omega, r)},
\end{split}\end{equation}
where $g_L = 1$, $g_T = 2$, and $\Gamma^{L,T}_{\Adp, \rm prod}$ and $\tau_{L,T}$ are the production rate and optical depth respectively. 
The exponential factor accounts for the absorption of dark photons by the medium and their decay. 
We separate the longitudinal (L) and transverse (T) modes because the medium effect modifies their dispersion relations and leads to different effective mixings of the two modes with the SM photon (see Appendix~\ref{app:plas}).
The production rate $\Gamma^{L,T}_{\Adp, \rm prod}$ is determined by the interactions involving the SM particles listed in Table~\ref{tab:dp_imfp}. 
By detailed balance, the rates of each production process and its inverse process are related by $\Gamma^{L,T}_{\Adp, \rm prod} = e^{-\omega/T} \Gamma^{L,T}_{\Adp, \rm abs}$, where $\Gamma^{L,T}_{\Adp, \rm abs}$ is the total absorption rate of the inverse process. 
Therefore,
\begin{equation}
    \Gamma^{L,T}_{\Adp, \rm prod} = e^{-\omega/T} (\Gamma^{L,T}_{\Adp np \rightarrow np} + \Gamma^{L,T}_{\Adp e^- \rightarrow e^- \gamma} + \Gamma^{L,T}_{\Adp \rightarrow e^- e^+}).
\end{equation}

For the optical depth, we include the absorption and decay of $A'$ by the same processes with SM particle as above and ignore those involving DS particles, to avoid double counting the total DS luminosity (see also later discussions in this subsection).\footnote{The pair-annihilation rate of $\Adp\Adp\rightarrow\chi\bar{\chi}$ is also ignored since the dark photon abundance inside the PNS is relatively small compared to that of SM particles in the nondiffuse regime.}
This gives 
\begin{equation}\label{eq:tau}
    \tau_{L,T}(\omega, r) = f(r) \int_r^{R_\nu} \frac{d\tilde{r}}{v} \Gamma_{\Adp, \rm abs}^{L,T}(\omega, \tilde{r}),
\end{equation}
where $v=|\vec{k}|/\omega$ is the dark photon velocity, and $f(r)$ is a geometric factor used in \cite{Chang:2016ntp} that effectively takes into account different path lengths of dark photons emitting locations to the neutrinosphere.
The explicit forms of these absorption rates are given in Appendix~\ref{app:int}.

Next, we compute the luminosity of the dark fermion ($\chi$) in the nondiffuse regime as
\begin{equation}\label{eq:Lchiemis}
    L_\chi = \int_0^{R_\nu} 4\pi r^2 dr \int \frac{d^3 p}{(2\pi)^3} g_\chi E \, \Gamma_{\chi, \rm prod}(E, r),
\end{equation}
where $g_\chi = 2$ is the physical degrees of freedom of $\chi$, and $\Gamma_{\chi, \rm prod}$ is the production rate of $\chi$ by the SM medium (see Table~\ref{tab:df_imfp}). 
Note that since $\chi$ and $\bar{\chi}$ are symmetric in our model, the total luminosity in the dark fermion pair is $L_\chi + L_{\bar{\chi}} = 2L_\chi$. 
Here we do not include the attenuation due to pair absorption. 
This is because in the nondiffuse regime, the dark fermion abundance below the neutrinosphere is very low, which suppresses the pair absorption of dark fermions.

If the dark fermions were in equilibrium with the medium, detailed balance could relate the production rates to the absorption rates by $\Gamma^{\rm eq}_{\chi, \rm prod} = e^{-E/T}\Gamma^{\rm eq}_{\chi, \rm abs}$. 
We take the equilibrium production rates $\Gamma^{\rm eq}_{\chi, \rm prod}$ as an approximation for the production rates $\Gamma_{\chi, \rm prod}$ used in Eq.~\eqref{eq:Lchiemis}. That is,
\begin{equation}
    \Gamma_{\chi, \rm prod} \simeq e^{-E/T} (\Gamma_{\chi\bar{\chi}np \rightarrow np} + \Gamma_{\chi\bar{\chi}\rightarrow e^- e^+} + \Gamma_{\chi\bar{\chi}\rightarrow \gamma^\ast}).
\end{equation}
This approximation in principle underestimates a bit the production rates of $\chi$ and $\bar\chi$ due to the assumed equilibrium occupation number, which effectively results in Pauli blocking. 
However, since the $\chi$ and $\bar\chi$ are always produced in pairs, the effective Pauli-blocking suppression of the rates is small due to their zero chemical potentials.

We note that when $\mdp \geq 2\mdm$, the dark fermion pair production is in fact dominated by the decay of on-shell dark photons.
Since in Eq.~\eqref{eq:LA'emis} we do not include the decay of $A'$ to $\chi$ and $\bar\chi$ when $A'$ are in nondiffuse regime, it leads to double counting of the total dark luminosity if we also include the dark fermion production.
Thus, we do not consider the contribution of Eq.~\eqref{eq:Lchiemis} when dark photons are in nondiffuse regime and when $\mdp \geq 2\mdm$.
We also note here that if dark photons are in the diffuse limit but dark fermions are not, then effectively the non-diffuse dark fermion luminosity 
can be determined by the decay rate of the trapped dark photons that are in thermal equilibrium with the SM medium
(see Appendix~\ref{app:nwa}).\footnote{Note that we consider separately the diffuse condition for the longitudinal and the transverse dark photons.
Thus, we further multiply the nondiffuse dark fermion luminosity by a factor of $1/3$ ($2/3$) if only the longitudinal (transverse) mode of the dark photons are trapped (see Appendix~\ref{app:scheme}).} 

Before we discuss the detailed numerical results of the DS luminosity in the nondiffuse regime, let us provide an analytic estimation for the relevant region of
the parameter space.
First, for most cases,
the production rates of longitudinal dark photons and dark fermions are suppressed by a factor of $\mdp^2/\omega^2$ (see Appendix~\ref{app:int}) and by the coupling constant $\alpha_D$, respectively, compared to the production rate of the transverse dark photons. 
We can thus approximate the nondiffuse luminosity of DS particles by considering transverse dark photons only.
Second, we consider for simplicity a homogeneous PNS with radius $R_c\simeq10\mbox{ km}$, temperature $T\simeq30\mbox{ MeV}$, density $\rho\simeq3\times10^{14}\mbox{ g}/\mbox{cm}^3$ and electron fraction $Y_e\simeq0.3$. 
We also assume that dark photons are relativistic. 
Taking the nucleon-nucleon bremsstrahlung $\Adp np\rightarrow np$ without the plasma effects (see Appendix~\ref{app:Anpnp}), the luminosity of DS particles, $L_D$, is approximately
\begin{equation}
    L_D \simeq L_\nu \times \left(\frac{\epsilon}{4\times10^{-10}}\right)^2\times\exp\left(-\frac{\mdp}{29\mbox{ MeV}}\right)
\end{equation}
for $m_{A'}\lesssim 1$~GeV.

\subsection{Diffuse regime}\label{subsec:trap}

In the diffuse regime, we assume that DS particles are in good thermal contact with the SM medium. 
Due to the temperature gradient, the DS phase space distributions are slightly anisotropic, which induces an outward energy flux. 
We use the radiative transfer equation for the DS particle energy flux through the neutrinosphere in Appendix \ref{app:rad}. 
The energy flux of a particle species $i$ is approximately given by 
\begin{equation}\begin{split}\label{eq:flux}
    L_i &= -\frac{2g_i R_\nu^2 T_\nu^3}{3\pi} \left.\frac{dT}{dr}\right|_{R_\nu} \frac{1}{\langle\lambda^{-1}_i(R_\nu)\rangle}\\
    &\quad\times \int_{m_i/T_\nu}^{\infty} \xi^3\sqrt{\xi^2 - \left( \frac{m_i}{T_\nu}\right)^2}\frac{e^\xi}{\left( e^\xi \pm 1 \right)^2} d\xi,
\end{split}\end{equation}
where the upper (lower) sign is for fermions (bosons), $g_i$ is the physical degrees of freedom of particle $i$, $T_\nu$ is the temperature at the neutrinosphere, $m_i$ is the mass of particle $i$, and $\langle\lambda^{-1}_i(r)\rangle$ is the thermally averaged inverse mean free path (IMFP)\footnote{The IMFP used here is rescaled by the relative abundances of the DS particles. See Sec.~\ref{subsec:crit} for the definition.} of particle $i$ at radius $r$ defined by
\begin{equation}\label{eq:avg_imfp}
    \langle\lambda^{-1}_i(r)\rangle \equiv \frac{\int d^3 p \, f_i(E, T(r)) \lambda^{-1}_i(E, r)}{\int d^3 p \, f_i(E, T(r))},
\end{equation}
where $f_i(E, T(r))$ is the distribution function of particle $i$ at radius $r$. 
We distinguish between the absorptive and scattering IMFP, $\lambda_{i,abs}^{-1}$ and $\lambda_{i,sca}^{-1}$,
and define the total IMFP as 
\begin{equation}\begin{split}\label{eq:imfp}
    \lambda_i^{-1}(E, r) &\equiv \lambda_{i,abs}^{-1}(E, r)(1 \pm e^{-E/T(r)})\\
    &\quad+\lambda_{i,sca}^{-1}(E, r).
\end{split}\end{equation}
As in the nondiffuse regime, we compute the energy fluxes of the longitudinal and transverse dark photon separately, and assume that the energy fluxes of $\chi$ and $\bar\chi$ are equal. 
Hence, the total energy loss rate in the DS particles is $L_{tot} = L_{\Adp,L} + L_{\Adp,T} + 2L_\chi$. 
The IMFP calculations can be found in Appendix~\ref{app:int}.

We now estimate the relevant parameter space region in the limit where the DS self-interaction is the dominant opacity source.
When $\mdp>2\mdm$, the dominant DS self-interaction is the (inverse) decay $\Adp\leftrightarrow\chi\bar{\chi}$. 
Given $T_\nu\simeq3.9\mbox{ MeV}$ and the temperature gradient $|dT/dr|_{R_\nu}\simeq6.2\times10^{-4}\mbox{ MeV}/\mbox{m}$, if we fix $\mdp/\mdm = 3$, the luminosities of the DS particles can be fitted by
\begin{equation}\begin{split}\label{eq:est_dif_decay_A}
    L_{\Adp} &\simeq L_\nu \times \left(\frac{6.3\times10^{-12}}{\alpha_D}\right) \left(\frac{\rm MeV}{\mdp}\right)^2\\
    &\quad\times \exp\left[-\sqrt{49.0 + \left(\frac{\mdp}{4.3~\rm MeV}\right)^2}\right] 
\end{split}\end{equation}
for $\mdp \lesssim 25$~MeV, and
\begin{equation}\begin{split}\label{eq:est_dif_decay_x}
    L_\chi + L_{\bar{\chi}} &\simeq L_\nu \times \left(\frac{2.1\times10^{-13}}{\alpha_D}\right) \left(\frac{\rm MeV}{\mdm}\right)^{2}\\
    &\quad\times \exp\left[\left(\frac{\mdm}{1.9~\rm MeV}\right)-\sqrt{49.0 + \left(\frac{\mdm}{4.3~\rm MeV}\right)^2}\right]
\end{split}\end{equation}
for $\mdm \lesssim 30$~MeV. 
If we take $\mdp = 9$~MeV and $\mdm = 3$~MeV, then the total DS luminosity is
\begin{equation}\label{eq:est1}
    L_D \simeq L_\nu \times \left(\frac{1.5\times10^{-16}}{\alpha_D}\right).
\end{equation}

The corresponding $\chi\bar\chi\rightarrow A'$ cross section is\footnote{The estimated value here is the thermally averaged cross section evaluated at the center of PNS $\sigma_{\chi\bar{\chi}\rightarrow A'}\equiv\langle\lambda^{-1}_{\chi\bar{\chi}\rightarrow A'}(r=0)\rangle/n_\chi(r=0)$.}
\begin{equation}\label{eq:sigma_xxA}
    \sigma_{\chi\bar{\chi}\rightarrow A'} \simeq 4.7\times10^{-41}\mbox{ cm}^2 \left(\frac{\alpha_D}{1.5\times10^{-16}}\right).
\end{equation}

When $\mdp < 2\mdm$, 
the diffuse luminosities of dark photons and dark fermions depend on
the values of $\epsilon$ and $\alpha_D$. 
For regions where $\epsilon$ is small enough such that the dominating opacity source for dark photons is DS self-interaction $\Adp\Adp\rightarrow\chi\bar{\chi}$, 
one may estimate the DS luminosity by considering the dark photon luminosity only. 
If we fix $\mdp/\mdm = 1/3$, then the dark photon luminosity can be fitted by
\begin{equation}\begin{split}\label{eq:est_dif_ndecay}
    L_{\Adp} &\simeq L_\nu \times \left(\frac{2.0\times10^{-7}}{\alpha_D}\right)^2\\
    &\quad\times\exp\left[\left(\frac{\mdp}{0.74~\rm MeV}\right) - \sqrt{49.0 + \left(\frac{\mdp}{4.3~\rm MeV}\right)^2}\right]
\end{split}\end{equation}
for $\mdp\lesssim 25$~MeV. 
Adopting $\mdp = 3\mbox{ MeV}$ and $\mdm = 9\mbox{ MeV}$, this gives
\begin{equation}\label{eq:est2}
    L_D \simeq L_\nu \times \left(\frac{4.5\times10^{-8}}{\alpha_D}\right)^2.
\end{equation}
The corresponding $\chi\chi\rightarrow\chi\chi$  cross section is\footnote{The cross section is evaluated in the same way as Eq.~(\ref{eq:sigma_xxA}).}
\begin{equation}\label{eq:sigma_xxxx}
    \sigma_{\chi\chi\rightarrow\chi\chi} \simeq 9.7\times10^{-37}\mbox{ cm}^2 \left(\frac{\alpha_D}{4.5\times10^{-8}}\right)^2,
\end{equation}
and the corresponding $A'A'\rightarrow\chi\bar{\chi}$ cross section is
\begin{equation}\label{eq:sigma_AAxx}
    \sigma_{A'A'\rightarrow\chi\bar{\chi}} \simeq 3.1\times10^{-39}\mbox{ cm}^2 \left(\frac{\alpha_D}{4.5\times10^{-8}}\right)^2.
\end{equation}

These analytical formulas Eqs.~\eqref{eq:est1}--\eqref{eq:sigma_AAxx} already illustrate that DS particles can be self-trapped by interaction cross sections of $\gtrsim \mathcal{O}(10^{-40})$~cm$^2$ such that the corresponding diffuse DS luminosity $L_D\lesssim L_\nu$.
Only very weakly interacting DS particle with $\alpha_D$ much smaller than those nominal values in Eqs.~\eqref{eq:est1}--\eqref{eq:sigma_AAxx} can result in $L_D>L_\nu$.

\subsection{Diffuse criteria}\label{subsec:crit}

Whether DS particles are in the nondiffuse or in the diffuse regime sensitively depends on their abundances in the PNS. 
For example, if the dark fermion abundance is significant, the contribution of their pair-absorption processes
(e.g., $\chi\bar{\chi}np\rightarrow np$, $\chi\bar{\chi}\rightarrow e^- e^+$, and $\chi\bar{\chi}\rightarrow\gamma^\ast$)
to the dark fermion optical depth cannot be neglected. 
Furthermore, the DS abundances lead to significant DS self-interactions (e.g.~$\chi\chi\rightarrow\chi\chi$, $\chi\bar{\chi}\rightarrow \Adp\Adp$, $\chi\Adp\rightarrow\Adp\chi$,...) that can trap themselves in the PNS. 
These DS self-interactions delay the escape times for both the dark photon and the dark fermion, which enhances their abundances and optical depths. 
Therefore, the abundances of DS particles is critical in determining whether self-trapping is important. 
Below, we formulate the diffuse criteria in terms of the DS abundances and their IMFP at the neutrinosphere.

To obtain the exact DS abundances, detailed transport incorporating their production, scattering, and absorption needs be solved. Here we roughly estimate the abundance of each particle species by their production rate and a relevant timescale.
The abundance of a DS particle species $i$ is approximated as
\begin{equation}\begin{split}\label{eq:DSabun}
    N_i &= \int_0^{R_\nu} 4\pi r^2 dr \int \frac{d^3 p_i}{(2\pi)^3} g_i \Gamma_{i,\mathrm{prod}}\Delta t_i,\\
\end{split}\end{equation}
where the quantity $\Delta t_i$ is subject to the following considerations.
First, the longest possible time for DS particles to accumulate is the cooling time of the PNS. We set this upper bound of $\Delta t_i$ as $t_{\rm cool}=1$~s. 
Second, the shortest possible timescale, i.e., the lower bound of $\Delta t_i$, is estimated by the free-escaping timescale $t_{\rm free} \equiv R_\nu/v_i$. 
Another characteristic timescale here in the limit where a DS particle is trapped is the diffusion timescale $t_{i,\rm diff} \simeq R_\nu^2/D_i$ where $D_i \simeq v_i/\lambda^{-1}_i$ is the diffusion coefficient (see below for details).
For $t_{\rm free} < t_{i,\rm diff} < t_{\rm cool}$, we then take $t_{i,\rm diff}$ as $\Delta t_i$.
Combining these criteria, the relevant timescale $\Delta t_i$ used in Eq.~\eqref{eq:DSabun} is given by
\begin{equation}\label{eq:deltat}
    \Delta t_i =
    \begin{cases}
        t_{\rm free}, \,{\rm if}~ t_{i,\rm diff} \leq t_{\rm free},\\
        t_{i,\rm diff}, \,{\rm if}~ t_{\rm free} < t_{i,\rm diff} < t_{\rm cool},\\
        t_{\rm cool}, \,{\rm if}~ t_{\rm cool} \leq t_{i,\rm diff}.
    \end{cases}
\end{equation}

Since the PNS is not homogeneous, the diffusion of DS particles cannot be described by a constant $D_i$. 
As shown in Tables~\ref{tab:dp_imfp} and \ref{tab:df_imfp}, the interactions relevant to the IMFP of DS particles can be categorized as SM-type or DS-type by particles involved in the interactions. 
We choose particular locations in the PNS where the IMFPs of each type are maximal to roughly estimate their contribution to the diffusion timescale.\footnote{Overestimating the IMFP makes Eq.~(\ref{eq:trapcrit}) easier to satisfy.
Because the luminosity in the diffuse regime is generally lower than that in the nondiffuse regime for the same parameters, our estimation leads to a conservative bound.} 
For the SM-type interactions, we estimate the IMFP at the center ($r_{\rm SM} = 0~\mathrm{km}$) of the PNS where the density of SM particles is the largest. 
As for the DS-type interactions, we estimate the IMFP at $r_{\rm DS}=11\mathrm{km}$ where the temperature is the highest. 
This location has the largest DS particle density, if DS particles are in thermal equilibrium with the SM medium.
Knowing the IMFPs of both types at these respective locations, we can then compute the diffusion timescale

\begin{equation}\label{eq:tdiff}
    t_{i,\rm diff} \equiv \frac{R^2_\nu}{v_i} \left[ \langle\tilde{\lambda}^{-1}_{i,SM}(r_{SM})\rangle + \langle\tilde{\lambda}^{-1}_{i,DS}(r_{DS})\rangle \right]
\end{equation}
where $\tilde{\lambda}^{-1}_{i,SM}$ and $\tilde{\lambda}^{-1}_{i,DS}$ are SM-type and DS-type contribution to the IMFP of particle $i$.
When computing the IMFP contribution from DS particles, $\tilde{\lambda}^{-1}_{i,DS}$ in Eq.~\eqref{eq:tdiff}, we take the assumption that DS particles are in full thermal equilibrium with the SM medium. 
Combining Eqs.~\eqref{eq:DSabun}--\eqref{eq:tdiff} then allows us to compute the DS abundance $N_i$ for particle $i$.

After deriving $N_i$, we then define 
a scaling factor $\eta_i \equiv N_i/N_i^{\rm eq}$ for each dark particle species $i$ for $N_i<N_i^{\rm eq}$ ($\eta_i=1$ if $N_i\geq N_{\rm eq}$), where 
\begin{equation}\begin{split}\label{eq:neq}
    N_i^{\rm eq}\equiv \int_0^{R_\nu} 4\pi r^2 dr \int \frac{d^3 p_i}{(2\pi)^3} g_i f_i(E_i, T(r))
\end{split}\end{equation}
is the corresponding total equilibrium abundance of $i$.
These scaling factors are then used to better estimate the IMFP, $\lambda^{-1}_{i,DS}$ due to the DS particles self-interaction, used in Eq.~\eqref{eq:avg_imfp} for computing the DS diffuse luminosities.
We also use $\eta_i$ to calculate the thermally averaged total IMFP of $i$ at $R_\nu$, $\langle{\lambda}^{-1}_{i}(R_\nu)\rangle$.

We are now finally ready to write down our diffuse criteria.
We say that a DS particle species $i$ can be treated as in diffuse limit if the following conditions are satisfied:
\begin{equation}\label{eq:trapcrit}
    N_i > N_i^{\rm eq}{,~\rm and~} 
    \langle{\lambda}^{-1}_{i}(R_\nu)\rangle > R_\nu^{-1}.
\end{equation}
The first condition ensures that a DS particle species $i$ is only considered to be diffusive when their production is efficient enough such that their amount can exceed the equilibrium number within the relevant timescale $\Delta t_i$.
The second condition requires that the IMFP of $i$ at $R_\nu$, where both the density and temperature are the lowest in the PNS, is large enough to trap particle $i$ inside the PNS (see Fig.~\ref{fig:SN_prof}).\footnote{Note that in computing the second criterion, we exclude the contribution from the dark photon decay via $\Adp\rightarrow e^- e^+$
This is because 
including this decay process leads to an artificially enhancement of the IMFP by several orders of magnitudes for dark photon heavier than $\simeq 30$~MeV, as the 
electron chemical potential $\mu_e\simeq 15$~MeV at $R_\nu$. However, this process should be strongly Pauli-blocked for $m_{A'}\lesssim 200$~MeV in most region inside the PNS (see Fig.~\ref{fig:SN_prof}) where dark photons are most abundant in the diffuse limit.}
When Eq.~\eqref{eq:trapcrit} is satisfied, we use Eq.~\eqref{eq:flux} to compute the DS diffuse luminosity for $i$.
Otherwise, we take the nondiffuse luminosities, Eqs.~\eqref{eq:LA'emis} and \eqref{eq:Lchiemis} for dark photons
and dark fermions, respectively.
Note that in the diffuse regime and when $\Delta t_i = t_{i,\rm diff}$, the first condition in Eq.~\eqref{eq:trapcrit} is approximately equivalent to having the characteristic thermalization IMFP~\cite{Keil:2002in,1983bhwd.book} of DS particles with SM medium,  $\sqrt{\lambda^{-1}_{i,SM,\rm abs} \times \lambda^{-1}_{i,\rm total}}$ averaged over the phase space and spatial volume inside $R_\nu$, being larger than $R_\nu^{-1}$.
This means that the energy exchange between the DS particles and the SM medium, enhanced by the DS self-interactions, is efficient enough to keep themselves in thermal contact with SM medium.

We provide in Appendix~\ref{app:scheme} a detailed work flow to additionally describe how we use the results in these sections to compute the dark sector luminosity for interested readers.

\subsection{Numerical calculations}\label{subsec:num}
\begin{figure*}[htb]
    \centering
    \begin{subfigure}[b]{0.5\textwidth}
        \centering
        \includegraphics[width=\textwidth]{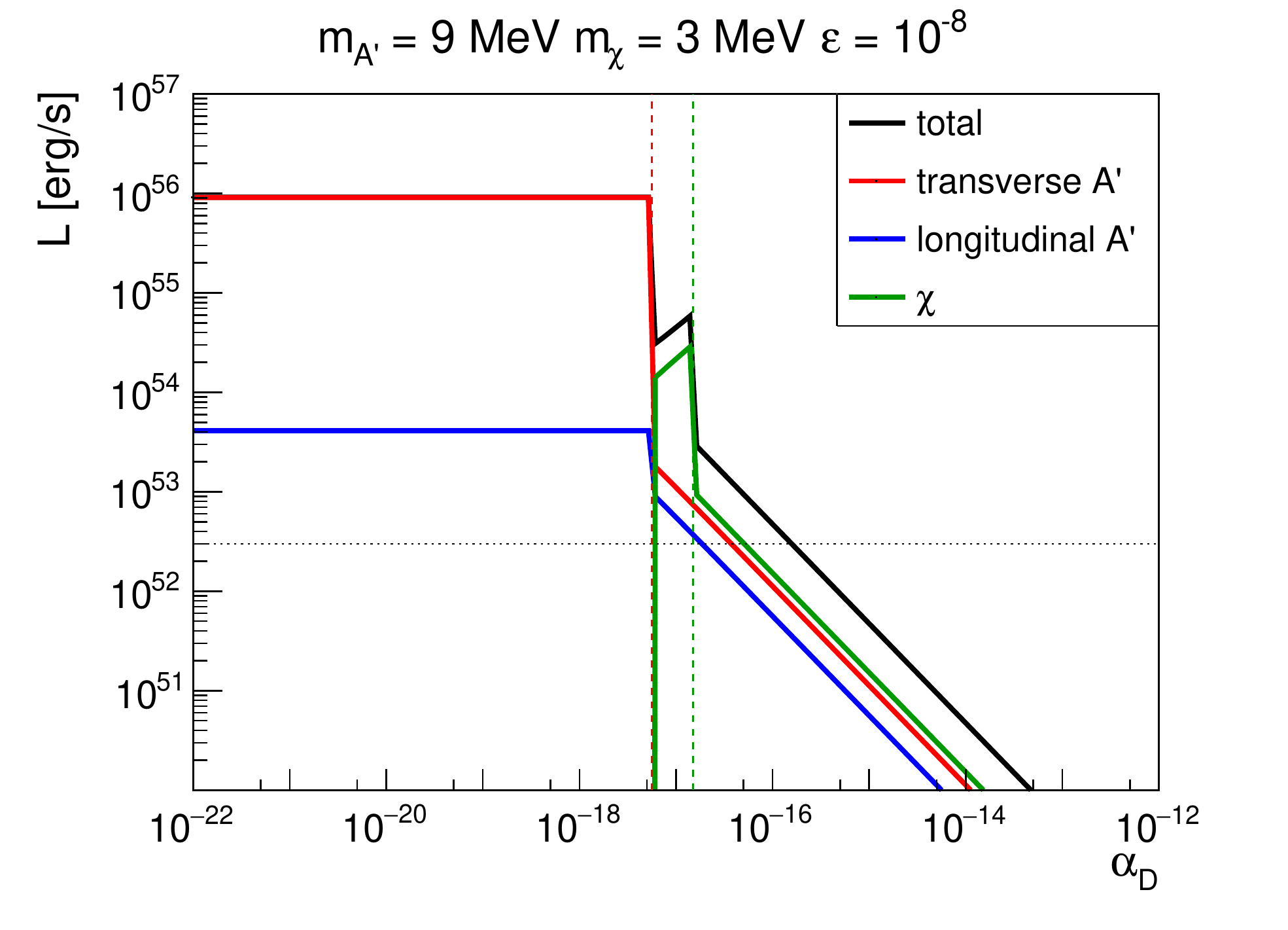}
    \end{subfigure}%
    \begin{subfigure}[b]{0.5\textwidth}
        \centering
        \includegraphics[width=\textwidth]{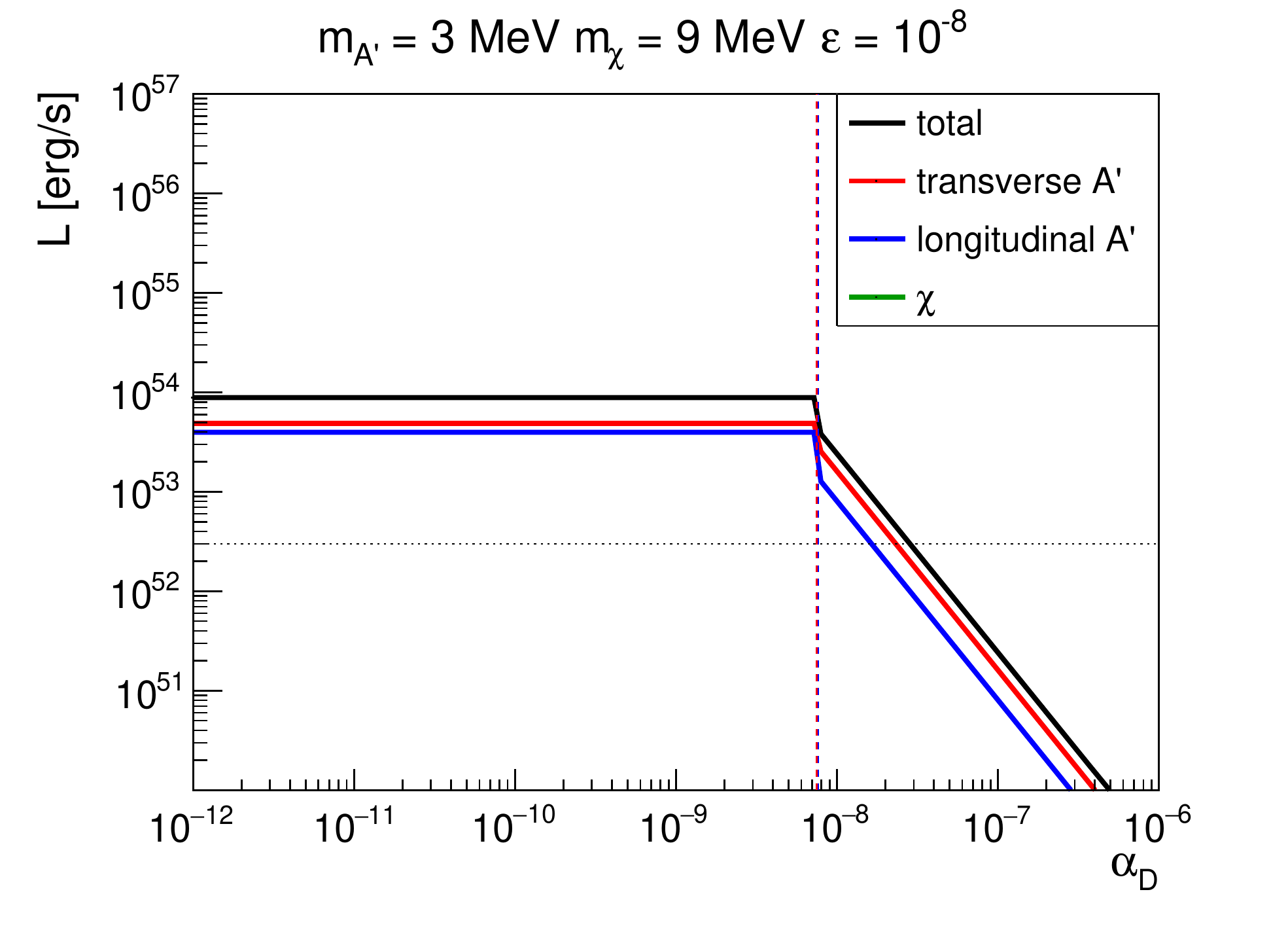}
    \end{subfigure}%
    \caption{Luminosity of different dark sector particles as a function of $\alpha_D$ for a fixed value of $\epsilon = 10^{-8}$. In the left (right) panel, $\mdp = 9\mbox{ MeV}$ and $\mdm = 3\mbox{ MeV}$ ($\mdp = 3\mbox{ MeV}$ and $\mdm = 9\mbox{ MeV}$). 
    The horizontal dotted lines label the neutrino luminosity $L_\nu = 3\times10^{52} \mbox{ erg/s}$ used to set the supernova bound [Eq.~\eqref{eq:L_crit}].
    The vertical dashed lines indicate the transition from the nondiffuse (smaller $\alpha_D)$ to the diffuse (larger $\alpha_D)$ regime (see text for details).
    Note that $\chi$ is in the nondiffuse limit for the range of $\alpha_D$ shown in the right panel.
    }
    \label{fig:L_alpha_d}
\end{figure*}
\begin{figure*}
    \begin{subfigure}[b]{0.5\textwidth}
        \centering
        \includegraphics[width=\textwidth]{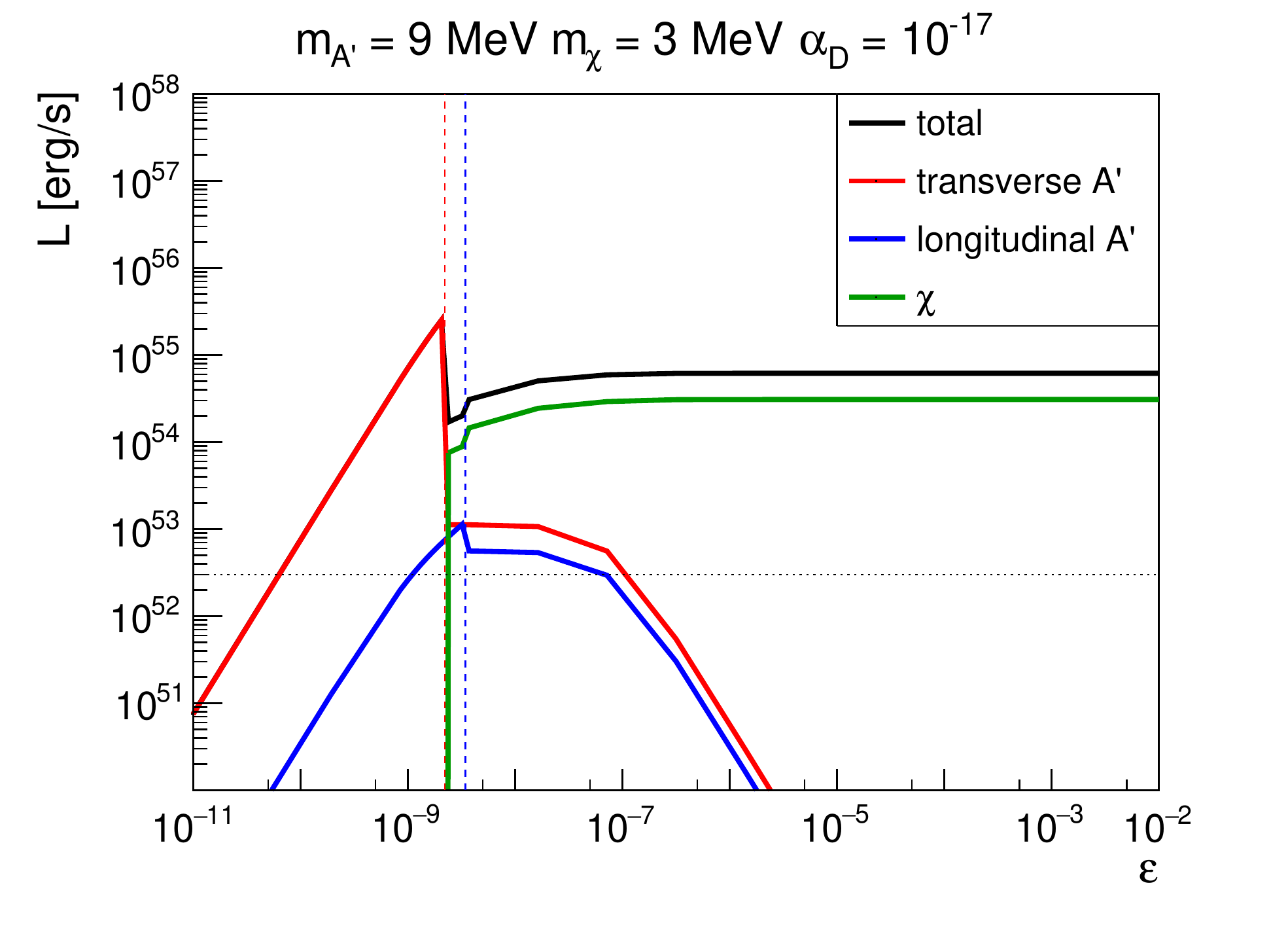}
    \end{subfigure}%
    \begin{subfigure}[b]{0.5\textwidth}
        \centering
        \includegraphics[width=\textwidth]{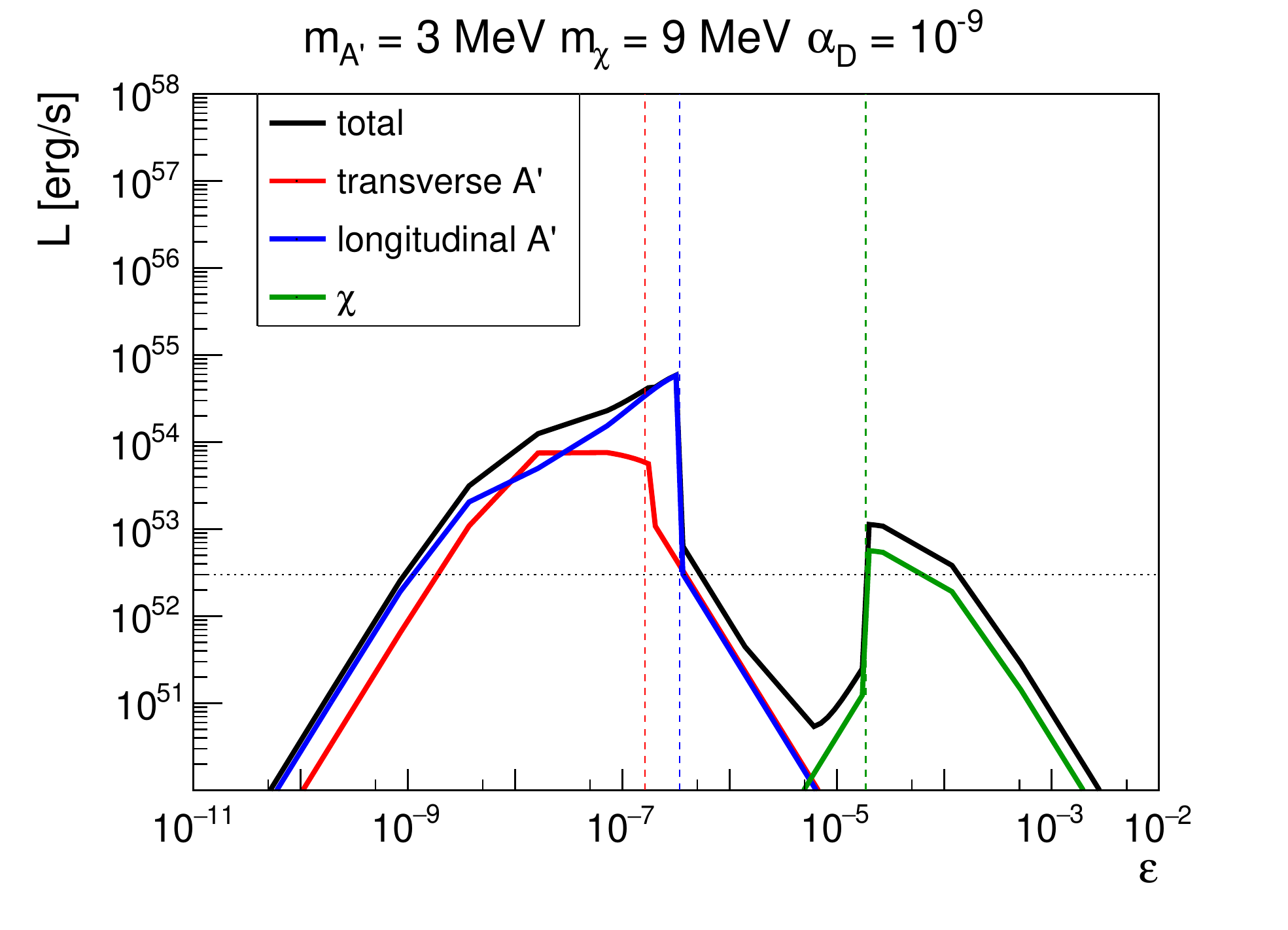}
    \end{subfigure}%
    \caption{Luminosity of different dark sector particles as a function of $\epsilon$ for fixed values of $\alpha_D = 10^{-17}$ (left), $10^{-9}$ (right). 
    In the left (right) panel, $\mdp = 9\mbox{ MeV}$ and $\mdm = 3\mbox{ MeV}$ ($\mdp = 3\mbox{ MeV}$ and $\mdm = 9\mbox{ MeV}$).
    The horizontal dotted lines label the neutrino luminosity $L_\nu = 3\times10^{52} \mbox{ erg/s}$ used to set the supernova bound [Eq.~\eqref{eq:L_crit}].
    The vertical dashed lines indicate the transition from the nondiffuse (smaller $\epsilon$) to the diffuse (larger $\epsilon$) regime (see text for details).
    Note that $\chi$ is in the nondiffuse limit for the range of $\epsilon$ shown in the left panel.
    }
    \label{fig:L_eps}
\end{figure*}

We now apply formulas derived in previous sections to compute the dark photon and dark fermion luminosities and examine how they depend on the coupling constants $\epsilon$ and $\alpha_D$. 
Figure~\ref{fig:L_alpha_d} shows the luminosities of different DS particles as functions of $\alpha_D$ with $\epsilon=10^{-8}$ for two different choices of DS masses. 
In the left panel, the masses are chosen such that the decay process $\Adp\rightarrow\chi\bar{\chi}$ is allowed. 
In this scenario, dark photons of different polarization modes are trapped diffusively when $\alpha_D\gtrsim 5\times 10^{-18}$, while the dark fermions are diffusive when $\alpha_D\gtrsim10^{-17}$. 
The nondiffuse luminosity of the dark photon for small $\alpha_D$ mainly depends on the interaction with the SM particles and thus is independent of $\alpha_D$.\footnote{The difference of $\sim 10^{2.5}$ between the luminosities of the longitudinal and transverse dark photon is due to the plasma effects.}. 
The dark fermion luminosity is proportional to $\alpha_D$ in the diffuse limit, as analyzed in Appendix~\ref{app:nwa} 
However, the dark fermion luminosity for $\alpha_D\lesssim5\times10^{-18}$ is set to zero to avoid double counting because they are predominantly produced through the decay of on-shell dark photons (see discussions in Sec.~\ref{subsec:free}).
In the diffuse regime, the luminosities of the DS particles are proportional to $\alpha_D^{-1}$ due to the self-trapping interactions effectively dominated by the (inverse) decay process $\Adp\leftrightarrow\chi\bar{\chi}$.

For the right panel in Fig.~\ref{fig:L_alpha_d}, we choose DS masses such that the decay process $\Adp\rightarrow\chi\bar{\chi}$ is not allowed. 
In this case, dark photons of different polarization modes are in diffuse regime when $\alpha_D\gtrsim8\times10^{-9}$, while dark fermions are always in the nondiffuse regime for the range of $\alpha_D$ shown in the plot. 
The main reason leading to several orders of magnitude larger difference in $\alpha_D$ here than the previous case is due to the extra $\alpha_D$ dependence in the IMFP of DS self-interactions [see, e.g., Table~\ref{tab:dp_imfp} or Eqs.~\eqref{eq:est1} and \eqref{eq:est2}].
Similar to the previous scenario, the nondiffuse luminosity of the dark photon is independent of $\alpha_D$. 
However, the luminosities in the diffuse regime scale as $\alpha_D^{-2}$. Once again, this is because the dominant interaction is responsible for trapping the dark photon being $\Adp\Adp\rightarrow\chi\bar{\chi}$, whose interaction rate is proportional to $\alpha_D^2$. 
The dark fermion luminosity is significantly smaller than the dark photon luminosity for the range of $\alpha_D$ in the plot, because the dark fermion production rate is proportional to $\epsilon^2\alpha_D$, suppressed by an extra factor of $\alpha_D$ when compared with the dark photon production rate.

In Fig.~\ref{fig:L_eps}, we show the luminosities of the DS particles as functions of $\epsilon$ with two choices of $\alpha_D$ and same DS masses as those in Fig.~\ref{fig:L_alpha_d}. 
For the case with $\alpha_D = 10^{-17}$ in the left panel where the decay process $\Adp\rightarrow\chi\bar{\chi}$ is again allowed, the longitudinal (transverse) dark photons are trapped when $\epsilon\gtrsim3\times10^{-9}$ ($\epsilon\gtrsim2\times10^{-9}$). 
For dark fermions, they just free stream independent of the value of $\epsilon$, because of the small value of $\alpha_D$ (cf., Fig.~\ref{fig:L_alpha_d}).
The nondiffuse luminosities of the dark photons of both modes are proportional to $\epsilon^2$ as determined by their production rates. 
The dark fermion luminosity for $\epsilon\lesssim2\times10^{-9}$ is set to zero for the same reason discussed above. 
The diffuse luminosities of dark photons are affected by both the interactions with the SM particles and the DS self-interactions. 
For $\epsilon\gtrsim10^{-7}$, dark photon--SM interactions dominate over DS self-interactions, so the dark photon luminosities are proportional to $\epsilon^{-2}$ as determined by the mean free path of the dark photon--SM interactions. 
Interestingly, the dark fermion luminosity becomes independent of $\epsilon$ and can thus remain larger than $L_\nu$ for large $\epsilon$.
This is because the branching ratio of the dark photon absorption processes approaches to $1$, such that dark fermions can be produced via the decay of trapped dark photons (see Appendix~\ref{app:nwa}).

For the right panel in Fig.~\ref{fig:L_eps}, we choose $\alpha_D = 10^{-9}$ for the case where $\Adp\rightarrow\chi\bar{\chi}$ is not allowed. 
There, longitudinal (transverse) dark photons are diffusively trapped when $\epsilon\gtrsim3\times10^{-7}$ ($\epsilon\gtrsim2\times10^{-7}$), while dark fermions are in the diffuse limit when $\epsilon\gtrsim2\times10^{-5}$. 
The dark photon luminosities are proportional to $\epsilon^2$ for $\epsilon\lesssim10^{-9}$, where the optical depths are negligible. 
The diffuse luminosities of dark photons scale as $\epsilon^{-2}$ because the absorption processes by the SM medium are dominant. 
The dark fermion luminosity is much smaller than that of the dark photons until $\epsilon\gtrsim10^{-5}$, because the dark fermion production rate is suppressed by an additional factor of $\alpha_D$ when compared with the dark photon production rate. 
The diffuse luminosity of the dark fermion is proportional to $\epsilon^{-2}$ for $\epsilon\gtrsim10^{-4}$, where 
$\chi e^- \rightarrow \chi e^-$ dominates the IMFP at $R_\nu$.

\begin{figure*}[htb]
    \centering
    \begin{subfigure}[b]{0.5\textwidth}
        \centering
        \includegraphics[width=\textwidth]{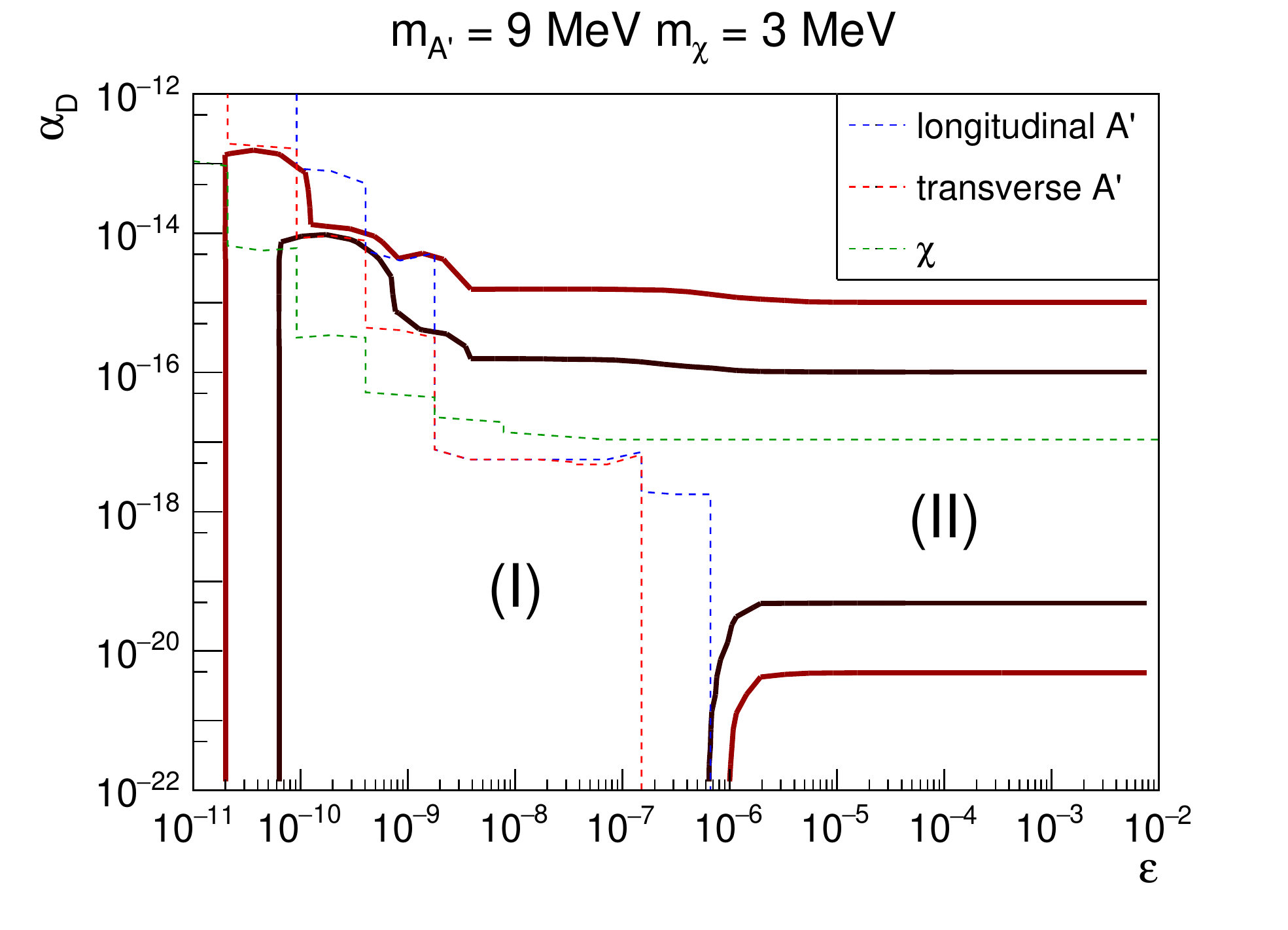}
    \end{subfigure}%
    \begin{subfigure}[b]{0.5\textwidth}
        \centering
        \includegraphics[width=\textwidth]{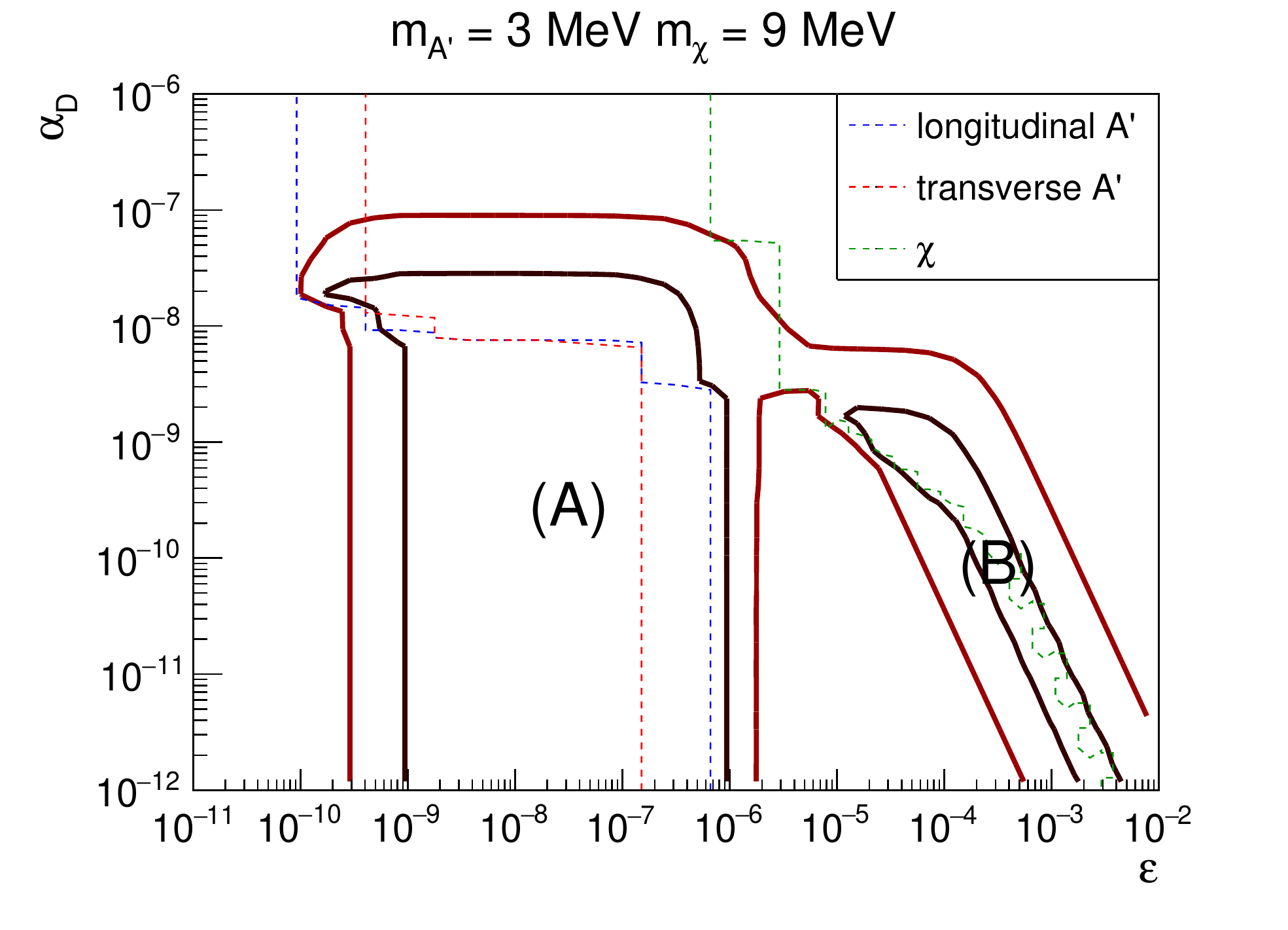}
    \end{subfigure}%
    \caption{Dark sector luminosity $L_D$ contours in the $\alpha_D$-$\epsilon$ plane for $\mdp = 9\mbox{ MeV}$, $\mdm = 3\mbox{ MeV}$ (left) and $\mdp = 3\mbox{ MeV}$, $\mdm = 9\mbox{ MeV}$ (right). 
    Solid black (brown) curves indicate where $L_D=L_\nu$ ($L_D=0.1 L_\nu$). 
    Regions inside the $L_D=L_\nu$ are excluded by the supernova bound. 
    Thin dash lines show where the transition from the nondiffuse to diffuse regime occur.
    See text for discussions on regions labeled (I), (II), and (A), (B).}
    \label{fig:alpha_d_eps}
\end{figure*}

\section{Cooling bounds on self-interacting dark sector particles}\label{sec:res}

In this section, we examine the excluded parameter space of the DS.
Since the model has four free parameters, $m_{A'}$, $m_\chi$, $\epsilon$, and $\alpha_D$, we choose to project the exclusion contours on various combinations of these parameters.
As discussed in the previous sections, whether the decay of $\Adp\rightarrow\chi\bar{\chi}$ is allowed affects the DS luminosity significantly. 
Here we choose two benchmark mass ratios $\mdp/\mdm = 3$ and $1/3$ to investigate the SN bounds for these two scenarios.

We first show the contours of the DS luminosity $L_D=L_\nu$ and $L_D=0.1 L_\nu$ on the $\alpha_D$--$\epsilon$ plane with fixed $m_{A'}$ and $m_\chi$ in Fig.~\ref{fig:alpha_d_eps}. 
The chosen masses are $m_{A'}=9$~MeV and $m_\chi=3$~MeV ($m_{A'}=3$~MeV and $m_\chi=9$~MeV) for the left (right) panel in the figure, which allows (forbids)  $\Adp\rightarrow\chi\bar{\chi}$.
Regions inside the $L_D=L_\nu$ contours indicate that the cooling bound [Eq.~\eqref{eq:L_crit}] is violated.
In the left panel where $\Adp\rightarrow\chi\bar{\chi}$ is allowed, 
the excluded region exhibits an L shape, covering the 
parameter ranges of (I) $\alpha_D\lesssim 10^{-16}$ for $10^{-10}\lesssim \epsilon \lesssim 10^{-6}$, as well as (II) $10^{-19}\lesssim \alpha_D\lesssim 10^{-16}$ for $\epsilon\gtrsim 10^{-6}$.
For the majority of the parameter space in (I) ($\alpha_D < 10^{-16}$ and $10^{-10} < \epsilon < 10^{-6}$), the DS luminosity is mainly contributed by the nondiffuse dark photons (see also Figs.~\ref{fig:L_alpha_d} and \ref{fig:L_eps}).
In (II) ($10^{-19}\lesssim \alpha_D\lesssim 10^{-16}$ for $\epsilon\gtrsim 10^{-6}$), $L_D$ is mainly contributed by the nondiffuse dark fermions (see also Fig.~\ref{fig:alpha_d_eps}).
Close to the upper edge of the $L_D=L_\nu$ contour in both regions, self-trapping of DS particle takes effect such that $L_D\propto \alpha_D^{-1}$ decreases with increasing $\alpha_D$ (see Fig.~\ref{fig:L_alpha_d}), leading to the horizontal edge at $\alpha_D \sim 10^{-16}$.
The lower bound of $\alpha_D\sim 10^{-19}$ for region (II) is due to the inefficient production of dark fermions.
For the right panel in Fig.~\ref{fig:alpha_d_eps} where $\Adp\rightarrow\chi\bar{\chi}$ is not allowed, two regions similarly exist. 
Region (A) with $10^{-9} \lesssim \epsilon \lesssim 10^{-6}$ and $\alpha_D < 10^{-8}$ receives dominant contribution from nondiffuse dark photons as region (I) in the left panel (see also Figs.~\ref{fig:L_alpha_d} and \ref{fig:L_eps}).
Similarly, the self-trapping of DS particles defines the upper edge of $L_D=L_\nu$ at $\alpha_D\sim 10^{-8}$. 
For the narrow diagonal-shape region (B) at $\epsilon\gtrsim 10^{-5}$, $L_D$ is dominated by dark fermions.
This shape is related to the fact that both the production and pair-absorption rates of dark fermions from the SM medium are proportional to $\epsilon^2\alpha_D$, as discussed also in Sec.~\ref{subsec:num}.

\begin{figure*}[htb]
    \centering
    \begin{subfigure}[b]{0.5\textwidth}
        \centering
        \includegraphics[width=\textwidth]{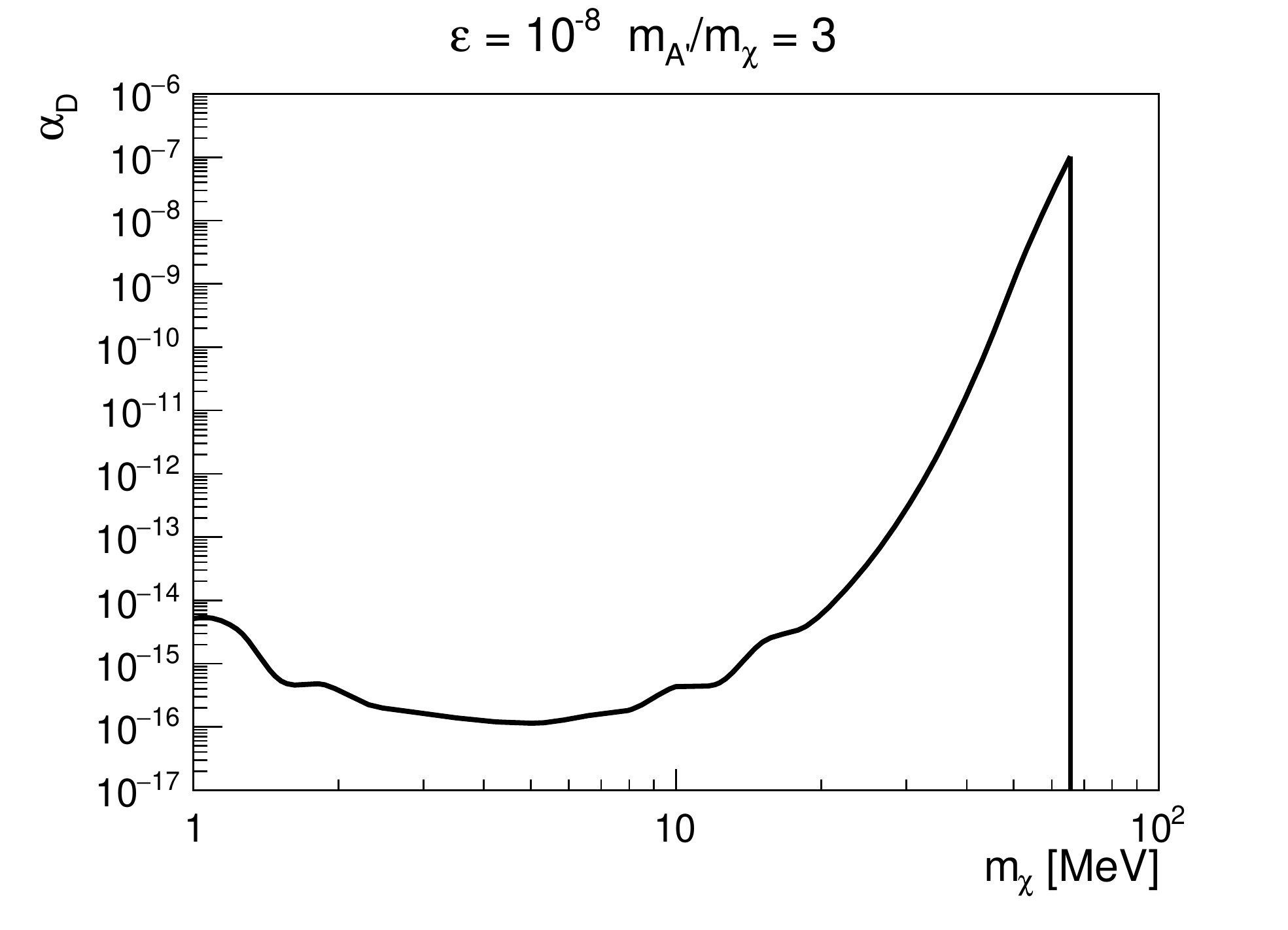}
    \end{subfigure}%
    \begin{subfigure}[b]{0.5\textwidth}
        \centering
        \includegraphics[width=\textwidth]{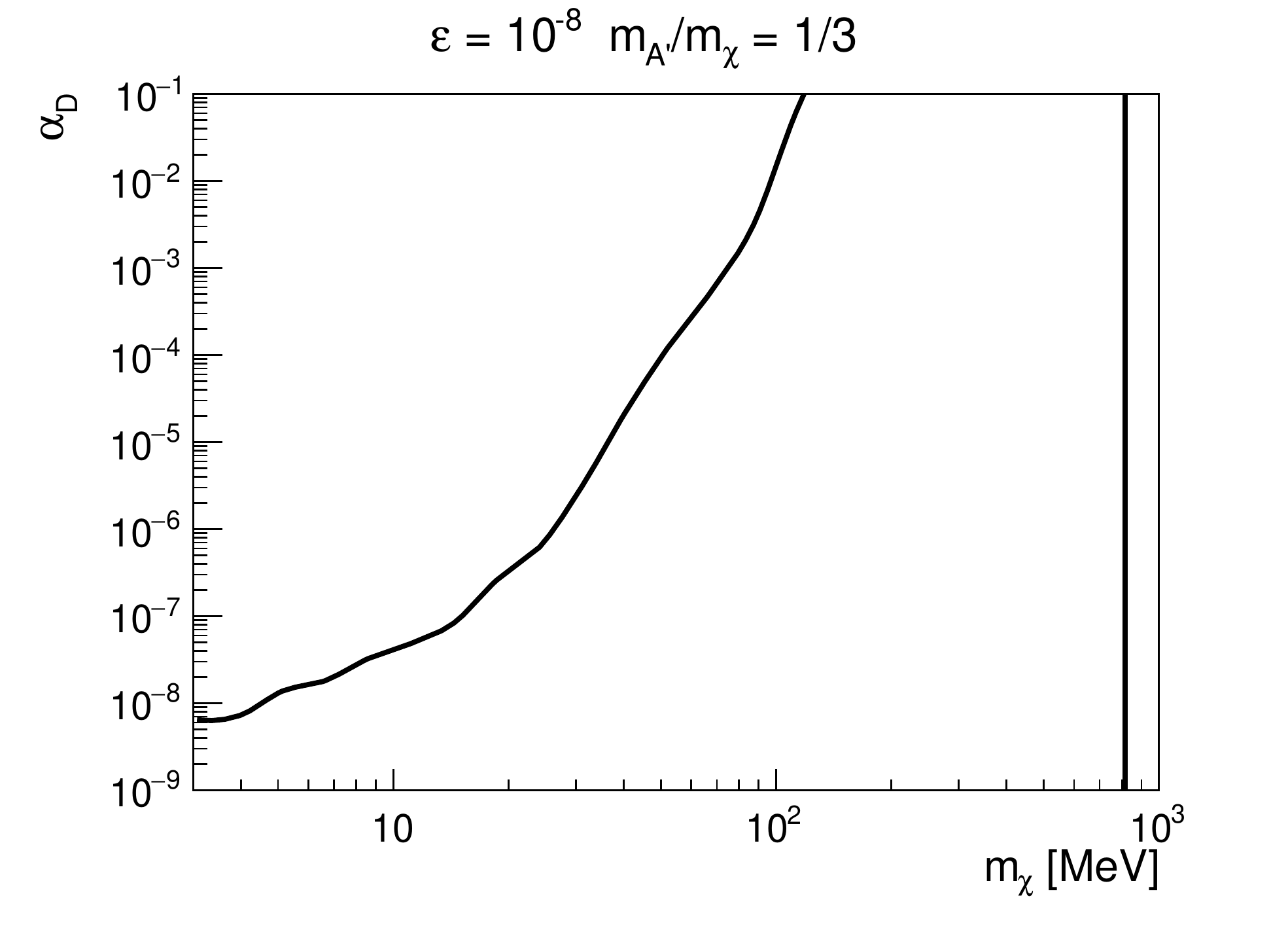}
    \end{subfigure}%
    \caption{The $L_D=L_\nu$ contour on the  $\alpha_D$-$\mdm$ plane for a fixed value of $\epsilon = 10^{-8}$ and mass ratios $\mdp/\mdm = 3$ (left), $1/3$ (right).
    Regions \emph{below} the contours are excluded by the SN bound. 
    Above the contours, the self-trapping of dark sector particles limit their luminosities and cannot be excluded by the supernova bound.}
    \label{fig:alpha_d_mx}
\end{figure*}

For both scenarios shown in Fig.~\ref{fig:alpha_d_eps}, there are specific values of $\alpha_D$ above which the cooling criterion gives no constraint due to DS self-trapping. 
We now investigate the dependence of these critical values of $\alpha_D$ on DS particle masses.
Figure~\ref{fig:alpha_d_mx} shows the excluded regions on the $\alpha_D$-$\mdm$ plane with $\epsilon=10^{-8}$ for fixed mass ratios $m_{A'}=3m_\chi$ (left panel) and $3m_{A'}=m_\chi$ (right panel).
Note that regions below the contours are excluded.
In the left panel where the decay $\Adp\rightarrow\chi\bar{\chi}$ is allowed, 
the SN bound only weakly depends on $\alpha_D$ for $m_\chi\lesssim 10$~MeV.
On the other hand, $\alpha_D$ increase sharply with $m_\chi$ for $m_\chi\gtrsim 10$~MeV.
The main reason is the DS particle abundances are insensitive to the mass for $m_\chi\ll 30$~MeV which is the typical temperature inside the PNS.
Larger $m_\chi$ (and $m_{A'})$ leads to smaller DS abundances, which in turn gives rise to a smaller IMFP of DS self-trapping for a given $\alpha_D$.
Thus, the suppression of $L_D$ due to DS self-trapping occurs at a larger $\alpha_D$ for larger $m_\chi$ (and $m_{A'})$.
For $m_\chi\gtrsim 70$~MeV ($m_{A'}\gtrsim 210$~MeV), no SN bounds can be placed due to the inefficient production of DS particles. 
In the right panel where $3 m_{A'}=m_\chi$, the bound is completely determined by the dark photon only because the dark fermion production is further suppressed by an extra factor of $\alpha_D$.
Here, the critical $\alpha_D$ also increases with $m_\chi$ sharply for $m_\chi\gtrsim 10$~MeV as in the left panel. 
Once again, this is resulting from smaller dark photon abundance for larger $m_\chi$ (thus $m_A'$) inside the PNS.
It thus requires a larger $\alpha_D$ for $\Adp\Adp\rightarrow\chi\bar{\chi}$ to self-trap the dark photons.
Note that the maximal $m_\chi\simeq 800$~MeV below which SN bound exists corresponds to a maximal $m_A'\simeq 250$~MeV, consistent with the maximal $m_A'$ in the left panel.

\begin{figure*}[htb]
    \centering
    \begin{subfigure}[b]{0.5\textwidth}
        \centering
        \includegraphics[width=\textwidth]{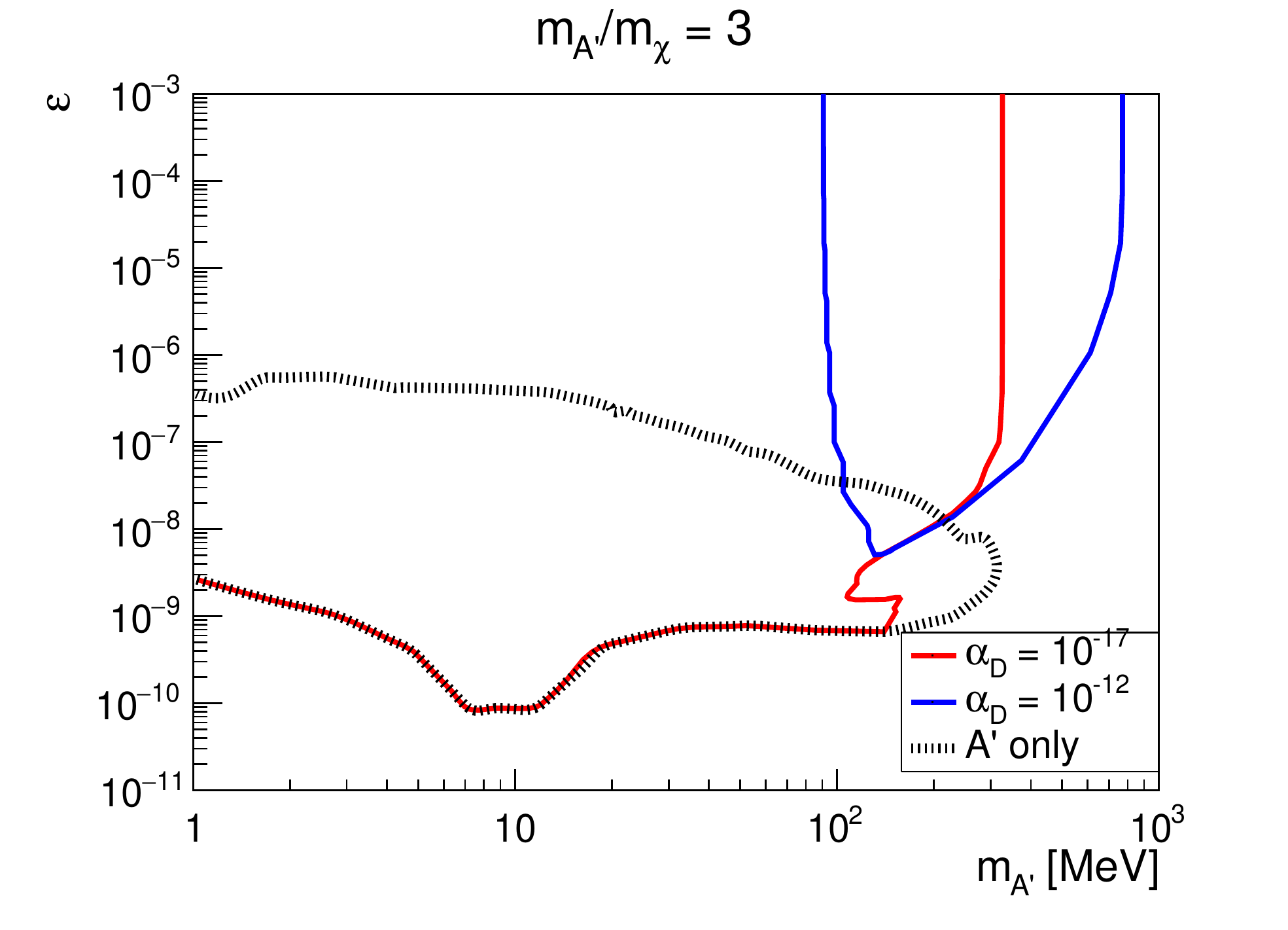}
    \end{subfigure}%
    \begin{subfigure}[b]{0.5\textwidth}
        \centering
        \includegraphics[width=\textwidth]{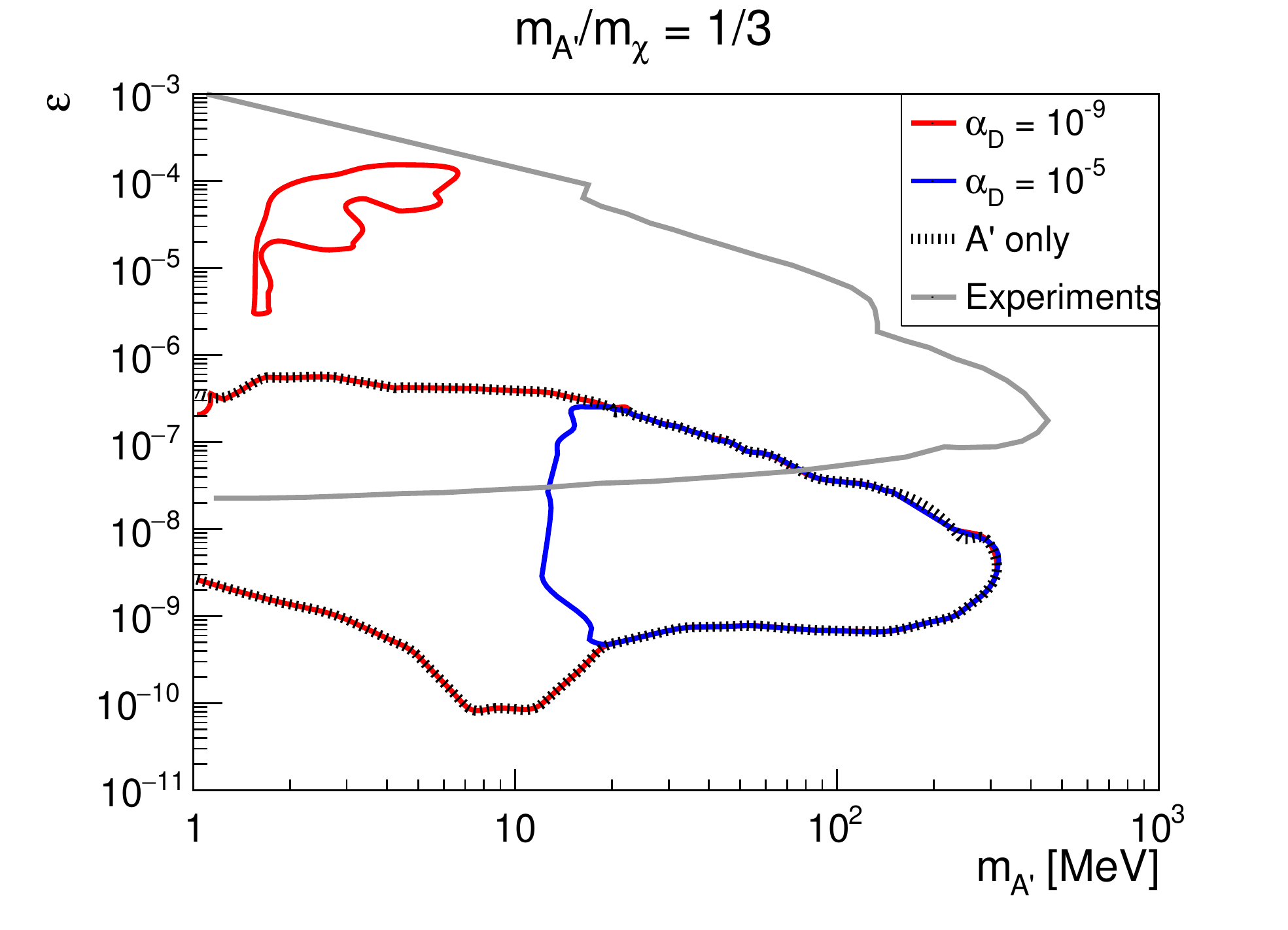}
    \end{subfigure}%
    \caption{The excluded regions (enclosed by the solid curves) on the $\epsilon$-$\mdp$ plane for different $\alpha_D$ values, with fixed mass ratios $\mdp/\mdm = 3$ (left), $1/3$ (right). 
    Note that we include bounds from considering the degree of freedom of dark photon only (dotted black curves) for comparison.
    The thin gray curve in the right panel shows regions excluded by the terrestrial experiments extracted from~\cite{Andreas:2012mt}.
    These curves show that even with tiny values of $\alpha_D$, supernova bounds on dark photon parameter space can be largely affected.
    }
    \label{fig:eps_mA}
\end{figure*}

The above examples clearly demonstrated how efficiently a small $\alpha_D$ within the self-interacting DS can affect the SN bound. 
In Fig.~\ref{fig:eps_mA}, we further show
the excluded regions in the $\epsilon$-$\mdp$ plane with two different values of $\alpha_D$ for each choice of the DS mass ratio. 
Also shown are the bounds derived by considering the dark photon degree of freedom only, as well as the existing experimental constraints on dark photon (extracted from Ref.~\cite{Andreas:2012mt}).
We refrain from showing other astrophysical or cosmological bounds on dark photon of similar masses; see e.g., Refs.~\cite{Jaeckel:2008fi,Mirizzi:2009iz,Redondo:2013lna,An:2013yfc,Fradette:2014sza,Sung:2019xie,DeRocco:2019njg,An:2020bxd,Li:2020roy,Sieverding:2021jfa}.
These plots show once again that even for small values of $\alpha_D=10^{-12}$ (left panel) and $10^{-5}$ (right panel), the SN exclusion regions shrink significantly due to the self-trapping effects when compared to results derived by considering only dark photons without self-trapping.
However, for very small values of $\alpha_D$, e.g., $10^{-17}$ and $10^{-9}$ shown in the left and right panels, respectively, the cooling bounds can be extended to larger $\epsilon$.
This is due to the contribution of dark fermions through the decay of trapped dark photons, as discussed in earlier sections.
Note that for $\alpha_D=10^{-12}$ in the left panel, the excluded region slightly extends to larger values of $\mdp$ for $\epsilon\gtrsim10^{-7}$.
This is because the dark fermion luminosity depends on both the decay rate of $\Adp\rightarrow\chi\bar{\chi}$ and the blackbody energy density of dark photons (see Appendix~\ref{app:nwa}). 
The larger decay rate can compensate the smaller dark photon energy density with increasing $\mdp$, provided that the branching ratio of the dark photon absorption processes is close to $1$.

\section{Conclusion}\label{sec:con}

In this work, we examined the SN bounds on self-interacting dark sector particles. 
Adopting a dark photon portal dark sector, we derived the relevant interaction cross sections and (inverse) decay rates for reactions listed in Tables~\ref{tab:dp_imfp} and \ref{tab:df_imfp}.
We then used these to compute the energy luminosities of dark sector particles in the nondiffuse and diffuse regimes separately, and formulated a simple criterion to connect these two regimes.
The self-interaction of dark sector particles can efficiently trap themselves inside the proto-neutron star and thus suppress their energy luminosities.

Comparing the dark sector luminosity with the neutrino luminosity inferred from the SN1987a event, we derived SN bounds for two assumed dark photon to dark fermion mass ratios, $m_{A'}/m_\chi=3$ and $1/3$, which represent scenarios where $A'\rightarrow \chi\bar\chi$ is allowed or not.
For the former (later) case with $m_{A'}/m_\chi=3$ ($m_{A'}/m_\chi=1/3$), SN bounds only apply to weakly interacting dark sectors whose dark fine structure constant $\alpha_D\lesssim 10^{-15}$ ($\lesssim 10^{-7}$), for $m_\chi\lesssim \mathcal{O}(20)$~MeV (see Figs.~\ref{fig:alpha_d_eps} and \ref{fig:alpha_d_mx}).
The dominating dark sector cross sections for these $\alpha_D$ values correspond to $\simeq 10^{-40}$~cm$^2$. 
In particular, there is no SN bound for the former case for $\alpha_D\gtrsim 10^{-7}$ (Fig.~\ref{fig:alpha_d_mx}).
Our results differ from the previous analysis~\cite{Chang:2018rso} considering similar models. 
This is because Ref.~\cite{Chang:2018rso} assumed that the DS particles decouple from the SM medium at the surface where the $\chi$-p scattering becomes inefficient, and ignored the DS self-interactions which can trap themselves and help thermalize the DS particles with the SM medium.

Although the exact excluded regions in the DS parameter space should also depend on the chosen value of $m_{A'}/m_\chi$, which are unexplored in this work, our results demonstrated that when applying the supernova bounds to dark sector particles, their self-interactions, which can evade the bounds, must be taken into considerations.
Our results here also imply that other stellar bounds, e.g., from the horizontal branch stars, tips of red giants, or white dwarfs, on dark sector particles may also be sensitive to the structure of the dark sector.
Similarly, our results also indicate that for nonstandard strongly self-interacting neutrinos proposed to resolve the Hubble tension~\cite{Kreisch:2019yzn}, the needed strong self-interaction, $\sim 10^6$ or $10^9$ times stronger than the SM weak interaction, will likely lead to self-trapping of neutrinos and results in inconsistency with the SN1987a observation [cf.~Eqs.~\eqref{eq:sigma_xxA} and \eqref{eq:sigma_AAxx}].
Moreover, we would like to point out that although the self-interaction of dark sector can completely evade the SN bound, new constraints may be further derived by considering their potential late-time heating to the remnant NS via decays or annihilations in a longer timescale.
Furthermore, such self-trapping effect might provide an efficient mechanism to produce DM-admixed NS, which might have implication for the GW detection of binary NS merger events~\cite{Ellis:2017jgp,Nelson:2018xtr,Bauswein:2020kor}.
All these aspects are beyond the scope of this paper and can be further investigated in future work.
We also note that our results indicate that self-interacting DM that can help solve the small-scale issues in galaxies cannot be excluded by the SN cooling bound, as the required $\chi$-$\chi$ self-interaction cross section $\sigma_{\chi\chi}/m_\chi\sim \mathcal{O}(1)$~cm$^2$~g$^{-1}$ largely exceeds values that can be constrained by SN cooling [cf.~Eq.~\eqref{eq:sigma_xxxx}].

Finally, we comment on potential caveats in this work.
First, our criterion of switching from the nondiffuse to diffuse regime is rather abrupt and can sometimes create non-negligible discontinuity in dark sector luminosities as shown in e.g., Figs.~\ref{fig:L_alpha_d} and \ref{fig:L_eps}.
In reality, the transition should be smooth and this can possibly introduce errors of a factor of a few in all our derived bounds.
For example, while evaluating the luminosities in the nondiffuse regime, we only included absorption and decay processes in the opacity.
In principle, scattering can somewhat reduce the energy luminosity of dark sector particles leaving the PNS, before the condition for diffusion is fully satisfied.
Also, we used a sharp neutrinosphere as a boundary to estimate the luminosities in the diffuse regime. 
This may lead to some errors when the dark sector particles are not fully in the diffuse limit.
Second, when we evaluated the diffusion timescale used to determine the diffuse criterion, we selected two specific locations where the IMFPs are largest for simplicity. 
This approximation may overestimate the diffusion time a bit.
All these sources of uncertainties can only be addressed by performing a full numerical calculation of multidimensional Boltzmann transport and can be pursued in future.
However, the main conclusion derived in this work -- self-interactions inside the dark sector can crucially affect stellar bounds -- should remain relatively solid and needs to be considered in all relevant studies.

\begin{acknowledgments}
We thank Jae Hyeok Chang, Yen-Hsun Lin, Huitzu Tu, and Tse-Chun Wang for useful discussions.
We also thank Luc Darme, Sam McDermott, Sunny Vagnozzi, and an anonymous referee for providing feedback during the reviewing process, which help improve this work.
This work was supported in part by the Ministry of Science and Technology, Taiwan under Grant No.~108-2112-M-001-010, No.~109-2112-M-001-004, 
the Academia Sinica under Project No. AS-CDA-109-M11, 
and the Physics Division of National Center for Theoretical Sciences.
\end{acknowledgments}

\begin{widetext}

\section{Narrow Width Approximation}\label{app:nwa}

Consider a process $\chi\bar{\chi}\rightarrow A^{\prime\ast} \rightarrow f_1 f_2 ...$ when $\mdp > 2\mdm$ so that the intermediate $\Adp$ can be on shell. Let $k = (\omega, \vec{k})$ be the momentum of the intermediate $\Adp$ and $p = (E, \vec{p})$, $p' = (E', \vec{p}')$ be the momentum of the initial $\chi$ and $\bar{\chi}$ respectively. The spin-averaged amplitude squared is given by
\begin{equation}\begin{split}
    \overline{|\mathcal{M}|^2} = g_D^2 \cdot \frac{k^2}{2} &\left\{ \frac{|\epsilon_L \cdot J|^2}{(k^2 - \mdp^2)^2 + (\omega\Gamma_L)^2} \left[ 1 - \left(\frac{\omega-2E}{|\vec{k}|}\right)^2 \right] \right.\\
    &\left. + \frac{\sum_\lambda |\epsilon_{T\lambda} \cdot J|^2}{(k^2 - \mdp^2)^2 + (\omega\Gamma_T)^2} \left[ \frac{1}{2} + \frac{2\mdm^2}{k^2} + \frac{1}{2}\left(\frac{\omega-2E}{|\vec{k}|}\right)^2 \right] \right\},\\
\end{split}\end{equation}
where $J^\mu$ is the final state current that couples to the dark photon, $\epsilon_L$, $\epsilon_{T\lambda}$ are the polarization vectors of the longitudinal and transverse dark photon with helicity index $\lambda = 1, 2$, and the thermal width of the dark photon $\Gamma_{L,T}$ is given by \cite{Weldon:1983jn}
\begin{equation}
    \Gamma_{L,T} = (1-e^{-\omega/T})(\Gamma^{L,T}_{\Adp,\rm abs} + \Gamma_{\Adp\rightarrow\chi\bar{\chi}}).
\end{equation}
In the limit $\Gamma_{L,T} \ll \mdp$, we can approximate the Breit-Wigner distribution by a $\delta$ function
\begin{equation}\label{eq:delta}
    \frac{1}{(k^2-\mdp^2)^2 + (\omega \Gamma_{L,T})^2} \rightarrow \frac{\pi}{\omega \Gamma_{L,T}}\delta(k^2 - \mdp^2).
\end{equation}
With the above approximation, the absorption rate of $\chi$ becomes

\begin{equation}\begin{split}
    \Gamma_{\chi\bar{\chi}\rightarrow A^{\prime\ast} \rightarrow f} &\equiv \frac{1}{2E} \int \frac{d^3p'}{(2\pi)^3} \frac{g_{\bar{\chi}}f_{\bar{\chi}}}{2E'} d\Pi_f (2\pi)^4 \delta^4(p+p'-p_f) \overline{|\mathcal{M}|^2}\\
    &\simeq \frac{1}{16\pi E |\vec{p}|} \int_{\omega_-}^{\omega_+} d\omega (1+f_{\Adp}) g_{\bar{\chi}} f_{\bar{\chi}} \left[ \mbox{Br}(\Adp_L \rightarrow f) \overline{|\mathcal{M}|^2}_{\chi\bar{\chi}\rightarrow \Adp_L} + \mbox{Br}(\Adp_T \rightarrow f)\overline{|\mathcal{M}|^2}_{\chi\bar{\chi}\rightarrow \Adp_T} \right],
\end{split}\end{equation}
where
\begin{gather}
    \omega_\pm = \frac{\mdp^2}{2\mdm^2}\left(E \pm |\vec{p}|\sqrt{1-\frac{4\mdm^2}{\mdp^2}}\right),\\
    \mbox{Br}(\Adp_{L,T}\rightarrow f) = \frac{\Gamma_{\Adp\rightarrow f}}{\Gamma^{L,T}_{\Adp,\rm abs} + \Gamma_{\Adp\rightarrow\chi\bar{\chi}}},\\
    \Gamma_{\Adp_L\rightarrow f} = \frac{1}{2\omega} \int d\Pi_f (2\pi)^4\delta^4(k-p_f) |\epsilon_L\cdot J|^2,\\
    \Gamma_{\Adp_T\rightarrow f} = \frac{1}{2\omega} \int d\Pi_f (2\pi)^4\delta^4(k-p_f) \frac{1}{2}\sum_{\lambda} |\epsilon_{T\lambda}\cdot J|^2,\\
    \overline{|\mathcal{M}|^2}_{\chi\bar{\chi}\rightarrow\Adp_L} = \frac{1}{2}g_D^2 \mdp^2 \left[1-\left(\frac{\omega-2E}{|\vec{k}|}\right)^2\right],\\
    \overline{|\mathcal{M}|^2}_{\chi\bar{\chi}\rightarrow\Adp_T} = g_D^2 \mdp^2 \left[\frac{1}{2} + \frac{2\mdm^2}{\mdp^2} + \frac{1}{2}\left(\frac{\omega-2E}{|\vec{k}|}\right)^2\right].
    \end{gather}
Note that the inverse decay rates into dark photons of different polarizations are
\begin{equation}
    \Gamma_{\chi\bar{\chi}\rightarrow \Adp_{L,T}} = \frac{1}{16\pi E |\vec{p}|} \int_{\omega_-}^{\omega_+} d\omega (1+f_{\Adp}) g_{\bar{\chi}} f_{\bar{\chi}} \overline{|\mathcal{M}|^2}_{\chi\bar{\chi}\rightarrow \Adp_{L,T}}.
\end{equation}
Thus, we can relate the differential absorption rate and inverse decay rates by
\begin{equation}\label{eq:xx_f_nwa}
    \frac{d\Gamma_{\chi\bar{\chi}\rightarrow A^{\prime\ast} \rightarrow f}}{d\omega} \simeq \mbox{Br}(\Adp_L \rightarrow f) \frac{d\Gamma_{\chi\bar{\chi}\rightarrow \Adp_L}}{d\omega} + \mbox{Br}(\Adp_T \rightarrow f) \frac{d\Gamma_{\chi\bar{\chi}\rightarrow \Adp_T}}{d\omega}.
\end{equation}
With the above equation, the total pair-absorption rate of $\chi$ is given by

\begin{equation}\begin{split}\label{eq:A17}
    \Gamma_{\chi, \rm pair} &= \sum_f \Gamma_{\chi\bar{\chi}\rightarrow A^{\prime\ast} \rightarrow f}\\
    &= \int_{\omega_-}^{\omega_+} d\omega \, \left[\frac{d\Gamma_{\chi\bar{\chi}\rightarrow \Adp_L}}{d\omega} \sum_f \mbox{Br}(\Adp_L \rightarrow f) + \frac{d\Gamma_{\chi\bar{\chi}\rightarrow \Adp_T}}{d\omega} \sum_f \mbox{Br}(\Adp_T \rightarrow f)\right]\\
    &= \int_{\omega_-}^{\omega_+} d\omega \, \left[\frac{d\Gamma_{\chi\bar{\chi}\rightarrow \Adp_L}}{d\omega} + \frac{d\Gamma_{\chi\bar{\chi}\rightarrow \Adp_T}}{d\omega}\right]\\
    &= \Gamma_{\chi\bar{\chi}\rightarrow \Adp}.
\end{split}\end{equation}
The total pair-absorption rate is equal to that of the inverse decay process $\chi\bar{\chi}\rightarrow \Adp$. Therefore, the inverse decay process $\chi\bar{\chi}\rightarrow\Adp$ accounts for all the pair-absorption processes with dark photon resonances when $\mdp > 2\mdm$. These processes include $\chi\bar{\chi}np \rightarrow np$, $\chi\bar{\chi}\rightarrow e^- e^+$, $\chi\bar{\chi}\rightarrow \gamma^\ast$, and $\chi\bar{\chi}\rightarrow\chi\bar{\chi}$. Note that the dark Compton scattering $\chi A'\rightarrow A' \chi$ also admits an on-shell dark fermion $\bar{\chi}$. With the approximation analogous to Eq.~(\ref{eq:delta}), one can show that the scattering rate of $\chi$ via $\chi A' \rightarrow A' \chi$ is equivalent to the inverse decay rate of $\chi\bar{\chi}\rightarrow A'$, and the scattering rate of $A'$ via the same process is equivalent to the decay rate of $A'\rightarrow\chi\bar{\chi}$. Thus, we do not include the dark Compton process in our IMFP calculations (see Tables~\ref{tab:dp_imfp} and \ref{tab:df_imfp}).

With Eq.~(\ref{eq:xx_f_nwa}), we can derive the luminosity of $\chi$ and $\bar{\chi}$ in the nondiffuse regime
\begin{equation}
    L_{\chi} + L_{\bar{\chi}} \simeq \int dV \int d\omega \frac{du_{\Adp}}{d\omega} \Gamma_{\Adp\rightarrow\chi\bar{\chi}} \left[ \frac{1}{3} \mbox{Br}(\Adp_L \rightarrow f) + \frac{2}{3} \mbox{Br}(\Adp_T \rightarrow f) \right],
\end{equation}
where $u_{\Adp}$ is the blackbody energy density of the dark photon. We shall make several comments on the above equation.
\begin{enumerate}
    \item It is assumed that dark photons are in thermal equilibrium with the SM medium while the dark fermion streams freely. (This corresponds to large $\epsilon$ and small $\alpha_D$.) In this scenario, $\mbox{Br}(\Adp_{L,T} \rightarrow SM) \approx 1$. So $1/3$ of the dark fermion is produced by the longitudinal dark photon, while $2/3$ of the dark fermion is produced by the transverse dark photon. However, if the dark photon escapes the PNS freely, this calculation is invalid, and the dark photon usually cannot decay into the dark fermions before escaping the PNS. Thus, the dark fermion luminosity would be negligible. Therefore, we compute the luminosity of $\chi$ according to Eq.~(\ref{eq:Lchi_free*}) for different scenarios.
    \item In the limit of large $\epsilon$ and small $\alpha_D$, the dark sector luminosity is mostly contributed by the dark fermion since the dark photon is trapped. The luminosity only depends on $\alpha_D$ and does not depend on $\epsilon$ in this limit. It is possible that the luminosity can exceed $L_\nu$ for some values of $\alpha_D$ even if $\epsilon$ is large, as shown in the left panels of Figs.~\ref{fig:L_eps}, \ref{fig:alpha_d_eps}, and \ref{fig:eps_mA}.
    \item The luminosity in this regime also depends on $\mdp$ through the blackbody energy density $u_{\Adp}$. Therefore, for a larger value of $\alpha_D$, the constraint can extends to larger $\mdp$ values, as shown in the left panel of Fig.~\ref{fig:eps_mA}.
\end{enumerate}

\section{Plasma Effects}\label{app:plas}
In the PNS, the hot ($T\sim30$ MeV) and dense ($\rho\sim10^{14}\mbox{ g}/\mbox{cm}^3$) plasma consisting of electrons and nucleons can modify the dispersion relations of the SM photons and give rise to a longitudinally polarized propagation mode called \textit{plasmon}. This plasma effect changes the SM photon propagator, so we must take it into account when calculating the interaction rates between SM and DS particles through photon--dark photon mixing. The \textit{effective mixing} $\epsilon_{k|L,T}$ and the \textit{plasma factor} $\beta_{k|L,T}$ for the SM current--dark photon interactions are defined by
\begin{equation}\label{eq:effmix}
    \beta_{k|L,T}^2 \equiv \frac{\epsilon_{k|L,T}^2}{\epsilon^2} \equiv \frac{(k^2)^2}{(k^2-\mbox{Re}\Pi_{L,T})^2 + (\mbox{Im}\Pi_{L,T})^2}, 
\end{equation}
where $k=(\omega, \vec{k})$ is the 4-momentum of the dark photon with $k^2 \equiv \omega^2 - |\vec{k}|^2$, and $\Pi_{L,T}$ is the longitudinal (L) and transverse (T) polarization functions and is related to the SM photon polarization tensor $\Pi^{\mu\nu} = e^2\langle A^\mu, A^\nu \rangle$ by
\begin{align}
    \Pi_L &= \frac{k^2}{|\vec{k}|^2} \Pi^{00},\\
    \Pi_T &= \frac{1}{2} \left(\delta^{ij} - \frac{k^i k^j}{|\vec{k}|^2}\right) \Pi^{ij}.
\end{align}
We would like to point out that our definition for $\Pi_L$ aligns with \cite{An:2013yfc} but differs from \cite{Braaten:1993jw}. For the on-shell dark photon, we replace the subscript ``$k$'' with ``$m$'' for the effective mixing, i.e., $\epsilon_{m|L,T}^2 = (\epsilon_{k|L,T}^2)|_{k^2 = \mdp^2}$, and likewise for the plasma factor. On the other hand, the DS current--SM photon counterpart of Eq.~(\ref{eq:effmix}) is
\begin{equation}\label{eq:effmix2}
    \frac{\bar{\epsilon}_{k|L,T}^2}{\epsilon^2} = \frac{(k^2)^2}{(k^2-\mdp^2)^2 + (\omega\Gamma_{L,T})^2} \equiv \bar{\beta}_{k|L,T}^2,    
\end{equation}
where $\Gamma_{L,T} = (1-e^{-\omega/T})\Gamma^{L,T}_{\Adp, \rm abs}$ is the thermal absorptive width of the dark photon \cite{Weldon:1983jn}. Similarly, for the on-shell transverse photon or plasmon, we replace the subscript $k$ with $m$ for the effective mixing and the plasma factor. Note that the dispersion relation of the SM photon $k^2 = \mathrm{Re}\Pi_{L,T}$ is a transcendental equation that can only be solved numerically. The derivation of Eqs.~\eqref{eq:effmix} and~\eqref{eq:effmix2} in diagonalized mass basis can be found in Appendix B of \cite{Lin:2019uvt}.

In the medium consisting of relativistic and degenerate electron, the real parts of the scalar polarization functions are 
\cite{Braaten:1993jw}
\begin{align}
    \mbox{Re}\Pi_L &= \frac{3\omega_p^2 (1-v^2)}{v^2}\left[ \frac{1}{2v}\ln{\left(\frac{1+v}{1-v}\right)} - 1 \right]\label{eq:RePiL},\\
    \mbox{Re}\Pi_T &= \frac{3\omega_p^2}{2v^2}\left[ 1 - \frac{1-v^2}{2v}\ln{\left(\frac{1+v}{1-v}\right)} \right]\label{eq:RePiT},
\end{align}
where $v\equiv |\vec{k}|/\omega$ and the plasma frequency in this limit is
\begin{equation}\label{eq:plasmafreq}
    \omega_p^2 = \frac{4\alpha_e}{3\pi} \left( \mu_e^2 + \frac{1}{3}\pi^2 T^2 \right).
\end{equation}
With detailed balance $\Gamma^{L,T}_{\rm prod} = e^{-\omega/T} \Gamma^{L,T}_{\rm abs}$, the imaginary part of the polarization functions is
\begin{equation}
    \mbox{Im}\Pi_{L,T} = -\omega(\Gamma^{L,T}_{\rm prod} - \Gamma^{L,T}_{\rm abs}) = -\omega \Gamma^{L,T}_{\rm abs} (1-e^{-\omega/T}),
\end{equation}
where $\Gamma^{L,T}_{\rm prod}$ ($\Gamma^{L,T}_{\rm abs}$) are the production (absorption) rates of the SM photon.
When calculating the plasmon decay, we should include the renormalization factor for the polarization vectors of the external photons,
\begin{equation}
    \tilde{\epsilon}_{L,T}^\mu = \sqrt{Z_{L,T}} \epsilon_{L,T}^\mu,
\end{equation}
where the renormalization factor is given by
\begin{equation}
    Z_{L,T}^{-1} = 1 - \left. \frac{\partial \mbox{Re}\Pi_{L,T}}{\partial \omega^2} \right|_{\rm pole},
\end{equation}
which can be calculated from Eqs.~\eqref{eq:RePiL} and \eqref{eq:RePiT},
\begin{align}
    Z_L^{-1} &= \frac{3\omega_p^2}{2\omega^2 v^2} \left[ 1 - \frac{1-v^2}{2v}\ln{\left(\frac{1+v}{1-v}\right)} \right]_{\rm pole},\\
    Z_T^{-1} &= 1 - \frac{3\omega_p^2}{2\omega^2 v^2} \left[ \frac{3}{2} - \frac{3-v^2}{4v}\ln{\left(\frac{1+v}{1-v}\right)} \right]_{
    \rm pole}.
\end{align}

\section{Interaction Rates and Inverse Mean Free Path}\label{app:int}
The interaction rate of particle $1$ via a process $1 + ... + n \rightarrow 1' + ... + m'$ is given by
\begin{equation}\label{eq:Gamma}
    \Gamma = \frac{1}{2E_1} \int \prod_{i=1}^n \frac{d^3p_{i}}{(2\pi)^3}\frac{g_if_i}{2E_i} \prod_{j=1'}^{m'} \frac{d^3p_j}{(2\pi)^3}\frac{1\pm f_j}{2E_j} \overline{|\mathcal{M}|^2} (2\pi)^4 \delta^4( \sum_{i=1}^n p_i - \sum_{j=1'}^{m'} p_j),
\end{equation}
where $g_i$ is the degrees of freedom of the initial state particles and $f_i$ and $f_j$ are the distribution functions of the initial and final state particles.
The upper (lower) sign in front of $f_j$ is for bosonic (fermionic) final states, which takes into account the bosonic enhancement and Pauli-blocking effects.
The above expression already assumes that all the particles are in thermal equilibrium.
The IMFP of the process is related to the interaction rate by
\begin{equation}
    \tilde{\lambda}^{-1} = \frac{\Gamma}{v_1} =  \frac{E_1}{|\vec{p}_1|}\Gamma.
\end{equation}
In the following, we denote the 4-momentum of $\Adp$ by $k = (\omega, \vec{k})$, that of $\chi$ by $p = (E, \vec{p})$, and that of $\bar{\chi}$ by $p^\prime = (E^\prime, \vec{p}^\prime)$, unless noted otherwise.

\subsection{$\Adp np\rightarrow np$}\label{app:Anpnp}
The absorption rate via inverse nucleon bremsstrahlung $\Adp np\rightarrow np$ is given by \cite{Chang:2016ntp}
\begin{equation}\label{eq:Anpnp}
    \Gamma^{L,T}_{\Adp np \rightarrow np} = \frac{32}{3\pi} \frac{\alpha_e \epsilon^2_{m|L,T} n_n n_p}{\omega^3} \left(\frac{\pi T}{m_N}\right)^{3/2} \left\langle\sigma_{np}^{(2)}(T)\right\rangle h_{L,T},
\end{equation}
where $n_{n,p}$ is the number density of neutrons and protons, respectively, $m_N$ is the nucleon mass, $\left\langle\sigma_{np}^{(2)}(T)\right\rangle$ is the thermally averaged $n$-$p$ cross section defined as \cite{Rrapaj:2015wgs}
\begin{equation}
    \left\langle \sigma^{(2)}_{np} (T) \right\rangle = \frac{1}{2} \int_{0}^{\infty} dx x^2 e^{-x} \int_{-1}^{1} d\cos{\theta_{cm}} ( 1 -  \cos{\theta_{cm}} ) \frac{d\sigma_{np}}{d\cos{\theta_{cm}}}\left( T_{cm} = xT, \cos{\theta_{cm}}\right),
\end{equation}
and $h_{L,T}$ is defined as
\begin{align}
    h_L &= \frac{\mdp^2}{\omega^2}, \label{eq:hL}\\
    h_T &= 1. \label{eq:hT}
\end{align}
In the derivation of Eq.~(\ref{eq:Anpnp}), the soft radiation approximation is used to connect the bremsstrahlung rate with the experimental $n$-$p$ scattering cross section. Additionally, the nucleons are assumed to follow Maxwell-Boltzmann distribution, and the Pauli-blocking effect is ignored.

\subsection{$\Adp e^- \rightarrow e^- \gamma$}\label{app:Aeer}
The absorption rate via the Compton-like process $\Adp e^- \rightarrow e^- \gamma$ is approximately \cite{Chang:2016ntp}
\begin{equation}
    \Gamma^{L,T}_{\Adp e^- \rightarrow e^- \gamma} = \frac{8\pi\alpha_e^2 \epsilon^2_{m|L,T} n_e}{3E_F^2} \sqrt{\frac{\omega_p}{\omega}} h_{L,T},
\end{equation}
where $n_e$ is the electron number density, $E_F = \sqrt{(3\pi^2 n_e)^{2/3} + m_e^2}$ is the electron Fermi energy, and $h_{L,T}$ is defined in Eqs.~(\ref{eq:hL}) and (\ref{eq:hT}). The approximated formula is valid for $\omega\lesssim 200$ MeV and $r \leq R_\nu$~\cite{Chang:2016ntp}.

\subsection{$\Adp\rightarrow e^- e^+$}\label{app:Aee}
The decay rates of dark photon via $\Adp\rightarrow e^- e^+$, given that $\mdp > 2m_e$, are given by
\begin{align}\label{eq:ALee}
    \Gamma_{\Adp\rightarrow e^- e^+}^L &= \frac{\epsilon_{m|L}^2 \alpha_e \mdp^2}{4 |\vec{k}|} \int_{-\xi_0}^{\xi_0} d\xi \left[ 1 + \exp{\left(\frac{\mu_e - \omega/2}{T} - \frac{\omega}{2T}\xi\right)} \right]^{-1} \left(1 - \frac{\omega^2}{|\vec{k}|^2} \xi^2 \right),\\
    \label{eq:ATee}
    \Gamma_{\Adp\rightarrow e^- e^+}^T &= \frac{\epsilon_{m|T}^2 \alpha_e \mdp^2}{4 |\vec{k}|} \int_{-\xi_0}^{\xi_0} d\xi \left[ 1 + \exp{\left(\frac{\mu_e - \omega/2}{T} - \frac{\omega}{2T}\xi\right)} \right]^{-1} \left(\frac{1}{2} + \frac{2m_e^2}{\mdp^2} + \frac{\omega^2}{2|\vec{k}|^2} \xi^2 \right),
\end{align}
where the integral limit is
\begin{equation}\label{eq:xi0}
    \xi_0 = \frac{|\vec{k}|}{\omega}\sqrt{1 - \frac{4m_e^2}{\mdp^2}}.
\end{equation}

\subsection{$\chi\bar{\chi} np \rightarrow np$}\label{app:xxnpnp}
The pair-annihilation rate via inverse nucleon bremsstrahlung $\chi\bar{\chi} np \rightarrow np$ is
\begin{equation}\label{eq:xxnpnp}
    \Gamma_{\chi\bar{\chi} np \rightarrow np} = \frac{16 \epsilon^2 \alpha_D \alpha_e n_n n_p}{3\pi^2 E} \left( \frac{\pi T}{m_N} \right)^{3/2} \left\langle \sigma_{np}^{(2)}(T) \right\rangle \int_0^\infty d|\vec{k}| \int_{-1}^{1} d\cos{\theta} \, \frac{|\vec{k}|^2 (\mathcal{F}^L_{\chi\bar{\chi} np \rightarrow np} + 2\mathcal{F}^T_{\chi\bar{\chi} np \rightarrow np})}{E^\prime \omega^2 k^2 (1 + e^{E^\prime/T})},
\end{equation}
where the momentum transfer is $k=p+p^\prime$ and $\cos{\theta} = \vec{k}\cdot\vec{p}/|\vec{k}||\vec{p}|$. Note that $E,p$ ($E',p'$) are for $\chi$ ($\bar\chi$). Thus $E^\prime= \sqrt{E^2 + |\vec{k}|^2 - 2|\vec{k}||\vec{p}|\cos{\theta}}$ and $\omega = E+E^\prime$. The two terms $\mathcal{F}^L_{\chi\bar{\chi} np \rightarrow np}$, $\mathcal{F}^T_{\chi\bar{\chi} np \rightarrow np}$ in the integrand are
\begin{align}
    \mathcal{F}^L_{\chi\bar{\chi} np \rightarrow np} &= \beta_{k|L}^2 \bar{\beta}_{k|L}^2\cdot\frac{k^2}{\omega^2}\left[ 1 - \left( \frac{\omega-2E}{|\vec{k}|} \right)^2 \right],\\\label{eq:xxnpnp_FT}
    \mathcal{F}^T_{\chi\bar{\chi} np \rightarrow np} &= \beta_{k|T}^2 \bar{\beta}_{k|T}^2\cdot\left[ \frac{1}{2} + \frac{2\mdm^2}{k^2} + \frac{1}{2}\left( \frac{\omega-2E}{|\vec{k}|} \right)^2 \right].
\end{align}
Similar to the calculation of $\Adp np \rightarrow np$, we use soft radiation approximation for the dark fermion pair absorption and ignore the Pauli-blocking effect of the nucleons. 

When $\mdp > 2\mdm$, we use the narrow width approximation (NWA) detailed in Appendix~\ref{app:nwa} to approximate the integral. The resulting formula is
\begin{equation}
    \Gamma^{\rm NWA}_{\chi\bar{\chi} np \rightarrow np} = \frac{8 \epsilon^2 \alpha_D \alpha_e \mdp^2 n_n n_p}{3\pi E |\vec{p}|} \left( \frac{\pi T}{m_N} \right)^{3/2} \left\langle \sigma_{np}^{(2)}(T) \right\rangle \int_{\omega_-}^{\omega_+} d\omega \, \frac{\mathcal{F}^{L\ast}_{\chi\bar{\chi} np \rightarrow np} + 2\mathcal{F}^{T\ast}_{\chi\bar{\chi} np \rightarrow np}}{\omega^3 [1+e^{(\omega-E)/T}]},
\end{equation}
where the two terms $\mathcal{F}^{L\ast}_{\chi\bar{\chi} np \rightarrow np}$, $\mathcal{F}^{T\ast}_{\chi\bar{\chi} np \rightarrow np}$ in the integrand are
\begin{align}
    \mathcal{F}^{L\ast}_{\chi\bar{\chi} np \rightarrow np} &= \frac{\beta_{m|L}^2}{\Gamma_L}\cdot\frac{\mdp^2}{\omega^2} \left[ 1 - \left(\frac{\omega - 2E}{|\vec{k}|} \right)^2\right], \label{eq:xxnpFL_nwa}\\
    \mathcal{F}^{T\ast}_{\chi\bar{\chi} np \rightarrow np} &= \frac{\beta_{m|T}^2}{\Gamma_T}\cdot\left[ \frac{1}{2} + \frac{2\mdm^2}{\mdp^2} + \frac{1}{2}\left( \frac{\omega - 2E}{|\vec{k}|} \right)^2 \right], \label{eq:xxnpFT_nwa}
\end{align}
and we now have $|\vec{k}| = \sqrt{\omega^2 - \mdp^2}$. The dark photon widths must include the contribution from the dark photon decay via $\Adp\rightarrow\chi\bar{\chi}$. Thus we have $\Gamma_{L,T} = (1-e^{-\omega/T})(\Gamma^{L,T}_{\Adp, \rm abs}+\Gamma_{\Adp\rightarrow\chi\bar{\chi}})$. 
Momentum conservation demands that $\omega$ is bounded by
\begin{equation}\label{eq:w+-}
    \omega_\pm = \frac{\mdp^2}{2\mdm^2} \left( E \pm |\vec{p}|\sqrt{1-\frac{4\mdm^2}{\mdp^2}} \right).
\end{equation}

\subsection{$\chi\bar{\chi}\rightarrow e^- e^+$}\label{app:xxee}
The pair-annihilation rate via $\chi(p) + \bar{\chi}(p^\prime)\rightarrow e^-(p_3) + e^+(p_4)$ is 
\begin{equation}\label{eq:xxee}
    \Gamma_{\chi\bar{\chi}\rightarrow e^- e^+} = \frac{\epsilon^2 \alpha_D \alpha_e}{4\pi E} \int_0^\infty d|\vec{k}| \int_{-1}^{1} d\cos{\theta} \, \int_{E_{3-}}^{E_{3+}} dE_3 \, \frac{|\vec{k}|}{E^\prime} f^\prime(1-f_3)(1-f_4)\, (\mathcal{F}^L_{\chi\bar{\chi}\rightarrow e^- e^+} + 2\mathcal{F}^T_{\chi\bar{\chi}\rightarrow e^- e^+}),
\end{equation}
where we have the same definitions for $\cos{\theta}$, $E^\prime$, and $\omega$ as in Eq.~(\ref{eq:xxnpnp}). Moreover, $f^\prime$, $f_3$, and $f_4$ are the distribution functions of $\bar{\chi}$, $e^-$, and $e^+$, respectively. That is,
\begin{align}
    f' &= \frac{1}{e^{E^\prime/T} + 1},\\
    f_3 &= \frac{1}{e^{(E_3 - \mu_e)/T} + 1},\\
    f_4 &= \frac{1}{e^{(E_4 + \mu_e)/T} + 1},
\end{align}
where the positron energy is expressed as $E_4 = \omega - E_3$ by energy conservation. With the electron and positron distribution function included, we take the Pauli-blocking effect into account, which is significant near the center of the PNS. The two terms $\mathcal{F}^L_{\chi\bar{\chi}\rightarrow e^- e^+}$, $\mathcal{F}^T_{\chi\bar{\chi}\rightarrow e^- e^+}$ in the integrand are
\begin{align}\label{eq:xxee_FL}
    \mathcal{F}^L_{\chi\bar{\chi}\rightarrow e^- e^+} &= \beta_{k|L}^2 \bar{\beta}_{k|L}^2 \left[ 1 - \left(\frac{\omega-2E}{|\vec{k}|}\right)^2 \right]\left[ 1 - \left(\frac{\omega-2E_3}{|\vec{k}|}\right)^2 \right],\\
    \label{eq:xxee_FT}
    \mathcal{F}^T_{\chi\bar{\chi}\rightarrow e^- e^+} &= \beta_{k|T}^2 \bar{\beta}_{k|T}^2\left[ \frac{1}{2} + \frac{2\mdm^2}{k^2} + \frac{1}{2}\left(\frac{\omega-2E}{|\vec{k}|}\right)^2 \right]\left[ \frac{1}{2} + \frac{2m_e^2}{k^2} + \frac{1}{2}\left(\frac{\omega-2E_3}{|\vec{k}|}\right)^2 \right].
\end{align}
Momentum conservation demands that $E_3$ is bounded by
\begin{equation}
    E_{3\pm} = \frac{1}{2}\left[\omega \pm |\vec{k}|\sqrt{1 - \frac{4m_e^2}{k^2}}\right].
\end{equation}

When $\mdp > 2\mdm$, we use NWA (see Appendix~\ref{app:nwa}) to approximate the annihilation rate. The resulting formula is
\begin{equation}
    \Gamma^{\rm NWA}_{\chi\bar{\chi}\rightarrow e^- e^+} = \frac{\epsilon^2 \alpha_e \alpha_D \mdp^4}{8 E |\vec{p}|} \int_{\omega_-}^{\omega_+} d\omega \int_{E_{3-}}^{E_{3+}} dE_3 \frac{f_2(1-f_3)(1-f_4)}{\omega |\vec{k}|} (\mathcal{F}^{L\ast}_{\chi\bar{\chi}\rightarrow e^- e^+} + 2\mathcal{F}^{T\ast}_{\chi\bar{\chi}\rightarrow e^- e^+}),
\end{equation}
where we now have $|\vec{k}| = \sqrt{\omega^2 - \mdp^2}$. The energies of $\bar{\chi}$ and $e^+$ on which their distribution functions depend are expressed as $E^\prime = \omega - E$ and $E_4 = \omega - E_3$. The two terms $\mathcal{F}^{L\ast}_{\chi\bar{\chi}\rightarrow e^- e^+}$, $\mathcal{F}^{T\ast}_{\chi\bar{\chi}\rightarrow e^- e^+}$ in the integrand now are
\begin{align}
    \mathcal{F}^{L\ast}_{\chi\bar{\chi}\rightarrow e^- e^+} &= \frac{\beta_{m|L}^2}{\Gamma_L}\left[1 - \left(\frac{\omega-2E}{|\vec{k}|}\right)^2\right]\left[1 - \left(\frac{\omega-2E_3}{|\vec{k}|}\right)^2\right],\\
    \mathcal{F}^{T\ast}_{\chi\bar{\chi}\rightarrow e^- e^+} &= \frac{\beta_{m|T}^2}{\Gamma_T}\left[\frac{1}{2} + \frac{2\mdm^2}{\mdp^2} + \frac{1}{2}\left(\frac{\omega-2E}{|\vec{k}|}\right)^2\right]\left[\frac{1}{2} + \frac{2m_e^2}{\mdp^2} + \frac{1}{2}\left(\frac{\omega-2E_3}{|\vec{k}|}\right)^2\right],
\end{align}
where $\Gamma_{L,T}$ is the same as in Eqs.~(\ref{eq:xxnpFL_nwa}) and (\ref{eq:xxnpFT_nwa}).
The limit of $E_3$ now becomes
\begin{equation}
    E_{3\pm} = \frac{1}{2}\left[\omega \pm |\vec{k}|\sqrt{1 - \frac{4m_e^2}{\mdp^2}}\right],
\end{equation}
while the limit of $\omega$ is the same as Eq.~(\ref{eq:w+-}).

\subsection{$\chi\bar{\chi}\rightarrow\gamma^\ast$}\label{app:plasmon}
Let $\omega_{L,T} = \omega_{L,T}(|\vec{k}|)$ be the energy of the longitudinal and transverse SM photon as functions of $|\vec{k}|$, and $m_{L,T}$ be the effective mass satisfying $m_{L,T}^2 = \omega_{L,T}^2 - |\vec{k}|^2 = \mathrm{Re}\Pi_{L,T}(\omega_{L,T}, |\vec{k}|)$. Then the pair-annihilation rate can be expressed as
\begin{equation}\label{eq:xxplasmon}
    \Gamma_{\chi\bar{\chi}\rightarrow\gamma^{\ast}} = \frac{\epsilon^2 \alpha_D}{4 E |\vec{p}|} \sum_{L,T} \int_0^\infty \frac{|\vec{k}| d|\vec{k}|}{\omega_{L,T}} \frac{g_{L,T}\mathcal{F}^{L,T}_{\chi\bar{\chi}\rightarrow\gamma^\ast}\Theta(1-\cos^2{\theta_{L,T}}) }{[1+e^{(\omega_{L,T}-E)/T}](1-e^{-\omega_{L,T}/T})},
\end{equation}
where $g_L = 1$, $g_T = 2$, and $\mathcal{F}^L_{\chi\bar{\chi}\rightarrow\gamma^\ast}$, $\mathcal{F}^T_{\chi\bar{\chi}\rightarrow\gamma^\ast}$ are defined as
\begin{align}
    \mathcal{F}^L_{\chi\bar{\chi}\rightarrow\gamma^\ast} &= \bar{\beta}^2_{m|L} Z_L m_L^2 \left[ 1 - \left( \frac{\omega_L-2E}{|\vec{k}|} \right)^2 \right],\\
    \mathcal{F}^T_{\chi\bar{\chi}\rightarrow\gamma^\ast} &= \bar{\beta}^2_{m|T} Z_T m_T^2 \left[ \frac{1}{2} + \frac{2\mdm^2}{m_T^2} + \frac{1}{2}\left( \frac{\omega_T-2E}{|\vec{k}|} \right)^2 \right],
\end{align}
where $Z_{L,T}$ is the renormalization factor for the longitudinal plasmon and transverse photon detailed in Appendix \ref{app:plas}. The bosonic enhancement for the plasmon is included in Eq.~(\ref{eq:xxplasmon}). The step function in Eq.~(\ref{eq:xxplasmon}) is to ensure that the momenta are kinematically allowed. The definition of $\cos{\theta_{L,T}}$ is
\begin{equation}
    \cos{\theta_{L,T}} = \frac{2\omega_{L,T}E - m_{L,T}^2}{2|\vec{k}||\vec{p}|}.
\end{equation}

We use NWA to approximate the annihilation rate when $\mdp > 2\mdm$. The resulting expression is
\begin{equation}
    \Gamma_{\chi\bar{\chi}\rightarrow\gamma^\ast} = \frac{\pi \epsilon^2 \alpha_D \mdp^6}{4 E |\vec{p}|} \sum_{L,T} \frac{g_{L,T} \mathcal{F}^{L,T\ast}_{\chi\bar{\chi}\rightarrow\gamma^\ast} \Theta(\omega_+ - \omega^\ast_{L,T})\Theta(\omega^\ast_{L,T} - \omega_-)}{[1+e^{(\omega^\ast_L-E)/T}](1-e^{-\omega^\ast_L/T})},
\end{equation}
where $\omega^\ast_{L,T}$ is the solution of the equation $\mbox{Re}\Pi_{L,T}(\omega, \sqrt{\omega^2 - \mdp^2}) = \mdp^2$. $\mathcal{F}^{L,T\ast}_{\chi\bar{\chi}\rightarrow\gamma^\ast}$ is now defined as
\begin{align}
    \mathcal{F}^{L\ast}_{\chi\bar{\chi}\rightarrow\gamma^\ast} &= \frac{1}{\omega^\ast_L \Gamma_L} \left[ 1 - \frac{(\omega^\ast_L-2E)^2}{\omega^{\ast 2}_L - \mdp^2} \right] \left|\left.\frac{d\mbox{Re}\Pi_L(\omega, \sqrt{\omega^2 - \mdp^2})}{d\omega}\right|_{\omega^\ast_L}\right|^{-1},\\
    \mathcal{F}^{T\ast}_{\chi\bar{\chi}\rightarrow\gamma^\ast} &= \frac{1}{\omega^\ast_T \Gamma_T} \left[ \frac{1}{2} + \frac{2\mdm^2}{\mdp^2} + \frac{1}{2}\frac{(\omega^\ast_T-2E)^2}{\omega^{\ast 2}_T - \mdp^2} \right] \left|\left.\frac{d\mbox{Re}\Pi_T(\omega, \sqrt{\omega^2 - \mdp^2})}{d\omega}\right|_{\omega^\ast_T}\right|^{-1}.
\end{align}
$\omega_\pm$ is the same as Eq.~(\ref{eq:w+-}) and $\Gamma_{L,T}$ is the same as in Eqs.~(\ref{eq:xxnpFL_nwa}) and (\ref{eq:xxnpFT_nwa}).

\subsection{$\chi e^- \rightarrow \chi e^-$ and $\chi p \rightarrow \chi p$}
The scattering rate of $\chi(p) + e^- (p_2) \rightarrow \chi(p^\prime) + e^- (p_4)$ is given by
\begin{equation}
    \Gamma_{\chi e^- \rightarrow \chi e^-} = \frac{\epsilon^2 \alpha_D \alpha_e}{4\pi E} \int d|\vec{k}| \, d\cos{\theta} \, \int_{E_{2+}}^{\infty} dE_2 \, \frac{|\vec{k}|}{E^\prime} f_2(1-f^\prime)(1-f_4)\, (\mathcal{F}^L_{\chi e^- \rightarrow \chi e^-} + 2\mathcal{F}^T_{\chi e^- \rightarrow \chi e^-}),
\end{equation}
where $k = p - p^\prime$ is the momentum transfer, $\cos{\theta}$ and $E^\prime$ are the same in Eq.~(\ref{eq:xxnpnp}) while $\omega = E - E^\prime$ instead. $f_2$, $f^\prime$, and $f_4$ are the distribution functions of the initial state $e^-$, final state $\chi$, and final state $e^-$, respectively. The final state electron energy is expressed as $E_4 = \omega + E_2$ by energy conservation. The two terms $\mathcal{F}^L_{\chi e^- \rightarrow \chi e^-}$, $\mathcal{F}^T_{\chi e^- \rightarrow \chi e^-}$ are the same as Eqs.~(\ref{eq:xxee_FL}) and (\ref{eq:xxee_FT}) except that $-E_3$ is replaced with $+E_2$, and $E_2$ is bounded below by
\begin{equation}
    E_{2+} = \frac{1}{2}\left( |\vec{k}|\sqrt{1-\frac{4m_e^2}{k^2}} - \omega \right).
\end{equation}
Note that the momentum transfer is spacelike ($k^2 < 0$) for the $\chi$-$e^-$ scattering. In this scenario, we take the dark photon width to be zero, while $\mbox{Im }\Pi_{L,T}$ is determined by the imaginary parts of Eqs.~(\ref{eq:RePiL}) and (\ref{eq:RePiT}) with $\omega\rightarrow\omega + i\varepsilon$ for an infinitesimal $\varepsilon > 0$ \cite{Kuznetsov:2013sea}. 

The scattering rate via $\chi p \rightarrow \chi p$ is the same as $\chi e^- \rightarrow \chi e^-$ if the differences in masses and distribution functions of the proton and the electron are taken into account. However, we can approximate the $\chi$-$p$ scattering as an elastic scattering due to the fact that $m_p$ is much larger than $\mdm$. We also ignore the Pauli-blocking effect of proton. Thus the scattering rate of $\chi$-$p$ scattering can be approximated by
\begin{equation}
    \Gamma_{\chi p \rightarrow \chi p} = \frac{4\pi\epsilon^2\alpha_e\alpha_D n_p |\vec{p}|}{E (1+e^{-E/T})} \int_{-1}^{1} d\cos{\theta} \beta_{k|L}^2 \cdot \frac{2E^2 - |\vec{p}|^2(1-\cos{\theta})}{(k^2 - \mdp^2)^2},
\end{equation}
where the momentum transfer is $k^2 \simeq -|\vec{k}|^2 \simeq -2|\vec{p}|^2(1-\cos{\theta})$. Note that in this static limit ($\omega\rightarrow 0$), the contribution from the transverse mode, corresponding to the classical magnetic field, is suppressed. And the plasma factor $\beta_{k|L}^2$ of the longitudinal mode, corresponding to the static electric field, accounts for the screening effect in medium \cite{kapusta_gale_2006}.

\subsection{$\Adp\chi\rightarrow\chi\Adp$ and $\Adp\Adp\leftrightarrow\chi\bar{\chi}$}
The spin-averaged matrix element squared of $\Adp(p_1) + \chi(p_2) \rightarrow \chi(p_3) + \Adp(p_4)$ is
\begin{equation}\begin{split}\label{eq:AxxA}
    \overline{|\mathcal{M}|^2}_{\Adp\chi\rightarrow\chi\Adp} &= \frac{64\pi^2 \alpha_D^2}{3}  \left\{ (\mdp^2 + 2 \mdm^2)^2 \left[ \frac{1}{(s - \mdm^2)^2} + \frac{1}{(t - \mdm^2)^2} \right] + \frac{8(\mdm^4 - \mdp^4)}{(s-\mdm^2)(t-\mdm^2)}\right.\\
    &\left. + \, 4 (\mdm^2 + \mdp^2) \left( \frac{1}{s-\mdm^2} + \frac{1}{t-\mdm^2} \right) - \left( \frac{s-\mdm^2}{t-\mdm^2} + \frac{t-\mdm^2}{s-\mdm^2} \right) \right\},
\end{split}\end{equation}
where $s=(p_1+p_2)^2$ and $t=(p_1-p_3)^2$ are the Mandelstam variables. By crossing symmetry, the matrix element of $\Adp(p_1) + \Adp(p_2) \rightarrow \chi(p_3) + \bar{\chi}(p_4)$ is the same as Eq.~(\ref{eq:AxxA}) except that $s$ is replaced by $u=(p_1-p_4)^2$ and the matrix element is multiplied by a factor of $-2/3$. (The minus sign is due to the crossing of one fermion state.) For $\chi(p_1) + \bar{\chi}(p_2) \rightarrow \Adp(p_3) + \Adp(p_4)$, we can again reuse Eq.~(\ref{eq:AxxA}), replace $s$ by $u$, and multiply by a factor of $-3/2$. 

\subsection{$\chi\bar{\chi}\rightarrow\chi\bar{\chi}$ and $\chi\chi\rightarrow\chi\chi$}
The spin-averaged matrix element squared of $\chi(p_1) + \bar{\chi}(p_2) \rightarrow \chi(p_3) + \bar{\chi}(p_4)$ is
\begin{equation}\begin{split}\label{eq:xx'xx'}
    &\overline{|\mathcal{M}|^2}_{\chi\bar{\chi}\rightarrow\chi\bar{\chi}} = 64 \pi^2 \alpha_D^2 \left\{ \left(\frac{5}{2}\mdp^4 - 4 \mdm^2\mdp^2 + 4 \mdm^4 \right) \left[ \frac{1}{(t-\mdp^2)^2} + \frac{1}{(s-\mdp^2)^2} \right] + \frac{4(\mdp^4 - \mdm^4)}{(t-\mdp^2)(s-\mdp^2)} \right.\\
    &\left. + (3\mdp^2 - 4\mdm^2) \left[ \frac{s-\mdp^2}{(t-\mdp^2)^2} + \frac{t-\mdp^2}{(s-\mdp^2)^2} \right] + 6\mdp^2 \left( \frac{1}{t-\mdp^2} + \frac{1}{s-\mdp^2} \right) + \left( \frac{s-\mdp^2}{t-\mdp^2} + 1 + \frac{t-\mdp^2}{s-\mdp^2} \right)^2 \right\}.
\end{split}\end{equation}
Due to crossing symmetry, we can reuse Eq.~(\ref{eq:xx'xx'}) for the scattering process $\chi(p_1) + \chi(p_2) \rightarrow \chi(p_3) + \chi(p_4)$ with $s$ replaced by $u$.

\subsection{$\Adp\leftrightarrow\chi\bar{\chi}$}
For the decay process $\Adp\rightarrow\chi\bar{\chi}$, we take into account the Pauli-blocking effect of the dark fermions. The decay rate is given by
\begin{equation}\label{eq:A'xx}
    \Gamma_{\Adp\rightarrow\chi\bar{\chi}} = \frac{\alpha_D (\mdp^2 + 2\mdm^2)}{3\omega|\vec{k}|} \frac{2T}{1-e^{-\omega/T}} \ln{\left\{\frac{\cosh{\left[(1+\xi_0)\omega/4T\right]}}{\cosh{\left[(1-\xi_0)\omega/4T\right]}}\right\}},
\end{equation}
where $\xi_0$ is similar to Eq.~(\ref{eq:xi0}),
\begin{equation}
    \xi_0 = \frac{|\vec{k}|}{\omega}\sqrt{1-\frac{4\mdm^2}{\mdp^2}}. 
\end{equation}
Note that in the limit $T\rightarrow0$, Eq.~(\ref{eq:A'xx}) reduces to the decay rate in vacuum given by Eq.~(2.6) in \cite{Chang:2018rso}.

The inverse decay rate of $\chi\bar{\chi}\rightarrow\Adp$, taking into account the bosonic enhancement of the dark photon, is given by
\begin{equation}
    \Gamma_{\chi\bar{\chi}\rightarrow\Adp} = \frac{\alpha_D (\mdp^2 + 2\mdm^2)}{2 E |\vec{p}|} \frac{T}{1 + e^{-E/T}} \ln{\left\{ \frac{\sinh{(\omega_+/2T)}\cosh{[(\omega_- - E)/2T]}}{\sinh{(\omega_-/2T)}\cosh{[(\omega_+ - E)/2T]}} \right\}},
\end{equation}
where $\omega_\pm$ is the same as Eq.~(\ref{eq:w+-}).

\section{Detailed Work Flow for Computing DS Luminosity}\label{app:scheme}
\begin{enumerate}
    \item We take the following steps to determine if a DS particle species is in the diffuse regime or not:
\begin{enumerate}[label=(\alph*)]
    \item Compute $\tilde{\lambda}^{-1}_i(E_i, r)$, the IMFP of each species $i$ assuming all DS particle species are in thermal equilibrium with the SM medium at temperature $T(r)$.
    \item Compute $\Delta t_i$, the escape timescale for each particle species $i$ with Eqs.~(\ref{eq:deltat}) and (\ref{eq:tdiff}).
    \item Compute $N_i$, the estimated abundance of each particle species with Eq.~(\ref{eq:DSabun}), and obtain the relative abundance $\eta_i \equiv N_i/N_i^{eq}$, where $N_i^{\rm eq}$ is the abundance of DS particle species $i$ assuming they are in thermal equilibrium with the SM medium. If $\eta_i > 1$, we simply set $\eta_i = 1$.
    \item Compute $\lambda^{-1}_i$, the IMFP of each species rescaled by the relative abundance as
    \begin{equation}
        \lambda^{-1}_i = \sum_{\rm possible\,final\,states} (\tilde{\lambda}^{-1}_{i\rightarrow \rm final\,states} + \sum_j \eta_j \tilde{\lambda}^{-1}_{i+j\rightarrow \rm final\,states}),
    \end{equation}
    where $j$ runs over all possible DS particle species. 
    \item Check the diffuse criteria for the dark photons: if $\eta_{\Adp_{L,T}} = 1$ and $\langle\lambda^{-1}_{\Adp_{L,T}}(R_\nu)\rangle - \langle\tilde{\lambda}^{-1}_{\Adp_{L,T}\rightarrow e^- e^+}(R_\nu)\rangle > R_\nu^{-1}$, then $\Adp_{L,T}$ is in the diffuse limit. Otherwise, $\Adp_{L,T}$ is treated as nondiffuse particles.
    \item Check the diffuse criteria for the dark fermion: if $\eta_\chi = 1$ and $\langle\lambda^{-1}_\chi(R_\nu)\rangle > R_\nu^{-1}$, then $\chi$ is in the diffuse regime. Otherwise, $\chi$ is treated as nondiffuse particles.
\end{enumerate}
    \item Depending on whether the DS particles are in the diffuse limit or not, we compute the luminosity of each particle species as follows:
\begin{enumerate}[label=(\alph*)]
    \item Regardless of the DS masses, the dark photon luminosity is given by
    \begin{equation}
        L_{\Adp_{L,T}} =
        \begin{cases}
            L_{\Adp_{L,T},\rm diff.}, \, \mbox{for diffuse } \Adp_{L,T} \mbox{ [use Eq.~\eqref{eq:flux}]},\\
            L_{\Adp_{L,T},\rm nondiff.}, \, \mbox{for nondiffuse } \Adp_{L,T} \mbox{ [use Eq.~\eqref{eq:LA'emis}]}.
        \end{cases}
    \end{equation}
    \item When $\mdp < 2\mdm$, the dark fermion luminosity is given by
    \begin{equation}
        L_\chi =
        \begin{cases}
            L_{\chi,\rm diff.}, \, \mbox{for diffuse } \chi \mbox{ [use Eq.~\eqref{eq:flux}]},\\
            L_{\chi,\rm nondiff.}, \, \mbox{for nondiffuse } \chi \mbox{ [use Eq.~\eqref{eq:Lchiemis}]}.
        \end{cases}
    \end{equation}
    \item When $\mdp > 2\mdm$, the dark fermion luminosity is given by
    \begin{equation}\label{eq:Lchi_free*}
        L_\chi = 
        \begin{cases}
            L_{\chi, \rm diff.}, \, \mbox{if } \chi \mbox{ is in the diffuse limit},\\
            L_{\chi, \rm nondiff.}, \, \mbox{if only } \Adp_L, \Adp_T \mbox{ are in the diffuse limit},\\
            \frac{2}{3} L_{\chi, \rm nondiff.}, \, \mbox{if only } \Adp_T \mbox{ is in the diffuse limit},\\
            \frac{1}{3} L_{\chi, \rm nondiff.}, \, \mbox{if only } \Adp_L \mbox{ is in the diffuse limit},\\
            0, \, \mbox{if no DS particle species is in the diffuse limit}.
        \end{cases}
    \end{equation}
    \item The total DS luminosity is $L_{\rm X} = L_{\Adp_L} + L_{\Adp_T} + 2L_\chi$.
\end{enumerate}
\end{enumerate}

\section{Radiative Transfer}\label{app:rad}
We now derive the energy flux carried by the DS particles through a surface of radius $r$ near the neutrinosphere, assuming they are in the diffuse limit. The derivation follows from Appendix I of \cite{Prialnik}. The equation of radiative transfer is given by
\begin{equation}\label{eq:rad}
    \frac{1}{\rho} \frac{\partial I}{\partial r} \cos{\theta} + \kappa I - j = 0,
\end{equation}
where $\rho$ is the matter density, $I$ is the intensity of radiation per unit solid angle per unit frequency, $\theta$ is the angle between the direction of radiation and the radial direction, $\kappa$ is the opacity, and $j$ is the total radiation power emitted per unit mass per unit frequency. We distinguish between the opacity contributions from scattering processes and absorption processes $\kappa = \kappa_s + \kappa_a$, and between the radiation by scattering and radiation emitted by matter $j = j_s + j_{em}$. In equilibrium and isotropic environment,
\begin{align}
    I^{\rm iso, eq} &\equiv B = \frac{g}{(2\pi)^3} \frac{E^2 p}{e^{E/T} \pm 1},\\
    j_{em}^{\rm iso, eq} &= \kappa_a B,
\end{align}
where $g$ is the degree of freedom, and the upper (lower) sign is for fermions (bosons). However, in an anisotropic environment, the relation between $j_{em}$ and $I$ is given by
\begin{equation}
    j_{em} = \kappa_a \left( 1 \pm e^{-E/T} \right) B \mp \kappa_a e^{-E/T} I.
\end{equation}
In the second term, the minus sign for fermion is due to Pauli blocking, while the plus sign for boson is due to stimulated emission. Substitute the above equation back into Eq.~(\ref{eq:rad}),
\begin{equation}\label{eq:rad2}
    \frac{1}{\rho} \frac{\partial I}{\partial r} \cos{\theta} + \kappa_s I - j_s + \kappa_a^\ast (I - B) = 0,
\end{equation}
where $\kappa_a^\ast \equiv \kappa_a (1 \pm e^{-E/T})$.
We assume that the radiation intensity $I$ is very close to $I^{\rm iso, eq}$.
Therefore, we can expand $I$ in terms of Legendre polynomials $P_n (\cos{\theta})$ and substitute it back into Eq.~(\ref{eq:rad2}). Keeping the terms up to $n=1$, we obtain
\begin{equation}
    I \simeq B - \frac{1}{\rho(\kappa_a^\ast + \kappa_s)} \frac{\partial B}{\partial r} \cos{\theta}.
\end{equation}
It follows that the energy flux through the spherical surface of radius $r$ is
\begin{equation}\begin{split}\label{eq:flux_app}
    L(r) &= 4\pi r^2 \int I \cos{\theta} dE d\Omega\\
    &= -\frac{16}{3}\pi^2 r^2 \frac{dT}{dr} \int \frac{1}{\rho(\kappa_a^\ast + \kappa_s)} \frac{\partial B}{\partial T} dE\\
    &= -\frac{2gr^2}{3\pi} \frac{T^3 dT}{dr} \int_{m/T}^{\infty} \frac{\xi^3\sqrt{\xi^2 - \left( \frac{m}{T}\right)^2}}{\lambda^{-1} (E = \xi T, r)} \frac{e^\xi}{\left( e^\xi \pm 1 \right)^2} d\xi,
\end{split}\end{equation}
where $m$ is the mass of the particle carrying the radiation, and $\lambda^{-1}$ is the effective IMFP defined as
\begin{equation}
    \lambda^{-1} = \rho(\kappa_a^\ast + \kappa_s).
\end{equation}
We can identify the absorptive IMFP $\lambda_{\rm abs}^{-1}$ with $\rho\kappa_a$ and the scattering IMFP $\lambda_{\rm sca}^{-1}$ with $\rho\kappa_s$. Thus, we can express the effective IMFP as
\begin{equation}
    \lambda^{-1}(E, r) \equiv \lambda_{\rm abs}^{-1}(E, r)(1 \pm e^{-E/T}) + \lambda_{\rm sca}^{-1}(E, r).
\end{equation}
We note that since the integral in Eq.~(\ref{eq:flux_app}) is dominated by the energies with small $\lambda^{-1}(E, r)$, it is possible that Eq.~(\ref{eq:flux_app}) could overestimate the energy flux if $\lambda^{-1}(E, r) \lesssim R_\nu$ for some energies $E$. We found that it occurs near the switching of diffuse and nondiffuse regimes, and thus it leads to orders of magnitude jumps of energy luminosity. To avoid this caveat, we approximate the IMFP $\lambda^{-1}(E, r)$ by the thermally averaged IMFP $\langle\lambda^{-1}(r)\rangle$ defined as
\begin{equation}
    \langle\lambda^{-1}(r)\rangle = \frac{\int d^3p\,f(E, T(r)) \lambda^{-1}(E, r)}{\int d^3p\,f(E, T(r))}.
\end{equation}
Thus, the energy flux is approximately given by
\begin{equation}
    L(r) \simeq -\frac{2gr^2}{3\pi} \frac{T^3 dT}{dr} \frac{1}{\langle\lambda^{-1}(r)\rangle} \int_{m/T}^{\infty} \xi^3\sqrt{\xi^2 - \left( \frac{m}{T}\right)^2} \frac{e^\xi}{\left( e^\xi \pm 1 \right)^2} d\xi.
\end{equation}
\end{widetext}

\bibliography{reference}

\begin{thebibliography}{100}%
\makeatletter
\providecommand \@ifxundefined [1]{%
 \@ifx{#1\undefined}
}%
\providecommand \@ifnum [1]{%
 \ifnum #1\expandafter \@firstoftwo
 \else \expandafter \@secondoftwo
 \fi
}%
\providecommand \@ifx [1]{%
 \ifx #1\expandafter \@firstoftwo
 \else \expandafter \@secondoftwo
 \fi
}%
\providecommand \natexlab [1]{#1}%
\providecommand \enquote  [1]{``#1''}%
\providecommand \bibnamefont  [1]{#1}%
\providecommand \bibfnamefont [1]{#1}%
\providecommand \citenamefont [1]{#1}%
\providecommand \href@noop [0]{\@secondoftwo}%
\providecommand \href [0]{\begingroup \@sanitize@url \@href}%
\providecommand \@href[1]{\@@startlink{#1}\@@href}%
\providecommand \@@href[1]{\endgroup#1\@@endlink}%
\providecommand \@sanitize@url [0]{\catcode `\\12\catcode `\$12\catcode
  `\&12\catcode `\#12\catcode `\^12\catcode `\_12\catcode `\%12\relax}%
\providecommand \@@startlink[1]{}%
\providecommand \@@endlink[0]{}%
\providecommand \url  [0]{\begingroup\@sanitize@url \@url }%
\providecommand \@url [1]{\endgroup\@href {#1}{\urlprefix }}%
\providecommand \urlprefix  [0]{URL }%
\providecommand \Eprint [0]{\href }%
\providecommand \doibase [0]{http://dx.doi.org/}%
\providecommand \selectlanguage [0]{\@gobble}%
\providecommand \bibinfo  [0]{\@secondoftwo}%
\providecommand \bibfield  [0]{\@secondoftwo}%
\providecommand \translation [1]{[#1]}%
\providecommand \BibitemOpen [0]{}%
\providecommand \bibitemStop [0]{}%
\providecommand \bibitemNoStop [0]{.\EOS\space}%
\providecommand \EOS [0]{\spacefactor3000\relax}%
\providecommand \BibitemShut  [1]{\csname bibitem#1\endcsname}%
\let\auto@bib@innerbib\@empty
\bibitem [{\citenamefont {Hirata}\ \emph {et~al.}(1987)\citenamefont {Hirata}
  \emph {et~al.}}]{Hirata:1987hu}%
  \BibitemOpen
  \bibfield  {author} {\bibinfo {author} {\bibfnamefont {K.}~\bibnamefont
  {Hirata}} \emph {et~al.} (\bibinfo {collaboration} {Kamiokande-II
  Collaboration}),\ }\href {\doibase 10.1103/PhysRevLett.58.1490} {\bibfield
  {journal} {\bibinfo  {journal} {Phys. Rev. Lett.}\ }\textbf {\bibinfo
  {volume} {58}},\ \bibinfo {pages} {1490} (\bibinfo {year}
  {1987})}\BibitemShut {NoStop}%
\bibitem [{\citenamefont {Bionta}\ \emph {et~al.}(1987)\citenamefont {Bionta}
  \emph {et~al.}}]{Bionta:1987qt}%
  \BibitemOpen
  \bibfield  {author} {\bibinfo {author} {\bibfnamefont {R.~M.}\ \bibnamefont
  {Bionta}} \emph {et~al.},\ }\href {\doibase 10.1103/PhysRevLett.58.1494}
  {\bibfield  {journal} {\bibinfo  {journal} {Phys. Rev. Lett.}\ }\textbf
  {\bibinfo {volume} {58}},\ \bibinfo {pages} {1494} (\bibinfo {year}
  {1987})}\BibitemShut {NoStop}%
\bibitem [{\citenamefont {Alekseev}\ \emph {et~al.}(1988)\citenamefont
  {Alekseev}, \citenamefont {Alekseeva}, \citenamefont {Krivosheina},\ and\
  \citenamefont {Volchenko}}]{Alekseev:1988gp}%
  \BibitemOpen
  \bibfield  {author} {\bibinfo {author} {\bibfnamefont {E.~N.}\ \bibnamefont
  {Alekseev}}, \bibinfo {author} {\bibfnamefont {L.~N.}\ \bibnamefont
  {Alekseeva}}, \bibinfo {author} {\bibfnamefont {I.~V.}\ \bibnamefont
  {Krivosheina}}, \ and\ \bibinfo {author} {\bibfnamefont {V.~I.}\ \bibnamefont
  {Volchenko}},\ }\href {\doibase 10.1016/0370-2693(88)91651-6} {\bibfield
  {journal} {\bibinfo  {journal} {Phys. Lett.}\ }\textbf {\bibinfo {volume}
  {B205}},\ \bibinfo {pages} {209} (\bibinfo {year} {1988})}\BibitemShut
  {NoStop}%
\bibitem [{\citenamefont {Sato}\ and\ \citenamefont
  {Suzuki}(1987)}]{Sato:1987rd}%
  \BibitemOpen
  \bibfield  {author} {\bibinfo {author} {\bibfnamefont {K.}~\bibnamefont
  {Sato}}\ and\ \bibinfo {author} {\bibfnamefont {H.}~\bibnamefont {Suzuki}},\
  }\href {\doibase 10.1103/PhysRevLett.58.2722} {\bibfield  {journal} {\bibinfo
   {journal} {Phys. Rev. Lett.}\ }\textbf {\bibinfo {volume} {58}},\ \bibinfo
  {pages} {2722} (\bibinfo {year} {1987})}\BibitemShut {NoStop}%
\bibitem [{\citenamefont {Spergel}\ \emph {et~al.}(1987)\citenamefont
  {Spergel}, \citenamefont {Piran}, \citenamefont {Loeb}, \citenamefont
  {Goodman},\ and\ \citenamefont {Bahcall}}]{Spergel:1987ch}%
  \BibitemOpen
  \bibfield  {author} {\bibinfo {author} {\bibfnamefont {D.~N.}\ \bibnamefont
  {Spergel}}, \bibinfo {author} {\bibfnamefont {T.}~\bibnamefont {Piran}},
  \bibinfo {author} {\bibfnamefont {A.}~\bibnamefont {Loeb}}, \bibinfo {author}
  {\bibfnamefont {J.}~\bibnamefont {Goodman}}, \ and\ \bibinfo {author}
  {\bibfnamefont {J.~N.}\ \bibnamefont {Bahcall}},\ }\href {\doibase
  10.1126/science.237.4821.1471} {\bibfield  {journal} {\bibinfo  {journal}
  {Science}\ }\textbf {\bibinfo {volume} {237}},\ \bibinfo {pages} {1471}
  (\bibinfo {year} {1987})}\BibitemShut {NoStop}%
\bibitem [{\citenamefont {{Bahcall}}\ \emph {et~al.}(1987)\citenamefont
  {{Bahcall}}, \citenamefont {{Piran}}, \citenamefont {{Press}},\ and\
  \citenamefont {{Spergel}}}]{Bahcall:1987}%
  \BibitemOpen
  \bibfield  {author} {\bibinfo {author} {\bibfnamefont {J.~N.}\ \bibnamefont
  {{Bahcall}}}, \bibinfo {author} {\bibfnamefont {T.}~\bibnamefont {{Piran}}},
  \bibinfo {author} {\bibfnamefont {W.~H.}\ \bibnamefont {{Press}}}, \ and\
  \bibinfo {author} {\bibfnamefont {D.~N.}\ \bibnamefont {{Spergel}}},\ }\href
  {\doibase 10.1038/327682a0} {\bibfield  {journal} {\bibinfo  {journal}
  {\nat}\ }\textbf {\bibinfo {volume} {327}},\ \bibinfo {pages} {682} (\bibinfo
  {year} {1987})}\BibitemShut {NoStop}%
\bibitem [{\citenamefont {{Burrows}}\ and\ \citenamefont
  {{Lattimer}}(1987)}]{Burrows:1987}%
  \BibitemOpen
  \bibfield  {author} {\bibinfo {author} {\bibfnamefont {A.}~\bibnamefont
  {{Burrows}}}\ and\ \bibinfo {author} {\bibfnamefont {J.~M.}\ \bibnamefont
  {{Lattimer}}},\ }\href {\doibase 10.1086/184938} {\bibfield  {journal}
  {\bibinfo  {journal} {Astrophys. J.}\ }\textbf {\bibinfo {volume} {318}},\
  \bibinfo {pages} {L63} (\bibinfo {year} {1987})}\BibitemShut {NoStop}%
\bibitem [{\citenamefont {Schramm}\ and\ \citenamefont
  {Truran}(1990)}]{Schramm:1990pf}%
  \BibitemOpen
  \bibfield  {author} {\bibinfo {author} {\bibfnamefont {D.}~\bibnamefont
  {Schramm}}\ and\ \bibinfo {author} {\bibfnamefont {J.}~\bibnamefont
  {Truran}},\ }\href {\doibase 10.1016/0370-1573(90)90020-3} {\bibfield
  {journal} {\bibinfo  {journal} {Phys. Rep.}\ }\textbf {\bibinfo {volume}
  {189}},\ \bibinfo {pages} {89} (\bibinfo {year} {1990})}\BibitemShut
  {NoStop}%
\bibitem [{\citenamefont {Loredo}\ and\ \citenamefont
  {Lamb}(2002)}]{Loredo:2001rx}%
  \BibitemOpen
  \bibfield  {author} {\bibinfo {author} {\bibfnamefont {T.~J.}\ \bibnamefont
  {Loredo}}\ and\ \bibinfo {author} {\bibfnamefont {D.~Q.}\ \bibnamefont
  {Lamb}},\ }\href {\doibase 10.1103/PhysRevD.65.063002} {\bibfield  {journal}
  {\bibinfo  {journal} {Phys. Rev. D}\ }\textbf {\bibinfo {volume} {65}},\
  \bibinfo {pages} {063002} (\bibinfo {year} {2002})},\ \Eprint
  {http://arxiv.org/abs/astro-ph/0107260} {arXiv:astro-ph/0107260 [astro-ph]}
  \BibitemShut {NoStop}%
\bibitem [{\citenamefont {Frieman}\ \emph {et~al.}(1988)\citenamefont
  {Frieman}, \citenamefont {Haber},\ and\ \citenamefont
  {Freese}}]{Frieman:1987as}%
  \BibitemOpen
  \bibfield  {author} {\bibinfo {author} {\bibfnamefont {J.~A.}\ \bibnamefont
  {Frieman}}, \bibinfo {author} {\bibfnamefont {H.~E.}\ \bibnamefont {Haber}},
  \ and\ \bibinfo {author} {\bibfnamefont {K.}~\bibnamefont {Freese}},\ }\href
  {\doibase 10.1016/0370-2693(88)91120-3} {\bibfield  {journal} {\bibinfo
  {journal} {Phys. Lett. B}\ }\textbf {\bibinfo {volume} {200}},\ \bibinfo
  {pages} {115} (\bibinfo {year} {1988})}\BibitemShut {NoStop}%
\bibitem [{\citenamefont {Kolb}\ and\ \citenamefont
  {Turner}(1989)}]{Kolb:1988pe}%
  \BibitemOpen
  \bibfield  {author} {\bibinfo {author} {\bibfnamefont {E.~W.}\ \bibnamefont
  {Kolb}}\ and\ \bibinfo {author} {\bibfnamefont {M.~S.}\ \bibnamefont
  {Turner}},\ }\href {\doibase 10.1103/PhysRevLett.62.509} {\bibfield
  {journal} {\bibinfo  {journal} {Phys. Rev. Lett.}\ }\textbf {\bibinfo
  {volume} {62}},\ \bibinfo {pages} {509} (\bibinfo {year} {1989})}\BibitemShut
  {NoStop}%
\bibitem [{\citenamefont {Berezhiani}\ and\ \citenamefont
  {Smirnov}(1989)}]{Berezhiani:1989za}%
  \BibitemOpen
  \bibfield  {author} {\bibinfo {author} {\bibfnamefont {Z.}~\bibnamefont
  {Berezhiani}}\ and\ \bibinfo {author} {\bibfnamefont {A.}~\bibnamefont
  {Smirnov}},\ }\href {\doibase 10.1016/0370-2693(89)90052-X} {\bibfield
  {journal} {\bibinfo  {journal} {Phys. Lett. B}\ }\textbf {\bibinfo {volume}
  {220}},\ \bibinfo {pages} {279} (\bibinfo {year} {1989})}\BibitemShut
  {NoStop}%
\bibitem [{\citenamefont {Farzan}(2003)}]{Farzan:2002wx}%
  \BibitemOpen
  \bibfield  {author} {\bibinfo {author} {\bibfnamefont {Y.}~\bibnamefont
  {Farzan}},\ }\href {\doibase 10.1103/PhysRevD.67.073015} {\bibfield
  {journal} {\bibinfo  {journal} {Phys. Rev. D}\ }\textbf {\bibinfo {volume}
  {67}},\ \bibinfo {pages} {073015} (\bibinfo {year} {2003})},\ \Eprint
  {http://arxiv.org/abs/hep-ph/0211375} {arXiv:hep-ph/0211375 [hep-ph]}
  \BibitemShut {NoStop}%
\bibitem [{\citenamefont {Raffelt}\ and\ \citenamefont
  {Seckel}(1988)}]{Raffelt:1987yt}%
  \BibitemOpen
  \bibfield  {author} {\bibinfo {author} {\bibfnamefont {G.}~\bibnamefont
  {Raffelt}}\ and\ \bibinfo {author} {\bibfnamefont {D.}~\bibnamefont
  {Seckel}},\ }\href {\doibase 10.1103/PhysRevLett.60.1793} {\bibfield
  {journal} {\bibinfo  {journal} {Phys. Rev. Lett.}\ }\textbf {\bibinfo
  {volume} {60}},\ \bibinfo {pages} {1793} (\bibinfo {year}
  {1988})}\BibitemShut {NoStop}%
\bibitem [{\citenamefont {Turner}(1988)}]{Turner:1987by}%
  \BibitemOpen
  \bibfield  {author} {\bibinfo {author} {\bibfnamefont {M.~S.}\ \bibnamefont
  {Turner}},\ }\href {\doibase 10.1103/PhysRevLett.60.1797} {\bibfield
  {journal} {\bibinfo  {journal} {Phys. Rev. Lett.}\ }\textbf {\bibinfo
  {volume} {60}},\ \bibinfo {pages} {1797} (\bibinfo {year}
  {1988})}\BibitemShut {NoStop}%
\bibitem [{\citenamefont {Mayle}\ \emph {et~al.}(1988)\citenamefont {Mayle},
  \citenamefont {Wilson}, \citenamefont {Ellis}, \citenamefont {Olive},
  \citenamefont {Schramm},\ and\ \citenamefont {Steigman}}]{Mayle:1987as}%
  \BibitemOpen
  \bibfield  {author} {\bibinfo {author} {\bibfnamefont {R.}~\bibnamefont
  {Mayle}}, \bibinfo {author} {\bibfnamefont {J.~R.}\ \bibnamefont {Wilson}},
  \bibinfo {author} {\bibfnamefont {J.~R.}\ \bibnamefont {Ellis}}, \bibinfo
  {author} {\bibfnamefont {K.~A.}\ \bibnamefont {Olive}}, \bibinfo {author}
  {\bibfnamefont {D.~N.}\ \bibnamefont {Schramm}}, \ and\ \bibinfo {author}
  {\bibfnamefont {G.}~\bibnamefont {Steigman}},\ }\href {\doibase
  10.1016/0370-2693(88)91595-X} {\bibfield  {journal} {\bibinfo  {journal}
  {Phys. Lett. B}\ }\textbf {\bibinfo {volume} {203}},\ \bibinfo {pages} {188}
  (\bibinfo {year} {1988})}\BibitemShut {NoStop}%
\bibitem [{\citenamefont {Brinkmann}\ and\ \citenamefont
  {Turner}(1988)}]{Brinkmann:1988vi}%
  \BibitemOpen
  \bibfield  {author} {\bibinfo {author} {\bibfnamefont {R.~P.}\ \bibnamefont
  {Brinkmann}}\ and\ \bibinfo {author} {\bibfnamefont {M.~S.}\ \bibnamefont
  {Turner}},\ }\href {\doibase 10.1103/PhysRevD.38.2338} {\bibfield  {journal}
  {\bibinfo  {journal} {Phys. Rev. D}\ }\textbf {\bibinfo {volume} {38}},\
  \bibinfo {pages} {2338} (\bibinfo {year} {1988})}\BibitemShut {NoStop}%
\bibitem [{\citenamefont {Janka}\ \emph {et~al.}(1996)\citenamefont {Janka},
  \citenamefont {Keil}, \citenamefont {Raffelt},\ and\ \citenamefont
  {Seckel}}]{Janka:1995ir}%
  \BibitemOpen
  \bibfield  {author} {\bibinfo {author} {\bibfnamefont {H.-T.}\ \bibnamefont
  {Janka}}, \bibinfo {author} {\bibfnamefont {W.}~\bibnamefont {Keil}},
  \bibinfo {author} {\bibfnamefont {G.}~\bibnamefont {Raffelt}}, \ and\
  \bibinfo {author} {\bibfnamefont {D.}~\bibnamefont {Seckel}},\ }\href
  {\doibase 10.1103/PhysRevLett.76.2621} {\bibfield  {journal} {\bibinfo
  {journal} {Phys. Rev. Lett.}\ }\textbf {\bibinfo {volume} {76}},\ \bibinfo
  {pages} {2621} (\bibinfo {year} {1996})},\ \Eprint
  {http://arxiv.org/abs/astro-ph/9507023} {arXiv:astro-ph/9507023 [astro-ph]}
  \BibitemShut {NoStop}%
\bibitem [{\citenamefont {Keil}\ \emph {et~al.}(1997)\citenamefont {Keil},
  \citenamefont {Janka}, \citenamefont {Schramm}, \citenamefont {Sigl},
  \citenamefont {Turner},\ and\ \citenamefont {Ellis}}]{Keil:1996ju}%
  \BibitemOpen
  \bibfield  {author} {\bibinfo {author} {\bibfnamefont {W.}~\bibnamefont
  {Keil}}, \bibinfo {author} {\bibfnamefont {H.-T.}\ \bibnamefont {Janka}},
  \bibinfo {author} {\bibfnamefont {D.~N.}\ \bibnamefont {Schramm}}, \bibinfo
  {author} {\bibfnamefont {G.}~\bibnamefont {Sigl}}, \bibinfo {author}
  {\bibfnamefont {M.~S.}\ \bibnamefont {Turner}}, \ and\ \bibinfo {author}
  {\bibfnamefont {J.~R.}\ \bibnamefont {Ellis}},\ }\href {\doibase
  10.1103/PhysRevD.56.2419} {\bibfield  {journal} {\bibinfo  {journal} {Phys.
  Rev. D}\ }\textbf {\bibinfo {volume} {56}},\ \bibinfo {pages} {2419}
  (\bibinfo {year} {1997})},\ \Eprint {http://arxiv.org/abs/astro-ph/9612222}
  {arXiv:astro-ph/9612222 [astro-ph]} \BibitemShut {NoStop}%
\bibitem [{\citenamefont {Fischer}\ \emph {et~al.}(2016)\citenamefont
  {Fischer}, \citenamefont {Chakraborty}, \citenamefont {Giannotti},
  \citenamefont {Mirizzi}, \citenamefont {Payez},\ and\ \citenamefont
  {Ringwald}}]{Fischer:2016cyd}%
  \BibitemOpen
  \bibfield  {author} {\bibinfo {author} {\bibfnamefont {T.}~\bibnamefont
  {Fischer}}, \bibinfo {author} {\bibfnamefont {S.}~\bibnamefont
  {Chakraborty}}, \bibinfo {author} {\bibfnamefont {M.}~\bibnamefont
  {Giannotti}}, \bibinfo {author} {\bibfnamefont {A.}~\bibnamefont {Mirizzi}},
  \bibinfo {author} {\bibfnamefont {A.}~\bibnamefont {Payez}}, \ and\ \bibinfo
  {author} {\bibfnamefont {A.}~\bibnamefont {Ringwald}},\ }\href {\doibase
  10.1103/PhysRevD.94.085012} {\bibfield  {journal} {\bibinfo  {journal} {Phys.
  Rev. D}\ }\textbf {\bibinfo {volume} {94}},\ \bibinfo {pages} {085012}
  (\bibinfo {year} {2016})},\ \Eprint {http://arxiv.org/abs/1605.08780}
  {arXiv:1605.08780 [astro-ph.HE]} \BibitemShut {NoStop}%
\bibitem [{\citenamefont {Carenza}\ \emph {et~al.}(2019)\citenamefont
  {Carenza}, \citenamefont {Fischer}, \citenamefont {Giannotti}, \citenamefont
  {Guo}, \citenamefont {Mart\'\i{}nez-Pinedo},\ and\ \citenamefont
  {Mirizzi}}]{Carenza:2019pxu}%
  \BibitemOpen
  \bibfield  {author} {\bibinfo {author} {\bibfnamefont {P.}~\bibnamefont
  {Carenza}}, \bibinfo {author} {\bibfnamefont {T.}~\bibnamefont {Fischer}},
  \bibinfo {author} {\bibfnamefont {M.}~\bibnamefont {Giannotti}}, \bibinfo
  {author} {\bibfnamefont {G.}~\bibnamefont {Guo}}, \bibinfo {author}
  {\bibfnamefont {G.}~\bibnamefont {Mart\'\i{}nez-Pinedo}}, \ and\ \bibinfo
  {author} {\bibfnamefont {A.}~\bibnamefont {Mirizzi}},\ }\href {\doibase
  10.1088/1475-7516/2019/10/016} {\bibfield  {journal} {\bibinfo  {journal} {J.
  Cosmol. Astropart. Phys.}\ }\textbf {\bibinfo {volume} {10}},\ \bibinfo
  {pages} {016} (\bibinfo {year} {2019})},\ \bibinfo {note} {[Erratum: JCAP 05,
  E01 (2020)]},\ \Eprint {http://arxiv.org/abs/1906.11844} {arXiv:1906.11844
  [hep-ph]} \BibitemShut {NoStop}%
\bibitem [{\citenamefont {Bollig}\ \emph {et~al.}(2020)\citenamefont {Bollig},
  \citenamefont {DeRocco}, \citenamefont {Graham},\ and\ \citenamefont
  {Janka}}]{Bollig:2020xdr}%
  \BibitemOpen
  \bibfield  {author} {\bibinfo {author} {\bibfnamefont {R.}~\bibnamefont
  {Bollig}}, \bibinfo {author} {\bibfnamefont {W.}~\bibnamefont {DeRocco}},
  \bibinfo {author} {\bibfnamefont {P.~W.}\ \bibnamefont {Graham}}, \ and\
  \bibinfo {author} {\bibfnamefont {H.-T.}\ \bibnamefont {Janka}},\ }\href
  {\doibase 10.1103/PhysRevLett.125.051104} {\bibfield  {journal} {\bibinfo
  {journal} {Phys. Rev. Lett.}\ }\textbf {\bibinfo {volume} {125}},\ \bibinfo
  {pages} {051104} (\bibinfo {year} {2020})},\ \Eprint
  {http://arxiv.org/abs/2005.07141} {arXiv:2005.07141 [hep-ph]} \BibitemShut
  {NoStop}%
\bibitem [{\citenamefont {Lucente}\ \emph {et~al.}(2020)\citenamefont
  {Lucente}, \citenamefont {Carenza}, \citenamefont {Fischer}, \citenamefont
  {Giannotti},\ and\ \citenamefont {Mirizzi}}]{Lucente:2020whw}%
  \BibitemOpen
  \bibfield  {author} {\bibinfo {author} {\bibfnamefont {G.}~\bibnamefont
  {Lucente}}, \bibinfo {author} {\bibfnamefont {P.}~\bibnamefont {Carenza}},
  \bibinfo {author} {\bibfnamefont {T.}~\bibnamefont {Fischer}}, \bibinfo
  {author} {\bibfnamefont {M.}~\bibnamefont {Giannotti}}, \ and\ \bibinfo
  {author} {\bibfnamefont {A.}~\bibnamefont {Mirizzi}},\ }\href {\doibase
  10.1088/1475-7516/2020/12/008} {\bibfield  {journal} {\bibinfo  {journal} {J.
  Cosmol. Astropart. Phys.}\ }\textbf {\bibinfo {volume} {12}},\ \bibinfo
  {pages} {008} (\bibinfo {year} {2020})},\ \Eprint
  {http://arxiv.org/abs/2008.04918} {arXiv:2008.04918 [hep-ph]} \BibitemShut
  {NoStop}%
\bibitem [{\citenamefont {Raffelt}\ and\ \citenamefont
  {Zhou}(2011)}]{Raffelt:2011nc}%
  \BibitemOpen
  \bibfield  {author} {\bibinfo {author} {\bibfnamefont {G.~G.}\ \bibnamefont
  {Raffelt}}\ and\ \bibinfo {author} {\bibfnamefont {S.}~\bibnamefont {Zhou}},\
  }\href {\doibase 10.1103/PhysRevD.83.093014} {\bibfield  {journal} {\bibinfo
  {journal} {Phys. Rev. D}\ }\textbf {\bibinfo {volume} {83}},\ \bibinfo
  {pages} {093014} (\bibinfo {year} {2011})},\ \Eprint
  {http://arxiv.org/abs/1102.5124} {arXiv:1102.5124 [hep-ph]} \BibitemShut
  {NoStop}%
\bibitem [{\citenamefont {Argüelles}\ \emph {et~al.}(2019)\citenamefont
  {Argüelles}, \citenamefont {Brdar},\ and\ \citenamefont
  {Kopp}}]{Arguelles:2016uwb}%
  \BibitemOpen
  \bibfield  {author} {\bibinfo {author} {\bibfnamefont {C.~A.}\ \bibnamefont
  {Argüelles}}, \bibinfo {author} {\bibfnamefont {V.}~\bibnamefont {Brdar}}, \
  and\ \bibinfo {author} {\bibfnamefont {J.}~\bibnamefont {Kopp}},\ }\href
  {\doibase 10.1103/PhysRevD.99.043012} {\bibfield  {journal} {\bibinfo
  {journal} {Phys. Rev. D}\ }\textbf {\bibinfo {volume} {99}},\ \bibinfo
  {pages} {043012} (\bibinfo {year} {2019})},\ \Eprint
  {http://arxiv.org/abs/1605.00654} {arXiv:1605.00654 [hep-ph]} \BibitemShut
  {NoStop}%
\bibitem [{\citenamefont {Suliga}\ \emph {et~al.}(2019)\citenamefont {Suliga},
  \citenamefont {Tamborra},\ and\ \citenamefont {Wu}}]{Suliga:2019bsq}%
  \BibitemOpen
  \bibfield  {author} {\bibinfo {author} {\bibfnamefont {A.~M.}\ \bibnamefont
  {Suliga}}, \bibinfo {author} {\bibfnamefont {I.}~\bibnamefont {Tamborra}}, \
  and\ \bibinfo {author} {\bibfnamefont {M.-R.}\ \bibnamefont {Wu}},\ }\href
  {\doibase 10.1088/1475-7516/2019/12/019} {\bibfield  {journal} {\bibinfo
  {journal} {J. Cosmol. Astropart. Phys.}\ }\textbf {\bibinfo {volume} {12}},\
  \bibinfo {pages} {019} (\bibinfo {year} {2019})},\ \Eprint
  {http://arxiv.org/abs/1908.11382} {arXiv:1908.11382 [astro-ph.HE]}
  \BibitemShut {NoStop}%
\bibitem [{\citenamefont {Syvolap}\ \emph {et~al.}(2019)\citenamefont
  {Syvolap}, \citenamefont {Ruchayskiy},\ and\ \citenamefont
  {Boyarsky}}]{Syvolap:2019dat}%
  \BibitemOpen
  \bibfield  {author} {\bibinfo {author} {\bibfnamefont {V.}~\bibnamefont
  {Syvolap}}, \bibinfo {author} {\bibfnamefont {O.}~\bibnamefont {Ruchayskiy}},
  \ and\ \bibinfo {author} {\bibfnamefont {A.}~\bibnamefont {Boyarsky}},\
  }\href@noop {} {\  (\bibinfo {year} {2019})},\ \Eprint
  {http://arxiv.org/abs/1909.06320} {arXiv:1909.06320 [hep-ph]} \BibitemShut
  {NoStop}%
\bibitem [{\citenamefont {Mastrototaro}\ \emph {et~al.}(2020)\citenamefont
  {Mastrototaro}, \citenamefont {Mirizzi}, \citenamefont {Serpico},\ and\
  \citenamefont {Esmaili}}]{Mastrototaro:2019vug}%
  \BibitemOpen
  \bibfield  {author} {\bibinfo {author} {\bibfnamefont {L.}~\bibnamefont
  {Mastrototaro}}, \bibinfo {author} {\bibfnamefont {A.}~\bibnamefont
  {Mirizzi}}, \bibinfo {author} {\bibfnamefont {P.~D.}\ \bibnamefont
  {Serpico}}, \ and\ \bibinfo {author} {\bibfnamefont {A.}~\bibnamefont
  {Esmaili}},\ }\href {\doibase 10.1088/1475-7516/2020/01/010} {\bibfield
  {journal} {\bibinfo  {journal} {J. Cosmol. Astropart. Phys.}\ }\textbf
  {\bibinfo {volume} {01}},\ \bibinfo {pages} {010} (\bibinfo {year} {2020})},\
  \Eprint {http://arxiv.org/abs/1910.10249} {arXiv:1910.10249 [hep-ph]}
  \BibitemShut {NoStop}%
\bibitem [{\citenamefont {Suliga}\ \emph {et~al.}(2020)\citenamefont {Suliga},
  \citenamefont {Tamborra},\ and\ \citenamefont {Wu}}]{Suliga:2020vpz}%
  \BibitemOpen
  \bibfield  {author} {\bibinfo {author} {\bibfnamefont {A.~M.}\ \bibnamefont
  {Suliga}}, \bibinfo {author} {\bibfnamefont {I.}~\bibnamefont {Tamborra}}, \
  and\ \bibinfo {author} {\bibfnamefont {M.-R.}\ \bibnamefont {Wu}},\ }\href
  {\doibase 10.1088/1475-7516/2020/08/018} {\bibfield  {journal} {\bibinfo
  {journal} {J. Cosmol. Astropart. Phys.}\ }\textbf {\bibinfo {volume} {08}},\
  \bibinfo {pages} {018} (\bibinfo {year} {2020})},\ \Eprint
  {http://arxiv.org/abs/2004.11389} {arXiv:2004.11389 [astro-ph.HE]}
  \BibitemShut {NoStop}%
\bibitem [{\citenamefont {Dent}\ \emph {et~al.}(2012)\citenamefont {Dent},
  \citenamefont {Ferrer},\ and\ \citenamefont {Krauss}}]{Dent:2012mx}%
  \BibitemOpen
  \bibfield  {author} {\bibinfo {author} {\bibfnamefont {J.~B.}\ \bibnamefont
  {Dent}}, \bibinfo {author} {\bibfnamefont {F.}~\bibnamefont {Ferrer}}, \ and\
  \bibinfo {author} {\bibfnamefont {L.~M.}\ \bibnamefont {Krauss}},\
  }\href@noop {} {\  (\bibinfo {year} {2012})},\ \Eprint
  {http://arxiv.org/abs/1201.2683} {arXiv:1201.2683 [astro-ph.CO]} \BibitemShut
  {NoStop}%
\bibitem [{\citenamefont {Rrapaj}\ and\ \citenamefont
  {Reddy}(2016)}]{Rrapaj:2015wgs}%
  \BibitemOpen
  \bibfield  {author} {\bibinfo {author} {\bibfnamefont {E.}~\bibnamefont
  {Rrapaj}}\ and\ \bibinfo {author} {\bibfnamefont {S.}~\bibnamefont {Reddy}},\
  }\href {\doibase 10.1103/PhysRevC.94.045805} {\bibfield  {journal} {\bibinfo
  {journal} {Phys. Rev. C}\ }\textbf {\bibinfo {volume} {94}},\ \bibinfo
  {pages} {045805} (\bibinfo {year} {2016})},\ \Eprint
  {http://arxiv.org/abs/1511.09136} {arXiv:1511.09136 [nucl-th]} \BibitemShut
  {NoStop}%
\bibitem [{\citenamefont {Mahoney}\ \emph {et~al.}(2017)\citenamefont
  {Mahoney}, \citenamefont {Leibovich},\ and\ \citenamefont
  {Zentner}}]{Mahoney:2017jqk}%
  \BibitemOpen
  \bibfield  {author} {\bibinfo {author} {\bibfnamefont {C.}~\bibnamefont
  {Mahoney}}, \bibinfo {author} {\bibfnamefont {A.~K.}\ \bibnamefont
  {Leibovich}}, \ and\ \bibinfo {author} {\bibfnamefont {A.~R.}\ \bibnamefont
  {Zentner}},\ }\href {\doibase 10.1103/PhysRevD.96.043018} {\bibfield
  {journal} {\bibinfo  {journal} {Phys. Rev. D}\ }\textbf {\bibinfo {volume}
  {96}},\ \bibinfo {pages} {043018} (\bibinfo {year} {2017})}\BibitemShut
  {NoStop}%
\bibitem [{\citenamefont {Chang}\ \emph {et~al.}(2017)\citenamefont {Chang},
  \citenamefont {Essig},\ and\ \citenamefont {McDermott}}]{Chang:2016ntp}%
  \BibitemOpen
  \bibfield  {author} {\bibinfo {author} {\bibfnamefont {J.~H.}\ \bibnamefont
  {Chang}}, \bibinfo {author} {\bibfnamefont {R.}~\bibnamefont {Essig}}, \ and\
  \bibinfo {author} {\bibfnamefont {S.~D.}\ \bibnamefont {McDermott}},\ }\href
  {\doibase 10.1007/JHEP01(2017)107} {\bibfield  {journal} {\bibinfo  {journal}
  {J. High Energy Phys.}\ }\textbf {\bibinfo {volume} {01}},\ \bibinfo {pages}
  {107} (\bibinfo {year} {2017})},\ \Eprint {http://arxiv.org/abs/1611.03864}
  {arXiv:1611.03864 [hep-ph]} \BibitemShut {NoStop}%
\bibitem [{\citenamefont {Hardy}\ and\ \citenamefont
  {Lasenby}(2017)}]{Hardy:2016kme}%
  \BibitemOpen
  \bibfield  {author} {\bibinfo {author} {\bibfnamefont {E.}~\bibnamefont
  {Hardy}}\ and\ \bibinfo {author} {\bibfnamefont {R.}~\bibnamefont
  {Lasenby}},\ }\href {\doibase 10.1007/JHEP02(2017)033} {\bibfield  {journal}
  {\bibinfo  {journal} {J. High Energy Phys.}\ }\textbf {\bibinfo {volume}
  {02}},\ \bibinfo {pages} {033} (\bibinfo {year} {2017})},\ \Eprint
  {http://arxiv.org/abs/1611.05852} {arXiv:1611.05852 [hep-ph]} \BibitemShut
  {NoStop}%
\bibitem [{\citenamefont {Chang}\ \emph {et~al.}(2018)\citenamefont {Chang},
  \citenamefont {Essig},\ and\ \citenamefont {McDermott}}]{Chang:2018rso}%
  \BibitemOpen
  \bibfield  {author} {\bibinfo {author} {\bibfnamefont {J.~H.}\ \bibnamefont
  {Chang}}, \bibinfo {author} {\bibfnamefont {R.}~\bibnamefont {Essig}}, \ and\
  \bibinfo {author} {\bibfnamefont {S.~D.}\ \bibnamefont {McDermott}},\ }\href
  {\doibase 10.1007/JHEP09(2018)051} {\bibfield  {journal} {\bibinfo  {journal}
  {J. High Energy Phys.}\ }\textbf {\bibinfo {volume} {09}},\ \bibinfo {pages}
  {051} (\bibinfo {year} {2018})},\ \Eprint {http://arxiv.org/abs/1803.00993}
  {arXiv:1803.00993 [hep-ph]} \BibitemShut {NoStop}%
\bibitem [{\citenamefont {Guha}\ \emph {et~al.}(2017)\citenamefont {Guha},
  \citenamefont {Selvaganapathy},\ and\ \citenamefont {Das}}]{Guha:2015kka}%
  \BibitemOpen
  \bibfield  {author} {\bibinfo {author} {\bibfnamefont {A.}~\bibnamefont
  {Guha}}, \bibinfo {author} {\bibfnamefont {J.}~\bibnamefont
  {Selvaganapathy}}, \ and\ \bibinfo {author} {\bibfnamefont {P.~K.}\
  \bibnamefont {Das}},\ }\href {\doibase 10.1103/PhysRevD.95.015001} {\bibfield
   {journal} {\bibinfo  {journal} {Phys. Rev. D}\ }\textbf {\bibinfo {volume}
  {95}},\ \bibinfo {pages} {015001} (\bibinfo {year} {2017})},\ \Eprint
  {http://arxiv.org/abs/1509.05901} {arXiv:1509.05901 [hep-ph]} \BibitemShut
  {NoStop}%
\bibitem [{\citenamefont {Guha}\ \emph {et~al.}(2019)\citenamefont {Guha},
  \citenamefont {Dev},\ and\ \citenamefont {Das}}]{Guha:2018mli}%
  \BibitemOpen
  \bibfield  {author} {\bibinfo {author} {\bibfnamefont {A.}~\bibnamefont
  {Guha}}, \bibinfo {author} {\bibfnamefont {P.~S.~B.}\ \bibnamefont {Dev}}, \
  and\ \bibinfo {author} {\bibfnamefont {P.~K.}\ \bibnamefont {Das}},\ }\href
  {\doibase 10.1088/1475-7516/2019/02/032} {\bibfield  {journal} {\bibinfo
  {journal} {J. Cosmol. Astropart. Phys.}\ }\textbf {\bibinfo {volume} {02}},\
  \bibinfo {pages} {032} (\bibinfo {year} {2019})},\ \Eprint
  {http://arxiv.org/abs/1810.00399} {arXiv:1810.00399 [hep-ph]} \BibitemShut
  {NoStop}%
\bibitem [{\citenamefont {Ishizuka}\ and\ \citenamefont
  {Yoshimura}(1990)}]{Ishizuka:1989ts}%
  \BibitemOpen
  \bibfield  {author} {\bibinfo {author} {\bibfnamefont {N.}~\bibnamefont
  {Ishizuka}}\ and\ \bibinfo {author} {\bibfnamefont {M.}~\bibnamefont
  {Yoshimura}},\ }\href {\doibase 10.1143/PTP.84.233} {\bibfield  {journal}
  {\bibinfo  {journal} {Prog. Theor. Phys.}\ }\textbf {\bibinfo {volume}
  {84}},\ \bibinfo {pages} {233} (\bibinfo {year} {1990})}\BibitemShut
  {NoStop}%
\bibitem [{\citenamefont {Arndt}\ and\ \citenamefont
  {Fox}(2003)}]{Arndt:2002yg}%
  \BibitemOpen
  \bibfield  {author} {\bibinfo {author} {\bibfnamefont {D.}~\bibnamefont
  {Arndt}}\ and\ \bibinfo {author} {\bibfnamefont {P.~J.}\ \bibnamefont
  {Fox}},\ }\href {\doibase 10.1088/1126-6708/2003/02/036} {\bibfield
  {journal} {\bibinfo  {journal} {J. High Energy Phys.}\ }\textbf {\bibinfo
  {volume} {02}},\ \bibinfo {pages} {036} (\bibinfo {year} {2003})},\ \Eprint
  {http://arxiv.org/abs/hep-ph/0207098} {arXiv:hep-ph/0207098 [hep-ph]}
  \BibitemShut {NoStop}%
\bibitem [{\citenamefont {Keung}\ \emph {et~al.}(2014)\citenamefont {Keung},
  \citenamefont {Ng}, \citenamefont {Tu},\ and\ \citenamefont
  {Yuan}}]{Keung:2013mfa}%
  \BibitemOpen
  \bibfield  {author} {\bibinfo {author} {\bibfnamefont {W.-Y.}\ \bibnamefont
  {Keung}}, \bibinfo {author} {\bibfnamefont {K.-W.}\ \bibnamefont {Ng}},
  \bibinfo {author} {\bibfnamefont {H.}~\bibnamefont {Tu}}, \ and\ \bibinfo
  {author} {\bibfnamefont {T.-C.}\ \bibnamefont {Yuan}},\ }\href {\doibase
  10.1103/PhysRevD.90.075014} {\bibfield  {journal} {\bibinfo  {journal} {Phys.
  Rev. D}\ }\textbf {\bibinfo {volume} {90}},\ \bibinfo {pages} {075014}
  (\bibinfo {year} {2014})},\ \Eprint {http://arxiv.org/abs/1312.3488}
  {arXiv:1312.3488 [hep-ph]} \BibitemShut {NoStop}%
\bibitem [{\citenamefont {Tu}\ and\ \citenamefont {Ng}(2017)}]{Tu:2017dhl}%
  \BibitemOpen
  \bibfield  {author} {\bibinfo {author} {\bibfnamefont {H.}~\bibnamefont
  {Tu}}\ and\ \bibinfo {author} {\bibfnamefont {K.-W.}\ \bibnamefont {Ng}},\
  }\href {\doibase 10.1007/JHEP07(2017)108} {\bibfield  {journal} {\bibinfo
  {journal} {J. High Energy Phys.}\ }\textbf {\bibinfo {volume} {07}},\
  \bibinfo {pages} {108} (\bibinfo {year} {2017})},\ \Eprint
  {http://arxiv.org/abs/1706.08340} {arXiv:1706.08340 [hep-ph]} \BibitemShut
  {NoStop}%
\bibitem [{\citenamefont {Dev}\ \emph {et~al.}(2020)\citenamefont {Dev},
  \citenamefont {Mohapatra},\ and\ \citenamefont {Zhang}}]{Dev:2020eam}%
  \BibitemOpen
  \bibfield  {author} {\bibinfo {author} {\bibfnamefont {P.~B.}\ \bibnamefont
  {Dev}}, \bibinfo {author} {\bibfnamefont {R.~N.}\ \bibnamefont {Mohapatra}},
  \ and\ \bibinfo {author} {\bibfnamefont {Y.}~\bibnamefont {Zhang}},\ }\href
  {\doibase 10.1088/1475-7516/2020/08/003} {\bibfield  {journal} {\bibinfo
  {journal} {J. Cosmol. Astropart. Phys.}\ }\textbf {\bibinfo {volume} {08}},\
  \bibinfo {pages} {003} (\bibinfo {year} {2020})},\ \bibinfo {note} {[Erratum:
  JCAP 11, E01 (2020)]},\ \Eprint {http://arxiv.org/abs/2005.00490}
  {arXiv:2005.00490 [hep-ph]} \BibitemShut {NoStop}%
\bibitem [{\citenamefont {Croon}\ \emph {et~al.}(2021)\citenamefont {Croon},
  \citenamefont {Elor}, \citenamefont {Leane},\ and\ \citenamefont
  {McDermott}}]{Croon:2020lrf}%
  \BibitemOpen
  \bibfield  {author} {\bibinfo {author} {\bibfnamefont {D.}~\bibnamefont
  {Croon}}, \bibinfo {author} {\bibfnamefont {G.}~\bibnamefont {Elor}},
  \bibinfo {author} {\bibfnamefont {R.~K.}\ \bibnamefont {Leane}}, \ and\
  \bibinfo {author} {\bibfnamefont {S.~D.}\ \bibnamefont {McDermott}},\ }\href
  {\doibase 10.1007/JHEP01(2021)107} {\bibfield  {journal} {\bibinfo  {journal}
  {J. High Energy Phys.}\ }\textbf {\bibinfo {volume} {01}},\ \bibinfo {pages}
  {107} (\bibinfo {year} {2021})},\ \Eprint {http://arxiv.org/abs/2006.13942}
  {arXiv:2006.13942 [hep-ph]} \BibitemShut {NoStop}%
\bibitem [{\citenamefont {Camalich}\ \emph {et~al.}(2020)\citenamefont
  {Camalich}, \citenamefont {Terol-Calvo}, \citenamefont {Tolos},\ and\
  \citenamefont {Ziegler}}]{Camalich:2020wac}%
  \BibitemOpen
  \bibfield  {author} {\bibinfo {author} {\bibfnamefont {J.~M.}\ \bibnamefont
  {Camalich}}, \bibinfo {author} {\bibfnamefont {J.}~\bibnamefont
  {Terol-Calvo}}, \bibinfo {author} {\bibfnamefont {L.}~\bibnamefont {Tolos}},
  \ and\ \bibinfo {author} {\bibfnamefont {R.}~\bibnamefont {Ziegler}},\
  }\href@noop {} {\  (\bibinfo {year} {2020})},\ \Eprint
  {http://arxiv.org/abs/2012.11632} {arXiv:2012.11632 [hep-ph]} \BibitemShut
  {NoStop}%
\bibitem [{\citenamefont {Hanhart}\ \emph
  {et~al.}(2001{\natexlab{a}})\citenamefont {Hanhart}, \citenamefont
  {Phillips}, \citenamefont {Reddy},\ and\ \citenamefont
  {Savage}}]{Hanhart:2000er}%
  \BibitemOpen
  \bibfield  {author} {\bibinfo {author} {\bibfnamefont {C.}~\bibnamefont
  {Hanhart}}, \bibinfo {author} {\bibfnamefont {D.~R.}\ \bibnamefont
  {Phillips}}, \bibinfo {author} {\bibfnamefont {S.}~\bibnamefont {Reddy}}, \
  and\ \bibinfo {author} {\bibfnamefont {M.~J.}\ \bibnamefont {Savage}},\
  }\href {\doibase 10.1016/S0550-3213(00)00667-2} {\bibfield  {journal}
  {\bibinfo  {journal} {Nucl. Phys.}\ }\textbf {\bibinfo {volume} {B595}},\
  \bibinfo {pages} {335} (\bibinfo {year} {2001}{\natexlab{a}})},\ \Eprint
  {http://arxiv.org/abs/nucl-th/0007016} {arXiv:nucl-th/0007016 [nucl-th]}
  \BibitemShut {NoStop}%
\bibitem [{\citenamefont {Hanhart}\ \emph
  {et~al.}(2001{\natexlab{b}})\citenamefont {Hanhart}, \citenamefont {Pons},
  \citenamefont {Phillips},\ and\ \citenamefont {Reddy}}]{Hanhart:2001fx}%
  \BibitemOpen
  \bibfield  {author} {\bibinfo {author} {\bibfnamefont {C.}~\bibnamefont
  {Hanhart}}, \bibinfo {author} {\bibfnamefont {J.~A.}\ \bibnamefont {Pons}},
  \bibinfo {author} {\bibfnamefont {D.~R.}\ \bibnamefont {Phillips}}, \ and\
  \bibinfo {author} {\bibfnamefont {S.}~\bibnamefont {Reddy}},\ }\href
  {\doibase 10.1016/S0370-2693(01)00544-5} {\bibfield  {journal} {\bibinfo
  {journal} {Phys. Lett.}\ }\textbf {\bibinfo {volume} {B509}},\ \bibinfo
  {pages} {1} (\bibinfo {year} {2001}{\natexlab{b}})},\ \Eprint
  {http://arxiv.org/abs/astro-ph/0102063} {arXiv:astro-ph/0102063 [astro-ph]}
  \BibitemShut {NoStop}%
\bibitem [{\citenamefont {Hannestad}\ \emph {et~al.}(2007)\citenamefont
  {Hannestad}, \citenamefont {Raffelt},\ and\ \citenamefont
  {Wong}}]{Hannestad:2007ys}%
  \BibitemOpen
  \bibfield  {author} {\bibinfo {author} {\bibfnamefont {S.}~\bibnamefont
  {Hannestad}}, \bibinfo {author} {\bibfnamefont {G.}~\bibnamefont {Raffelt}},
  \ and\ \bibinfo {author} {\bibfnamefont {Y.~Y.~Y.}\ \bibnamefont {Wong}},\
  }\href {\doibase 10.1103/PhysRevD.76.121701} {\bibfield  {journal} {\bibinfo
  {journal} {Phys. Rev. D}\ }\textbf {\bibinfo {volume} {76}},\ \bibinfo
  {pages} {121701} (\bibinfo {year} {2007})},\ \Eprint
  {http://arxiv.org/abs/0708.1404} {arXiv:0708.1404 [hep-ph]} \BibitemShut
  {NoStop}%
\bibitem [{\citenamefont {Freitas}\ and\ \citenamefont
  {Wyler}(2007)}]{Freitas:2007ip}%
  \BibitemOpen
  \bibfield  {author} {\bibinfo {author} {\bibfnamefont {A.}~\bibnamefont
  {Freitas}}\ and\ \bibinfo {author} {\bibfnamefont {D.}~\bibnamefont
  {Wyler}},\ }\href {\doibase 10.1088/1126-6708/2007/12/033} {\bibfield
  {journal} {\bibinfo  {journal} {J. High Energy Phys.}\ }\textbf {\bibinfo
  {volume} {12}},\ \bibinfo {pages} {033} (\bibinfo {year} {2007})},\ \Eprint
  {http://arxiv.org/abs/0708.4339} {arXiv:0708.4339 [hep-ph]} \BibitemShut
  {NoStop}%
\bibitem [{\citenamefont {Bar}\ \emph {et~al.}(2020)\citenamefont {Bar},
  \citenamefont {Blum},\ and\ \citenamefont {D'Amico}}]{Bar:2019ifz}%
  \BibitemOpen
  \bibfield  {author} {\bibinfo {author} {\bibfnamefont {N.}~\bibnamefont
  {Bar}}, \bibinfo {author} {\bibfnamefont {K.}~\bibnamefont {Blum}}, \ and\
  \bibinfo {author} {\bibfnamefont {G.}~\bibnamefont {D'Amico}},\ }\href
  {\doibase 10.1103/PhysRevD.101.123025} {\bibfield  {journal} {\bibinfo
  {journal} {Phys. Rev. D}\ }\textbf {\bibinfo {volume} {101}},\ \bibinfo
  {pages} {123025} (\bibinfo {year} {2020})},\ \Eprint
  {http://arxiv.org/abs/1907.05020} {arXiv:1907.05020 [hep-ph]} \BibitemShut
  {NoStop}%
\bibitem [{\citenamefont {Sung}\ \emph {et~al.}(2019)\citenamefont {Sung},
  \citenamefont {Tu},\ and\ \citenamefont {Wu}}]{Sung:2019xie}%
  \BibitemOpen
  \bibfield  {author} {\bibinfo {author} {\bibfnamefont {A.}~\bibnamefont
  {Sung}}, \bibinfo {author} {\bibfnamefont {H.}~\bibnamefont {Tu}}, \ and\
  \bibinfo {author} {\bibfnamefont {M.-R.}\ \bibnamefont {Wu}},\ }\href
  {\doibase 10.1103/PhysRevD.99.121305} {\bibfield  {journal} {\bibinfo
  {journal} {Phys. Rev. D}\ }\textbf {\bibinfo {volume} {99}},\ \bibinfo
  {pages} {121305} (\bibinfo {year} {2019})},\ \Eprint
  {http://arxiv.org/abs/1903.07923} {arXiv:1903.07923 [hep-ph]} \BibitemShut
  {NoStop}%
\bibitem [{\citenamefont {Kazanas}\ \emph {et~al.}(2014)\citenamefont
  {Kazanas}, \citenamefont {Mohapatra}, \citenamefont {Nussinov}, \citenamefont
  {Teplitz},\ and\ \citenamefont {Zhang}}]{Kazanas:2014mca}%
  \BibitemOpen
  \bibfield  {author} {\bibinfo {author} {\bibfnamefont {D.}~\bibnamefont
  {Kazanas}}, \bibinfo {author} {\bibfnamefont {R.~N.}\ \bibnamefont
  {Mohapatra}}, \bibinfo {author} {\bibfnamefont {S.}~\bibnamefont {Nussinov}},
  \bibinfo {author} {\bibfnamefont {V.~L.}\ \bibnamefont {Teplitz}}, \ and\
  \bibinfo {author} {\bibfnamefont {Y.}~\bibnamefont {Zhang}},\ }\href
  {\doibase 10.1016/j.nuclphysb.2014.11.009} {\bibfield  {journal} {\bibinfo
  {journal} {Nucl. Phys.}\ }\textbf {\bibinfo {volume} {B890}},\ \bibinfo
  {pages} {17} (\bibinfo {year} {2014})},\ \Eprint
  {http://arxiv.org/abs/1410.0221} {arXiv:1410.0221 [hep-ph]} \BibitemShut
  {NoStop}%
\bibitem [{\citenamefont {DeRocco}\ \emph
  {et~al.}(2019{\natexlab{a}})\citenamefont {DeRocco}, \citenamefont {Graham},
  \citenamefont {Kasen}, \citenamefont {Marques-Tavares},\ and\ \citenamefont
  {Rajendran}}]{DeRocco:2019njg}%
  \BibitemOpen
  \bibfield  {author} {\bibinfo {author} {\bibfnamefont {W.}~\bibnamefont
  {DeRocco}}, \bibinfo {author} {\bibfnamefont {P.~W.}\ \bibnamefont {Graham}},
  \bibinfo {author} {\bibfnamefont {D.}~\bibnamefont {Kasen}}, \bibinfo
  {author} {\bibfnamefont {G.}~\bibnamefont {Marques-Tavares}}, \ and\ \bibinfo
  {author} {\bibfnamefont {S.}~\bibnamefont {Rajendran}},\ }\href {\doibase
  10.1007/JHEP02(2019)171} {\bibfield  {journal} {\bibinfo  {journal} {J. High
  Energy Phys.}\ }\textbf {\bibinfo {volume} {02}},\ \bibinfo {pages} {171}
  (\bibinfo {year} {2019}{\natexlab{a}})},\ \Eprint
  {http://arxiv.org/abs/1901.08596} {arXiv:1901.08596 [hep-ph]} \BibitemShut
  {NoStop}%
\bibitem [{\citenamefont {DeRocco}\ \emph
  {et~al.}(2019{\natexlab{b}})\citenamefont {DeRocco}, \citenamefont {Graham},
  \citenamefont {Kasen}, \citenamefont {Marques-Tavares},\ and\ \citenamefont
  {Rajendran}}]{DeRocco:2019jti}%
  \BibitemOpen
  \bibfield  {author} {\bibinfo {author} {\bibfnamefont {W.}~\bibnamefont
  {DeRocco}}, \bibinfo {author} {\bibfnamefont {P.~W.}\ \bibnamefont {Graham}},
  \bibinfo {author} {\bibfnamefont {D.}~\bibnamefont {Kasen}}, \bibinfo
  {author} {\bibfnamefont {G.}~\bibnamefont {Marques-Tavares}}, \ and\ \bibinfo
  {author} {\bibfnamefont {S.}~\bibnamefont {Rajendran}},\ }\href {\doibase
  10.1103/PhysRevD.100.075018} {\bibfield  {journal} {\bibinfo  {journal}
  {Phys. Rev. D}\ }\textbf {\bibinfo {volume} {100}},\ \bibinfo {pages}
  {075018} (\bibinfo {year} {2019}{\natexlab{b}})},\ \Eprint
  {http://arxiv.org/abs/1905.09284} {arXiv:1905.09284 [hep-ph]} \BibitemShut
  {NoStop}%
\bibitem [{\citenamefont {Darm\'e}\ \emph {et~al.}(2020)\citenamefont
  {Darm\'e}, \citenamefont {Giacchino}, \citenamefont {Nardi},\ and\
  \citenamefont {Raggi}}]{Darme:2020sjf}%
  \BibitemOpen
  \bibfield  {author} {\bibinfo {author} {\bibfnamefont {L.}~\bibnamefont
  {Darm\'e}}, \bibinfo {author} {\bibfnamefont {F.}~\bibnamefont {Giacchino}},
  \bibinfo {author} {\bibfnamefont {E.}~\bibnamefont {Nardi}}, \ and\ \bibinfo
  {author} {\bibfnamefont {M.}~\bibnamefont {Raggi}},\ }\href@noop {} {\
  (\bibinfo {year} {2020})},\ \Eprint {http://arxiv.org/abs/2012.07894}
  {arXiv:2012.07894 [hep-ph]} \BibitemShut {NoStop}%
\bibitem [{\citenamefont {Spergel}\ and\ \citenamefont
  {Steinhardt}(2000)}]{Spergel:1999mh}%
  \BibitemOpen
  \bibfield  {author} {\bibinfo {author} {\bibfnamefont {D.~N.}\ \bibnamefont
  {Spergel}}\ and\ \bibinfo {author} {\bibfnamefont {P.~J.}\ \bibnamefont
  {Steinhardt}},\ }\href {\doibase 10.1103/PhysRevLett.84.3760} {\bibfield
  {journal} {\bibinfo  {journal} {Phys. Rev. Lett.}\ }\textbf {\bibinfo
  {volume} {84}},\ \bibinfo {pages} {3760} (\bibinfo {year} {2000})},\ \Eprint
  {http://arxiv.org/abs/astro-ph/9909386} {arXiv:astro-ph/9909386} \BibitemShut
  {NoStop}%
\bibitem [{\citenamefont {{Vogelsberger}}\ \emph {et~al.}(2012)\citenamefont
  {{Vogelsberger}}, \citenamefont {{Zavala}},\ and\ \citenamefont
  {{Loeb}}}]{Vogelsberger2012}%
  \BibitemOpen
  \bibfield  {author} {\bibinfo {author} {\bibfnamefont {M.}~\bibnamefont
  {{Vogelsberger}}}, \bibinfo {author} {\bibfnamefont {J.}~\bibnamefont
  {{Zavala}}}, \ and\ \bibinfo {author} {\bibfnamefont {A.}~\bibnamefont
  {{Loeb}}},\ }\href {\doibase 10.1111/j.1365-2966.2012.21182.x} {\bibfield
  {journal} {\bibinfo  {journal} {Mon. Not. R. Astron. Soc.}\ }\textbf
  {\bibinfo {volume} {423}},\ \bibinfo {pages} {3740} (\bibinfo {year}
  {2012})},\ \Eprint {http://arxiv.org/abs/1201.5892} {arXiv:1201.5892
  [astro-ph.CO]} \BibitemShut {NoStop}%
\bibitem [{\citenamefont {Rocha}\ \emph {et~al.}(2013)\citenamefont {Rocha},
  \citenamefont {Peter}, \citenamefont {Bullock}, \citenamefont {Kaplinghat},
  \citenamefont {Garrison-Kimmel}, \citenamefont {Onorbe},\ and\ \citenamefont
  {Moustakas}}]{Rocha:2012jg}%
  \BibitemOpen
  \bibfield  {author} {\bibinfo {author} {\bibfnamefont {M.}~\bibnamefont
  {Rocha}}, \bibinfo {author} {\bibfnamefont {A.~H.~G.}\ \bibnamefont {Peter}},
  \bibinfo {author} {\bibfnamefont {J.~S.}\ \bibnamefont {Bullock}}, \bibinfo
  {author} {\bibfnamefont {M.}~\bibnamefont {Kaplinghat}}, \bibinfo {author}
  {\bibfnamefont {S.}~\bibnamefont {Garrison-Kimmel}}, \bibinfo {author}
  {\bibfnamefont {J.}~\bibnamefont {Onorbe}}, \ and\ \bibinfo {author}
  {\bibfnamefont {L.~A.}\ \bibnamefont {Moustakas}},\ }\href {\doibase
  10.1093/mnras/sts514} {\bibfield  {journal} {\bibinfo  {journal} {Mon. Not.
  R. Astron. Soc.}\ }\textbf {\bibinfo {volume} {430}},\ \bibinfo {pages} {81}
  (\bibinfo {year} {2013})},\ \Eprint {http://arxiv.org/abs/1208.3025}
  {arXiv:1208.3025 [astro-ph.CO]} \BibitemShut {NoStop}%
\bibitem [{\citenamefont {Tulin}\ and\ \citenamefont
  {Yu}(2018)}]{Tulin:2017ara}%
  \BibitemOpen
  \bibfield  {author} {\bibinfo {author} {\bibfnamefont {S.}~\bibnamefont
  {Tulin}}\ and\ \bibinfo {author} {\bibfnamefont {H.-B.}\ \bibnamefont {Yu}},\
  }\href {\doibase 10.1016/j.physrep.2017.11.004} {\bibfield  {journal}
  {\bibinfo  {journal} {Phys. Rep.}\ }\textbf {\bibinfo {volume} {730}},\
  \bibinfo {pages} {1} (\bibinfo {year} {2018})},\ \Eprint
  {http://arxiv.org/abs/1705.02358} {arXiv:1705.02358 [hep-ph]} \BibitemShut
  {NoStop}%
\bibitem [{\citenamefont {Tulin}\ \emph {et~al.}(2013)\citenamefont {Tulin},
  \citenamefont {Yu},\ and\ \citenamefont {Zurek}}]{Tulin:2013teo}%
  \BibitemOpen
  \bibfield  {author} {\bibinfo {author} {\bibfnamefont {S.}~\bibnamefont
  {Tulin}}, \bibinfo {author} {\bibfnamefont {H.-B.}\ \bibnamefont {Yu}}, \
  and\ \bibinfo {author} {\bibfnamefont {K.~M.}\ \bibnamefont {Zurek}},\ }\href
  {\doibase 10.1103/PhysRevD.87.115007} {\bibfield  {journal} {\bibinfo
  {journal} {Phys. Rev. D}\ }\textbf {\bibinfo {volume} {87}},\ \bibinfo
  {pages} {115007} (\bibinfo {year} {2013})},\ \Eprint
  {http://arxiv.org/abs/1302.3898} {arXiv:1302.3898 [hep-ph]} \BibitemShut
  {NoStop}%
\bibitem [{\citenamefont {Bernal}\ \emph {et~al.}(2016)\citenamefont {Bernal},
  \citenamefont {Chu}, \citenamefont {Garcia-Cely}, \citenamefont {Hambye},\
  and\ \citenamefont {Zaldivar}}]{Bernal:2015ova}%
  \BibitemOpen
  \bibfield  {author} {\bibinfo {author} {\bibfnamefont {N.}~\bibnamefont
  {Bernal}}, \bibinfo {author} {\bibfnamefont {X.}~\bibnamefont {Chu}},
  \bibinfo {author} {\bibfnamefont {C.}~\bibnamefont {Garcia-Cely}}, \bibinfo
  {author} {\bibfnamefont {T.}~\bibnamefont {Hambye}}, \ and\ \bibinfo {author}
  {\bibfnamefont {B.}~\bibnamefont {Zaldivar}},\ }\href {\doibase
  10.1088/1475-7516/2016/03/018} {\bibfield  {journal} {\bibinfo  {journal} {J.
  Cosmol. Astropart. Phys.}\ }\textbf {\bibinfo {volume} {03}},\ \bibinfo
  {pages} {018} (\bibinfo {year} {2016})},\ \Eprint
  {http://arxiv.org/abs/1510.08063} {arXiv:1510.08063 [hep-ph]} \BibitemShut
  {NoStop}%
\bibitem [{\citenamefont {Kawasaki}\ \emph {et~al.}(2015)\citenamefont
  {Kawasaki}, \citenamefont {Kohri}, \citenamefont {Moroi},\ and\ \citenamefont
  {Takaesu}}]{Kawasaki:2015yya}%
  \BibitemOpen
  \bibfield  {author} {\bibinfo {author} {\bibfnamefont {M.}~\bibnamefont
  {Kawasaki}}, \bibinfo {author} {\bibfnamefont {K.}~\bibnamefont {Kohri}},
  \bibinfo {author} {\bibfnamefont {T.}~\bibnamefont {Moroi}}, \ and\ \bibinfo
  {author} {\bibfnamefont {Y.}~\bibnamefont {Takaesu}},\ }\href {\doibase
  10.1016/j.physletb.2015.10.048} {\bibfield  {journal} {\bibinfo  {journal}
  {Phys. Lett. B}\ }\textbf {\bibinfo {volume} {751}},\ \bibinfo {pages} {246}
  (\bibinfo {year} {2015})},\ \Eprint {http://arxiv.org/abs/1509.03665}
  {arXiv:1509.03665 [hep-ph]} \BibitemShut {NoStop}%
\bibitem [{\citenamefont {Bringmann}\ \emph {et~al.}(2017)\citenamefont
  {Bringmann}, \citenamefont {Kahlhoefer}, \citenamefont {Schmidt-Hoberg},\
  and\ \citenamefont {Walia}}]{Bringmann:2016din}%
  \BibitemOpen
  \bibfield  {author} {\bibinfo {author} {\bibfnamefont {T.}~\bibnamefont
  {Bringmann}}, \bibinfo {author} {\bibfnamefont {F.}~\bibnamefont
  {Kahlhoefer}}, \bibinfo {author} {\bibfnamefont {K.}~\bibnamefont
  {Schmidt-Hoberg}}, \ and\ \bibinfo {author} {\bibfnamefont {P.}~\bibnamefont
  {Walia}},\ }\href {\doibase 10.1103/PhysRevLett.118.141802} {\bibfield
  {journal} {\bibinfo  {journal} {Phys. Rev. Lett.}\ }\textbf {\bibinfo
  {volume} {118}},\ \bibinfo {pages} {141802} (\bibinfo {year} {2017})},\
  \Eprint {http://arxiv.org/abs/1612.00845} {arXiv:1612.00845 [hep-ph]}
  \BibitemShut {NoStop}%
\bibitem [{\citenamefont {Elor}\ \emph {et~al.}(2016)\citenamefont {Elor},
  \citenamefont {Rodd}, \citenamefont {Slatyer},\ and\ \citenamefont
  {Xue}}]{Elor:2015bho}%
  \BibitemOpen
  \bibfield  {author} {\bibinfo {author} {\bibfnamefont {G.}~\bibnamefont
  {Elor}}, \bibinfo {author} {\bibfnamefont {N.~L.}\ \bibnamefont {Rodd}},
  \bibinfo {author} {\bibfnamefont {T.~R.}\ \bibnamefont {Slatyer}}, \ and\
  \bibinfo {author} {\bibfnamefont {W.}~\bibnamefont {Xue}},\ }\href {\doibase
  10.1088/1475-7516/2016/06/024} {\bibfield  {journal} {\bibinfo  {journal} {J.
  Cosmol. Astropart. Phys.}\ }\textbf {\bibinfo {volume} {06}},\ \bibinfo
  {pages} {024} (\bibinfo {year} {2016})},\ \Eprint
  {http://arxiv.org/abs/1511.08787} {arXiv:1511.08787 [hep-ph]} \BibitemShut
  {NoStop}%
\bibitem [{\citenamefont {Ackermann}\ \emph {et~al.}(2015)\citenamefont
  {Ackermann} \emph {et~al.}}]{Ackermann:2015tah}%
  \BibitemOpen
  \bibfield  {author} {\bibinfo {author} {\bibfnamefont {M.}~\bibnamefont
  {Ackermann}} \emph {et~al.} (\bibinfo {collaboration} {Fermi-LAT
  Collaboration}),\ }\href {\doibase 10.1088/1475-7516/2015/09/008} {\bibfield
  {journal} {\bibinfo  {journal} {J. Cosmol. Astropart. Phys.}\ }\textbf
  {\bibinfo {volume} {09}},\ \bibinfo {pages} {008} (\bibinfo {year} {2015})},\
  \Eprint {http://arxiv.org/abs/1501.05464} {arXiv:1501.05464 [astro-ph.CO]}
  \BibitemShut {NoStop}%
\bibitem [{\citenamefont {Abdallah}\ \emph {et~al.}(2016)\citenamefont
  {Abdallah} \emph {et~al.}}]{Abdallah:2016ygi}%
  \BibitemOpen
  \bibfield  {author} {\bibinfo {author} {\bibfnamefont {H.}~\bibnamefont
  {Abdallah}} \emph {et~al.} (\bibinfo {collaboration} {H.E.S.S.
  Collaboration}),\ }\href {\doibase 10.1103/PhysRevLett.117.111301} {\bibfield
   {journal} {\bibinfo  {journal} {Phys. Rev. Lett.}\ }\textbf {\bibinfo
  {volume} {117}},\ \bibinfo {pages} {111301} (\bibinfo {year} {2016})},\
  \Eprint {http://arxiv.org/abs/1607.08142} {arXiv:1607.08142 [astro-ph.HE]}
  \BibitemShut {NoStop}%
\bibitem [{\citenamefont {Feng}\ \emph {et~al.}(2016)\citenamefont {Feng},
  \citenamefont {Smolinsky},\ and\ \citenamefont {Tanedo}}]{Feng:2016ijc}%
  \BibitemOpen
  \bibfield  {author} {\bibinfo {author} {\bibfnamefont {J.~L.}\ \bibnamefont
  {Feng}}, \bibinfo {author} {\bibfnamefont {J.}~\bibnamefont {Smolinsky}}, \
  and\ \bibinfo {author} {\bibfnamefont {P.}~\bibnamefont {Tanedo}},\ }\href
  {\doibase 10.1103/PhysRevD.93.115036, 10.1103/PhysRevD.96.099903} {\bibfield
  {journal} {\bibinfo  {journal} {Phys. Rev. D}\ }\textbf {\bibinfo {volume}
  {93}},\ \bibinfo {pages} {115036} (\bibinfo {year} {2016})},\ \bibinfo {note}
  {[Erratum: Phys. Rev.D96,no.9,099903(2017)]},\ \Eprint
  {http://arxiv.org/abs/1602.01465} {arXiv:1602.01465 [hep-ph]} \BibitemShut
  {NoStop}%
\bibitem [{\citenamefont {Leane}\ \emph {et~al.}(2017)\citenamefont {Leane},
  \citenamefont {Ng},\ and\ \citenamefont {Beacom}}]{Leane:2017vag}%
  \BibitemOpen
  \bibfield  {author} {\bibinfo {author} {\bibfnamefont {R.~K.}\ \bibnamefont
  {Leane}}, \bibinfo {author} {\bibfnamefont {K.~C.~Y.}\ \bibnamefont {Ng}}, \
  and\ \bibinfo {author} {\bibfnamefont {J.~F.}\ \bibnamefont {Beacom}},\
  }\href {\doibase 10.1103/PhysRevD.95.123016} {\bibfield  {journal} {\bibinfo
  {journal} {Phys. Rev. D}\ }\textbf {\bibinfo {volume} {95}},\ \bibinfo
  {pages} {123016} (\bibinfo {year} {2017})},\ \Eprint
  {http://arxiv.org/abs/1703.04629} {arXiv:1703.04629 [astro-ph.HE]}
  \BibitemShut {NoStop}%
\bibitem [{\citenamefont {Kouvaris}\ \emph {et~al.}(2016)\citenamefont
  {Kouvaris}, \citenamefont {Lang\ae{}ble},\ and\ \citenamefont
  {Nielsen}}]{Kouvaris:2016ltf}%
  \BibitemOpen
  \bibfield  {author} {\bibinfo {author} {\bibfnamefont {C.}~\bibnamefont
  {Kouvaris}}, \bibinfo {author} {\bibfnamefont {K.}~\bibnamefont
  {Lang\ae{}ble}}, \ and\ \bibinfo {author} {\bibfnamefont {N.~G.}\
  \bibnamefont {Nielsen}},\ }\href {\doibase 10.1088/1475-7516/2016/10/012}
  {\bibfield  {journal} {\bibinfo  {journal} {J. Cosmol. Astropart. Phys.}\
  }\textbf {\bibinfo {volume} {10}},\ \bibinfo {pages} {012} (\bibinfo {year}
  {2016})},\ \Eprint {http://arxiv.org/abs/1607.00374} {arXiv:1607.00374
  [hep-ph]} \BibitemShut {NoStop}%
\bibitem [{\citenamefont {Essig}\ \emph {et~al.}(2019)\citenamefont {Essig},
  \citenamefont {McDermott}, \citenamefont {Yu},\ and\ \citenamefont
  {Zhong}}]{Essig:2018pzq}%
  \BibitemOpen
  \bibfield  {author} {\bibinfo {author} {\bibfnamefont {R.}~\bibnamefont
  {Essig}}, \bibinfo {author} {\bibfnamefont {S.~D.}\ \bibnamefont
  {McDermott}}, \bibinfo {author} {\bibfnamefont {H.-B.}\ \bibnamefont {Yu}}, \
  and\ \bibinfo {author} {\bibfnamefont {Y.-M.}\ \bibnamefont {Zhong}},\ }\href
  {\doibase 10.1103/PhysRevLett.123.121102} {\bibfield  {journal} {\bibinfo
  {journal} {Phys. Rev. Lett.}\ }\textbf {\bibinfo {volume} {123}},\ \bibinfo
  {pages} {121102} (\bibinfo {year} {2019})},\ \Eprint
  {http://arxiv.org/abs/1809.01144} {arXiv:1809.01144 [hep-ph]} \BibitemShut
  {NoStop}%
\bibitem [{\citenamefont {Chang}\ \emph {et~al.}(2019)\citenamefont {Chang},
  \citenamefont {Egana-Ugrinovic}, \citenamefont {Essig},\ and\ \citenamefont
  {Kouvaris}}]{Chang:2018bgx}%
  \BibitemOpen
  \bibfield  {author} {\bibinfo {author} {\bibfnamefont {J.~H.}\ \bibnamefont
  {Chang}}, \bibinfo {author} {\bibfnamefont {D.}~\bibnamefont
  {Egana-Ugrinovic}}, \bibinfo {author} {\bibfnamefont {R.}~\bibnamefont
  {Essig}}, \ and\ \bibinfo {author} {\bibfnamefont {C.}~\bibnamefont
  {Kouvaris}},\ }\href {\doibase 10.1088/1475-7516/2019/03/036} {\bibfield
  {journal} {\bibinfo  {journal} {J. Cosmol. Astropart. Phys.}\ }\textbf
  {\bibinfo {volume} {03}},\ \bibinfo {pages} {036} (\bibinfo {year} {2019})},\
  \Eprint {http://arxiv.org/abs/1812.07000} {arXiv:1812.07000 [hep-ph]}
  \BibitemShut {NoStop}%
\bibitem [{\citenamefont {Depta}\ \emph {et~al.}(2019)\citenamefont {Depta},
  \citenamefont {Hufnagel}, \citenamefont {Schmidt-Hoberg},\ and\ \citenamefont
  {Wild}}]{Depta:2019lbe}%
  \BibitemOpen
  \bibfield  {author} {\bibinfo {author} {\bibfnamefont {P.~F.}\ \bibnamefont
  {Depta}}, \bibinfo {author} {\bibfnamefont {M.}~\bibnamefont {Hufnagel}},
  \bibinfo {author} {\bibfnamefont {K.}~\bibnamefont {Schmidt-Hoberg}}, \ and\
  \bibinfo {author} {\bibfnamefont {S.}~\bibnamefont {Wild}},\ }\href {\doibase
  10.1088/1475-7516/2019/04/029} {\bibfield  {journal} {\bibinfo  {journal} {J.
  Cosmol. Astropart. Phys.}\ }\textbf {\bibinfo {volume} {04}},\ \bibinfo
  {pages} {029} (\bibinfo {year} {2019})},\ \Eprint
  {http://arxiv.org/abs/1901.06944} {arXiv:1901.06944 [hep-ph]} \BibitemShut
  {NoStop}%
\bibitem [{\citenamefont {Bernal}\ \emph {et~al.}(2020)\citenamefont {Bernal},
  \citenamefont {Chu}, \citenamefont {Kulkarni},\ and\ \citenamefont
  {Pradler}}]{Bernal:2019uqr}%
  \BibitemOpen
  \bibfield  {author} {\bibinfo {author} {\bibfnamefont {N.}~\bibnamefont
  {Bernal}}, \bibinfo {author} {\bibfnamefont {X.}~\bibnamefont {Chu}},
  \bibinfo {author} {\bibfnamefont {S.}~\bibnamefont {Kulkarni}}, \ and\
  \bibinfo {author} {\bibfnamefont {J.}~\bibnamefont {Pradler}},\ }\href
  {\doibase 10.1103/PhysRevD.101.055044} {\bibfield  {journal} {\bibinfo
  {journal} {Phys. Rev. D}\ }\textbf {\bibinfo {volume} {101}},\ \bibinfo
  {pages} {055044} (\bibinfo {year} {2020})},\ \Eprint
  {http://arxiv.org/abs/1912.06681} {arXiv:1912.06681 [hep-ph]} \BibitemShut
  {NoStop}%
\bibitem [{\citenamefont {Foot}\ and\ \citenamefont
  {Vagnozzi}(2015)}]{Foot:2014uba}%
  \BibitemOpen
  \bibfield  {author} {\bibinfo {author} {\bibfnamefont {R.}~\bibnamefont
  {Foot}}\ and\ \bibinfo {author} {\bibfnamefont {S.}~\bibnamefont
  {Vagnozzi}},\ }\href {\doibase 10.1103/PhysRevD.91.023512} {\bibfield
  {journal} {\bibinfo  {journal} {Phys. Rev. D}\ }\textbf {\bibinfo {volume}
  {91}},\ \bibinfo {pages} {023512} (\bibinfo {year} {2015})},\ \Eprint
  {http://arxiv.org/abs/1409.7174} {arXiv:1409.7174 [hep-ph]} \BibitemShut
  {NoStop}%
\bibitem [{\citenamefont {Mezzacappa}\ \emph {et~al.}(2020)\citenamefont
  {Mezzacappa}, \citenamefont {Endeve}, \citenamefont {Messer},\ and\
  \citenamefont {Bruenn}}]{Mezzacappa:2020oyq}%
  \BibitemOpen
  \bibfield  {author} {\bibinfo {author} {\bibfnamefont {A.}~\bibnamefont
  {Mezzacappa}}, \bibinfo {author} {\bibfnamefont {E.}~\bibnamefont {Endeve}},
  \bibinfo {author} {\bibfnamefont {O.~B.}\ \bibnamefont {Messer}}, \ and\
  \bibinfo {author} {\bibfnamefont {S.~W.}\ \bibnamefont {Bruenn}},\
  }\href@noop {} {\  (\bibinfo {year} {2020})},\ \Eprint
  {http://arxiv.org/abs/2010.09013} {arXiv:2010.09013 [astro-ph.HE]}
  \BibitemShut {NoStop}%
\bibitem [{\citenamefont {Holdom}(1986)}]{Holdom:1985ag}%
  \BibitemOpen
  \bibfield  {author} {\bibinfo {author} {\bibfnamefont {B.}~\bibnamefont
  {Holdom}},\ }\href {\doibase 10.1016/0370-2693(86)91377-8} {\bibfield
  {journal} {\bibinfo  {journal} {Phys. Lett. B}\ }\textbf {\bibinfo {volume}
  {166}},\ \bibinfo {pages} {196} (\bibinfo {year} {1986})}\BibitemShut
  {NoStop}%
\bibitem [{\citenamefont {Okun}(1982)}]{Okun:1982xi}%
  \BibitemOpen
  \bibfield  {author} {\bibinfo {author} {\bibfnamefont {L.}~\bibnamefont
  {Okun}},\ }\href@noop {} {\bibfield  {journal} {\bibinfo  {journal} {Sov.
  Phys. JETP}\ }\textbf {\bibinfo {volume} {56}},\ \bibinfo {pages} {502}
  (\bibinfo {year} {1982})}\BibitemShut {NoStop}%
\bibitem [{\citenamefont {Pospelov}\ \emph {et~al.}(2008)\citenamefont
  {Pospelov}, \citenamefont {Ritz},\ and\ \citenamefont
  {Voloshin}}]{Pospelov:2007mp}%
  \BibitemOpen
  \bibfield  {author} {\bibinfo {author} {\bibfnamefont {M.}~\bibnamefont
  {Pospelov}}, \bibinfo {author} {\bibfnamefont {A.}~\bibnamefont {Ritz}}, \
  and\ \bibinfo {author} {\bibfnamefont {M.~B.}\ \bibnamefont {Voloshin}},\
  }\href {\doibase 10.1016/j.physletb.2008.02.052} {\bibfield  {journal}
  {\bibinfo  {journal} {Phys. Lett. B}\ }\textbf {\bibinfo {volume} {662}},\
  \bibinfo {pages} {53} (\bibinfo {year} {2008})},\ \Eprint
  {http://arxiv.org/abs/0711.4866} {arXiv:0711.4866 [hep-ph]} \BibitemShut
  {NoStop}%
\bibitem [{\citenamefont {Fischer}\ \emph {et~al.}(2010)\citenamefont
  {Fischer}, \citenamefont {Whitehouse}, \citenamefont {Mezzacappa},
  \citenamefont {Thielemann},\ and\ \citenamefont
  {Liebendorfer}}]{Fischer:2009af}%
  \BibitemOpen
  \bibfield  {author} {\bibinfo {author} {\bibfnamefont {T.}~\bibnamefont
  {Fischer}}, \bibinfo {author} {\bibfnamefont {S.~C.}\ \bibnamefont
  {Whitehouse}}, \bibinfo {author} {\bibfnamefont {A.}~\bibnamefont
  {Mezzacappa}}, \bibinfo {author} {\bibfnamefont {F.~K.}\ \bibnamefont
  {Thielemann}}, \ and\ \bibinfo {author} {\bibfnamefont {M.}~\bibnamefont
  {Liebendorfer}},\ }\href {\doibase 10.1051/0004-6361/200913106} {\bibfield
  {journal} {\bibinfo  {journal} {Astron. Astrophys.}\ }\textbf {\bibinfo
  {volume} {517}},\ \bibinfo {pages} {A80} (\bibinfo {year} {2010})},\ \Eprint
  {http://arxiv.org/abs/0908.1871} {arXiv:0908.1871 [astro-ph.HE]} \BibitemShut
  {NoStop}%
\bibitem [{\citenamefont {Raffelt}(1990)}]{Raffelt:1990yz}%
  \BibitemOpen
  \bibfield  {author} {\bibinfo {author} {\bibfnamefont {G.~G.}\ \bibnamefont
  {Raffelt}},\ }\href {\doibase 10.1016/0370-1573(90)90054-6} {\bibfield
  {journal} {\bibinfo  {journal} {Phys. Rep.}\ }\textbf {\bibinfo {volume}
  {198}},\ \bibinfo {pages} {1} (\bibinfo {year} {1990})}\BibitemShut {NoStop}%
\bibitem [{\citenamefont {An}\ \emph {et~al.}(2013)\citenamefont {An},
  \citenamefont {Pospelov},\ and\ \citenamefont {Pradler}}]{An:2013yfc}%
  \BibitemOpen
  \bibfield  {author} {\bibinfo {author} {\bibfnamefont {H.}~\bibnamefont
  {An}}, \bibinfo {author} {\bibfnamefont {M.}~\bibnamefont {Pospelov}}, \ and\
  \bibinfo {author} {\bibfnamefont {J.}~\bibnamefont {Pradler}},\ }\href
  {\doibase 10.1016/j.physletb.2013.07.008} {\bibfield  {journal} {\bibinfo
  {journal} {Phys. Lett. B}\ }\textbf {\bibinfo {volume} {725}},\ \bibinfo
  {pages} {190} (\bibinfo {year} {2013})},\ \Eprint
  {http://arxiv.org/abs/1302.3884} {arXiv:1302.3884 [hep-ph]} \BibitemShut
  {NoStop}%
\bibitem [{\citenamefont {Keil}\ \emph {et~al.}(2003)\citenamefont {Keil},
  \citenamefont {Raffelt},\ and\ \citenamefont {Janka}}]{Keil:2002in}%
  \BibitemOpen
  \bibfield  {author} {\bibinfo {author} {\bibfnamefont {M.~T.}\ \bibnamefont
  {Keil}}, \bibinfo {author} {\bibfnamefont {G.~G.}\ \bibnamefont {Raffelt}}, \
  and\ \bibinfo {author} {\bibfnamefont {H.-T.}\ \bibnamefont {Janka}},\ }\href
  {\doibase 10.1086/375130} {\bibfield  {journal} {\bibinfo  {journal}
  {Astrophys. J.}\ }\textbf {\bibinfo {volume} {590}},\ \bibinfo {pages} {971}
  (\bibinfo {year} {2003})},\ \Eprint {http://arxiv.org/abs/astro-ph/0208035}
  {arXiv:astro-ph/0208035} \BibitemShut {NoStop}%
\bibitem [{\citenamefont {{Shapiro}}\ and\ \citenamefont
  {{Teukolsky}}(1983)}]{1983bhwd.book}%
  \BibitemOpen
  \bibfield  {author} {\bibinfo {author} {\bibfnamefont {S.~L.}\ \bibnamefont
  {{Shapiro}}}\ and\ \bibinfo {author} {\bibfnamefont {S.~A.}\ \bibnamefont
  {{Teukolsky}}},\ }\href@noop {} {\emph {\bibinfo {title} {{Black Holes, White
  Dwarfs, and Neutron Stars : The Physics of Compact Objects}}}}\ (\bibinfo
  {publisher} {Wiley},\ \bibinfo {address} {New York},\ \bibinfo {year}
  {1983})\BibitemShut {NoStop}%
\bibitem [{\citenamefont {Andreas}\ \emph {et~al.}(2012)\citenamefont
  {Andreas}, \citenamefont {Niebuhr},\ and\ \citenamefont
  {Ringwald}}]{Andreas:2012mt}%
  \BibitemOpen
  \bibfield  {author} {\bibinfo {author} {\bibfnamefont {S.}~\bibnamefont
  {Andreas}}, \bibinfo {author} {\bibfnamefont {C.}~\bibnamefont {Niebuhr}}, \
  and\ \bibinfo {author} {\bibfnamefont {A.}~\bibnamefont {Ringwald}},\ }\href
  {\doibase 10.1103/PhysRevD.86.095019} {\bibfield  {journal} {\bibinfo
  {journal} {Phys. Rev. D}\ }\textbf {\bibinfo {volume} {86}},\ \bibinfo
  {pages} {095019} (\bibinfo {year} {2012})},\ \Eprint
  {http://arxiv.org/abs/1209.6083} {arXiv:1209.6083 [hep-ph]} \BibitemShut
  {NoStop}%
\bibitem [{\citenamefont {Jaeckel}\ \emph {et~al.}(2008)\citenamefont
  {Jaeckel}, \citenamefont {Redondo},\ and\ \citenamefont
  {Ringwald}}]{Jaeckel:2008fi}%
  \BibitemOpen
  \bibfield  {author} {\bibinfo {author} {\bibfnamefont {J.}~\bibnamefont
  {Jaeckel}}, \bibinfo {author} {\bibfnamefont {J.}~\bibnamefont {Redondo}}, \
  and\ \bibinfo {author} {\bibfnamefont {A.}~\bibnamefont {Ringwald}},\ }\href
  {\doibase 10.1103/PhysRevLett.101.131801} {\bibfield  {journal} {\bibinfo
  {journal} {Phys. Rev. Lett.}\ }\textbf {\bibinfo {volume} {101}},\ \bibinfo
  {pages} {131801} (\bibinfo {year} {2008})},\ \Eprint
  {http://arxiv.org/abs/0804.4157} {arXiv:0804.4157 [astro-ph]} \BibitemShut
  {NoStop}%
\bibitem [{\citenamefont {Mirizzi}\ \emph {et~al.}(2009)\citenamefont
  {Mirizzi}, \citenamefont {Redondo},\ and\ \citenamefont
  {Sigl}}]{Mirizzi:2009iz}%
  \BibitemOpen
  \bibfield  {author} {\bibinfo {author} {\bibfnamefont {A.}~\bibnamefont
  {Mirizzi}}, \bibinfo {author} {\bibfnamefont {J.}~\bibnamefont {Redondo}}, \
  and\ \bibinfo {author} {\bibfnamefont {G.}~\bibnamefont {Sigl}},\ }\href
  {\doibase 10.1088/1475-7516/2009/03/026} {\bibfield  {journal} {\bibinfo
  {journal} {J. Cosmol. Astropart. Phys.}\ }\textbf {\bibinfo {volume} {03}},\
  \bibinfo {pages} {026} (\bibinfo {year} {2009})},\ \Eprint
  {http://arxiv.org/abs/0901.0014} {arXiv:0901.0014 [hep-ph]} \BibitemShut
  {NoStop}%
\bibitem [{\citenamefont {Redondo}\ and\ \citenamefont
  {Raffelt}(2013)}]{Redondo:2013lna}%
  \BibitemOpen
  \bibfield  {author} {\bibinfo {author} {\bibfnamefont {J.}~\bibnamefont
  {Redondo}}\ and\ \bibinfo {author} {\bibfnamefont {G.}~\bibnamefont
  {Raffelt}},\ }\href {\doibase 10.1088/1475-7516/2013/08/034} {\bibfield
  {journal} {\bibinfo  {journal} {J. Cosmol. Astropart. Phys.}\ }\textbf
  {\bibinfo {volume} {08}},\ \bibinfo {pages} {034} (\bibinfo {year} {2013})},\
  \Eprint {http://arxiv.org/abs/1305.2920} {arXiv:1305.2920 [hep-ph]}
  \BibitemShut {NoStop}%
\bibitem [{\citenamefont {Fradette}\ \emph {et~al.}(2014)\citenamefont
  {Fradette}, \citenamefont {Pospelov}, \citenamefont {Pradler},\ and\
  \citenamefont {Ritz}}]{Fradette:2014sza}%
  \BibitemOpen
  \bibfield  {author} {\bibinfo {author} {\bibfnamefont {A.}~\bibnamefont
  {Fradette}}, \bibinfo {author} {\bibfnamefont {M.}~\bibnamefont {Pospelov}},
  \bibinfo {author} {\bibfnamefont {J.}~\bibnamefont {Pradler}}, \ and\
  \bibinfo {author} {\bibfnamefont {A.}~\bibnamefont {Ritz}},\ }\href {\doibase
  10.1103/PhysRevD.90.035022} {\bibfield  {journal} {\bibinfo  {journal} {Phys.
  Rev. D}\ }\textbf {\bibinfo {volume} {90}},\ \bibinfo {pages} {035022}
  (\bibinfo {year} {2014})},\ \Eprint {http://arxiv.org/abs/1407.0993}
  {arXiv:1407.0993 [hep-ph]} \BibitemShut {NoStop}%
\bibitem [{\citenamefont {An}\ \emph {et~al.}(2020)\citenamefont {An},
  \citenamefont {Pospelov}, \citenamefont {Pradler},\ and\ \citenamefont
  {Ritz}}]{An:2020bxd}%
  \BibitemOpen
  \bibfield  {author} {\bibinfo {author} {\bibfnamefont {H.}~\bibnamefont
  {An}}, \bibinfo {author} {\bibfnamefont {M.}~\bibnamefont {Pospelov}},
  \bibinfo {author} {\bibfnamefont {J.}~\bibnamefont {Pradler}}, \ and\
  \bibinfo {author} {\bibfnamefont {A.}~\bibnamefont {Ritz}},\ }\href {\doibase
  10.1103/PhysRevD.102.115022} {\bibfield  {journal} {\bibinfo  {journal}
  {Phys. Rev. D}\ }\textbf {\bibinfo {volume} {102}},\ \bibinfo {pages}
  {115022} (\bibinfo {year} {2020})},\ \Eprint
  {http://arxiv.org/abs/2006.13929} {arXiv:2006.13929 [hep-ph]} \BibitemShut
  {NoStop}%
\bibitem [{\citenamefont {Li}\ \emph {et~al.}(2020)\citenamefont {Li},
  \citenamefont {Fuller},\ and\ \citenamefont {Grohs}}]{Li:2020roy}%
  \BibitemOpen
  \bibfield  {author} {\bibinfo {author} {\bibfnamefont {J.-T.}\ \bibnamefont
  {Li}}, \bibinfo {author} {\bibfnamefont {G.~M.}\ \bibnamefont {Fuller}}, \
  and\ \bibinfo {author} {\bibfnamefont {E.}~\bibnamefont {Grohs}},\ }\href
  {\doibase 10.1088/1475-7516/2020/12/049} {\bibfield  {journal} {\bibinfo
  {journal} {J. Cosmol. Astropart. Phys.}\ }\textbf {\bibinfo {volume} {12}},\
  \bibinfo {pages} {049} (\bibinfo {year} {2020})},\ \Eprint
  {http://arxiv.org/abs/2009.14325} {arXiv:2009.14325 [astro-ph.CO]}
  \BibitemShut {NoStop}%
\bibitem [{\citenamefont {Sieverding}\ \emph {et~al.}(2021)\citenamefont
  {Sieverding}, \citenamefont {Rrapaj}, \citenamefont {Guo},\ and\
  \citenamefont {Qian}}]{Sieverding:2021jfa}%
  \BibitemOpen
  \bibfield  {author} {\bibinfo {author} {\bibfnamefont {A.}~\bibnamefont
  {Sieverding}}, \bibinfo {author} {\bibfnamefont {E.}~\bibnamefont {Rrapaj}},
  \bibinfo {author} {\bibfnamefont {G.}~\bibnamefont {Guo}}, \ and\ \bibinfo
  {author} {\bibfnamefont {Y.~Z.}\ \bibnamefont {Qian}},\ }\href@noop {} {\
  (\bibinfo {year} {2021})},\ \Eprint {http://arxiv.org/abs/2101.08672}
  {arXiv:2101.08672 [astro-ph.SR]} \BibitemShut {NoStop}%
\bibitem [{\citenamefont {Kreisch}\ \emph {et~al.}(2020)\citenamefont
  {Kreisch}, \citenamefont {Cyr-Racine},\ and\ \citenamefont
  {Dor\'e}}]{Kreisch:2019yzn}%
  \BibitemOpen
  \bibfield  {author} {\bibinfo {author} {\bibfnamefont {C.~D.}\ \bibnamefont
  {Kreisch}}, \bibinfo {author} {\bibfnamefont {F.-Y.}\ \bibnamefont
  {Cyr-Racine}}, \ and\ \bibinfo {author} {\bibfnamefont {O.}~\bibnamefont
  {Dor\'e}},\ }\href {\doibase 10.1103/PhysRevD.101.123505} {\bibfield
  {journal} {\bibinfo  {journal} {Phys. Rev. D}\ }\textbf {\bibinfo {volume}
  {101}},\ \bibinfo {pages} {123505} (\bibinfo {year} {2020})},\ \Eprint
  {http://arxiv.org/abs/1902.00534} {arXiv:1902.00534 [astro-ph.CO]}
  \BibitemShut {NoStop}%
\bibitem [{\citenamefont {Ellis}\ \emph {et~al.}(2018)\citenamefont {Ellis},
  \citenamefont {Hektor}, \citenamefont {H\"utsi}, \citenamefont {Kannike},
  \citenamefont {Marzola}, \citenamefont {Raidal},\ and\ \citenamefont
  {Vaskonen}}]{Ellis:2017jgp}%
  \BibitemOpen
  \bibfield  {author} {\bibinfo {author} {\bibfnamefont {J.}~\bibnamefont
  {Ellis}}, \bibinfo {author} {\bibfnamefont {A.}~\bibnamefont {Hektor}},
  \bibinfo {author} {\bibfnamefont {G.}~\bibnamefont {H\"utsi}}, \bibinfo
  {author} {\bibfnamefont {K.}~\bibnamefont {Kannike}}, \bibinfo {author}
  {\bibfnamefont {L.}~\bibnamefont {Marzola}}, \bibinfo {author} {\bibfnamefont
  {M.}~\bibnamefont {Raidal}}, \ and\ \bibinfo {author} {\bibfnamefont
  {V.}~\bibnamefont {Vaskonen}},\ }\href {\doibase
  10.1016/j.physletb.2018.04.048} {\bibfield  {journal} {\bibinfo  {journal}
  {Phys. Lett. B}\ }\textbf {\bibinfo {volume} {781}},\ \bibinfo {pages} {607}
  (\bibinfo {year} {2018})},\ \Eprint {http://arxiv.org/abs/1710.05540}
  {arXiv:1710.05540 [astro-ph.CO]} \BibitemShut {NoStop}%
\bibitem [{\citenamefont {Nelson}\ \emph {et~al.}(2019)\citenamefont {Nelson},
  \citenamefont {Reddy},\ and\ \citenamefont {Zhou}}]{Nelson:2018xtr}%
  \BibitemOpen
  \bibfield  {author} {\bibinfo {author} {\bibfnamefont {A.}~\bibnamefont
  {Nelson}}, \bibinfo {author} {\bibfnamefont {S.}~\bibnamefont {Reddy}}, \
  and\ \bibinfo {author} {\bibfnamefont {D.}~\bibnamefont {Zhou}},\ }\href
  {\doibase 10.1088/1475-7516/2019/07/012} {\bibfield  {journal} {\bibinfo
  {journal} {J. Cosmol. Astropart. Phys.}\ }\textbf {\bibinfo {volume} {07}},\
  \bibinfo {pages} {012} (\bibinfo {year} {2019})},\ \Eprint
  {http://arxiv.org/abs/1803.03266} {arXiv:1803.03266 [hep-ph]} \BibitemShut
  {NoStop}%
\bibitem [{\citenamefont {Bauswein}\ \emph {et~al.}(2020)\citenamefont
  {Bauswein}, \citenamefont {Guo}, \citenamefont {Lien}, \citenamefont {Lin},\
  and\ \citenamefont {Wu}}]{Bauswein:2020kor}%
  \BibitemOpen
  \bibfield  {author} {\bibinfo {author} {\bibfnamefont {A.}~\bibnamefont
  {Bauswein}}, \bibinfo {author} {\bibfnamefont {G.}~\bibnamefont {Guo}},
  \bibinfo {author} {\bibfnamefont {J.-H.}\ \bibnamefont {Lien}}, \bibinfo
  {author} {\bibfnamefont {Y.-H.}\ \bibnamefont {Lin}}, \ and\ \bibinfo
  {author} {\bibfnamefont {M.-R.}\ \bibnamefont {Wu}},\ }\href@noop {} {\
  (\bibinfo {year} {2020})},\ \Eprint {http://arxiv.org/abs/2012.11908}
  {arXiv:2012.11908 [astro-ph.HE]} \BibitemShut {NoStop}%
\bibitem [{\citenamefont {Weldon}(1983)}]{Weldon:1983jn}%
  \BibitemOpen
  \bibfield  {author} {\bibinfo {author} {\bibfnamefont {H.}~\bibnamefont
  {Weldon}},\ }\href {\doibase 10.1103/PhysRevD.28.2007} {\bibfield  {journal}
  {\bibinfo  {journal} {Phys. Rev. D}\ }\textbf {\bibinfo {volume} {28}},\
  \bibinfo {pages} {2007} (\bibinfo {year} {1983})}\BibitemShut {NoStop}%
\bibitem [{\citenamefont {Braaten}\ and\ \citenamefont
  {Segel}(1993)}]{Braaten:1993jw}%
  \BibitemOpen
  \bibfield  {author} {\bibinfo {author} {\bibfnamefont {E.}~\bibnamefont
  {Braaten}}\ and\ \bibinfo {author} {\bibfnamefont {D.}~\bibnamefont
  {Segel}},\ }\href {\doibase 10.1103/PhysRevD.48.1478} {\bibfield  {journal}
  {\bibinfo  {journal} {Phys. Rev. D}\ }\textbf {\bibinfo {volume} {48}},\
  \bibinfo {pages} {1478} (\bibinfo {year} {1993})},\ \Eprint
  {http://arxiv.org/abs/hep-ph/9302213} {arXiv:hep-ph/9302213 [hep-ph]}
  \BibitemShut {NoStop}%
\bibitem [{\citenamefont {Lin}(2019)}]{Lin:2019uvt}%
  \BibitemOpen
  \bibfield  {author} {\bibinfo {author} {\bibfnamefont {T.}~\bibnamefont
  {Lin}},\ }\href {\doibase 10.22323/1.333.0009} {\bibfield  {journal}
  {\bibinfo  {journal} {Proc. Sci.}\ }\textbf {\bibinfo {volume} {333}},\
  \bibinfo {pages} {009} (\bibinfo {year} {2019})},\ \Eprint
  {http://arxiv.org/abs/1904.07915} {arXiv:1904.07915 [hep-ph]} \BibitemShut
  {NoStop}%
\bibitem [{\citenamefont {Kuznetsov}\ and\ \citenamefont
  {Mikheev}(2013)}]{Kuznetsov:2013sea}%
  \BibitemOpen
  \bibfield  {author} {\bibinfo {author} {\bibfnamefont {A.}~\bibnamefont
  {Kuznetsov}}\ and\ \bibinfo {author} {\bibfnamefont {N.}~\bibnamefont
  {Mikheev}},\ }\href {\doibase 10.1007/978-3-642-36226-2} {\emph {\bibinfo
  {title} {{Electroweak Processes in External Active Media}}}},\ \bibinfo
  {series} {Springer Tracts in Modern Physics}, Vol.\ \bibinfo {volume} {252}\
  (\bibinfo  {publisher} {Springer},\ \bibinfo {address} {New York},\ \bibinfo
  {year} {2013})\BibitemShut {NoStop}%
\bibitem [{\citenamefont {Kapusta}\ and\ \citenamefont
  {Gale}(2006)}]{kapusta_gale_2006}%
  \BibitemOpen
  \bibfield  {author} {\bibinfo {author} {\bibfnamefont {J.~I.}\ \bibnamefont
  {Kapusta}}\ and\ \bibinfo {author} {\bibfnamefont {C.}~\bibnamefont {Gale}},\
  }\href {\doibase 10.1017/CBO9780511535130} {\emph {\bibinfo {title}
  {Finite-Temperature Field Theory: Principles and Applications}}},\ \bibinfo
  {edition} {2nd}\ ed.,\ Cambridge Monographs on Mathematical Physics\
  (\bibinfo  {publisher} {Cambridge University Press},\ \bibinfo {address}
  {Cambridge, United Kingdom},\ \bibinfo {year} {2006})\BibitemShut {NoStop}%
\bibitem [{\citenamefont {{Prialnik}}(2000)}]{Prialnik}%
  \BibitemOpen
  \bibfield  {author} {\bibinfo {author} {\bibfnamefont {D.}~\bibnamefont
  {{Prialnik}}},\ }\href@noop {} {\emph {\bibinfo {title} {{An Introduction to
  the Theory of Stellar Structure and Evolution}}}}\ (\bibinfo  {publisher}
  {Cambridge University Press},\ \bibinfo {address} {Cambridge, England},\
  \bibinfo {year} {2000})\BibitemShut {NoStop}%
\end{thebibliography}%

\end{document}